\documentclass[twocolumn]{aastex631}
\usepackage{mathtools}

\begin{document}

\title{Reassessing Sub-Neptune Structure, Radii, and Thermal Evolution}

\author[0000-0003-3980-7808]{Yao Tang}
\affiliation{Department of Astronomy and Astrophysics, University of California, Santa Cruz \\
1156 High Street, Santa Cruz, CA 95064, USA}

\author[0000-0002-9843-4354]{Jonathan J. Fortney}
\affiliation{Department of Astronomy and Astrophysics, University of California, Santa Cruz \\
1156 High Street, Santa Cruz, CA 95064, USA}

\author[0000-0003-3573-5915]{Francis Nimmo}
\affiliation{Department of Earth and Planetary Science, University of California, Santa Cruz \\
1156 High Street, Santa Cruz, CA 95064, USA}

\author[0000-0002-5113-8558]{Daniel Thorngren}
\affiliation{Department of Physics and Astronomy, Johns Hopkins University, Baltimore, MD 21218, USA}

\author[0000-0003-3290-6758]{Kazumasa Ohno}
\affiliation{Division of Science, National Astronomical Observatory of Japan, 2-21-1 Osawa, Mitaka-shi, Tokyo, Japan}

\author[0000-0001-5061-0462]{Ruth Murray-Clay}
\affiliation{Department of Astronomy and Astrophysics, University of California, Santa Cruz \\
1156 High Street, Santa Cruz, CA 95064, USA}

\begin{abstract}
We present a novel python-based 1D sub-Neptune evolution model that emphasizes the thermal evolution and potential solidification of the rock/iron core and the structure of the radiative atmosphere. This model explores planetary structure from the molten center to nbar pressure levels. Treating the radiative atmosphere is crucial for sub-Neptunes, due to the large scale height and low gravity, which contributes up to 40\% of their observed radius, especially for low-mass, highly irradiated planets. Consequently, we generically find that lower H/He mass fractions are needed to match a given planetary radius, compared to previous work. While the presence of metal-enrichment in the H/He layers (here modeled as 50$\times$ solar) does not substantially influence the size of the convective envelope, it notably reduces the transit radius by shrinking the radiative atmospheric scale height. Sub-Neptunes cool differently from terrestrial planets, with the rock/iron core's cooling rate limited by the envelope, leading to longer solidification timescales. Complete solidification of the silicate mantle by 10 Gyr is found only for planets with very low masses ($\leq 1M_\oplus$) and small H/He envelopes ($\leq$ 0.1\%). Dynamo action in sub-Neptune iron cores persists as long as the mantle surface remains molten, often exceeding 10 Gyr, and becomes sensitive to core thermal conductivity after solidification.  We examine aspects of ``boil-off," which sets the maximum allowed H/He mass and planetary radius for subsequent evolution. The rock/iron's cooling energy moderately decreases the post-boil-off H/He mass fraction in planets with large atmospheric scale heights only.
\end{abstract}


\section{Introduction} \label{sec:intro}
The \emph{Kepler} and \emph{TESS} missions have revealed that close-in exoplanets smaller than Neptune but larger than Earth are ubiquitous in our galaxy. These planets are typically categorized into two groups based on their interior compositions: super-Earths, which have thin or negligible atmospheres over rocky cores, and sub-Neptunes, which have much lower bulk densities, suggesting they possess thick H/He envelopes. There is a notable scarcity of planets between $1.5 R_\oplus$ and $2 R_\oplus$, likely due to the absence of planets with very thin H/He envelopes \citep{Lopez12,Owen13,Lopez13,Lee21,Lee22}. Atmospheric escape is a key factor explaining this gap, as it can lead to a rapid transformation into super-Earths. Additionally, a decrease in H/He material can significantly shrink a small planet's radius \citep{Lopez14}. Consequently, planets that undergo atmospheric loss become substantially smaller, creating a clear separation between the two populations. 

Modeling suggests that for a sub-Neptune with a H/He envelope, the envelope mass fraction is typically a few percent \citep{Lopez12,Wolfgang15}. This means the high-density core, presumably made of rock and iron, serves as a significant energy reservoir due to its dominance in planetary mass. As the planet cools, the heat flux from the rock/iron core keeps the envelope warm, delaying thermal contraction \citep{Nettelmann11,Lopez12}. Consequently, the increased planetary radius plays a crucial role in enhancing mass loss by setting a higher density at the hydrodynamic wind base \citep{Lopez12,Ginzburg16,Owen23,Tang24}. As the envelope loses mass and the composition changes, this feedback mechanism in turn affects the thermal evolution of the planet because the envelope's contraction rate is closely linked to its energy budget. Additionally, there is likely a significant interaction between the molten rock/iron surface and the bottom of the envelope, leading to volatile exchange and chemical reactions \citep{Chachan18,Kite19,Kite20,Schlichting22,Seo2024}. Furthermore, \citet {Vazan18a} shows that the coupling between the solidification of the rock/iron and the envelope evolution leads to a prolonged cooling timescale compared to terrestrial planets. Therefore, the presence of the rock/iron core significantly impacts major features of a sub-Neptune.

Since the pioneering \citet{Nettelmann11} sub-Neptune model for GJ 1214b, many subsequent models have adopted similar simplifications for the rock/iron core \citep{Lopez12,Owen13,Ginzburg16,Chen16,Vazan18a}. These models often assume an isothermal temperature profile for the rock/iron core, set by the temperature at the bottom of the envelope. This simplification contrasts with more realistic adiabatic temperature profiles. This assumption underestimates the rock/iron core's role as an energy reservoir because, in reality, an adiabatic deep center would typically be several times hotter than the mantle surface. As the H/He envelope cools, these models treat the thermal energy stored in the isothermal rock/iron core as being released instantly, leading to an overestimation of the core luminosity. This overestimation is particularly significant when the rocky interior solidifies, as solid convection is much less efficient than magma convection. Furthermore, the latent heat, which substantially contributes to the energy reservoir \citep{Vazan18a}, is often ignored. These treatments introduce significant uncertainty in predicting the radius evolution and cooling timescales of sub-Neptunes.

To improve our understanding of sub-Neptune interiors, more sophisticated models are necessary. This requires a more accurate treatment of energy transport mechanisms, including convection, conduction, and latent heat release, as well as a detailed understanding of the phase transitions within the rock/iron core. These refinements would reduce the uncertainties and lead to more reliable predictions of planetary radius and cooling timescales.

The radiative atmosphere is another layer that is often treated with simplified conditions, or completely ignored, in structure calculations due to its negligible mass and being decoupled from the thermal evolution. However, \citet[hereafter TFM24]{Tang24} showed a very large contrast in the population distributions of sub-Neptunes after Gyrs of thermal evolution between  models with and without the radiative atmosphere in the structure calculation, strongly suggesting that more care is needed. Simplifying assumptions include an isothermal temperature profile, a lower boundary condition fixed at a constant pressure, ignoring the radiative atmospheric mass, and using a constant surface gravity throughout the whole radiative atmosphere. While these assumptions are considered to be reasonable for many giant planets and Earth-like terrestrial planets, they may potentially be invalid for sub-Neptunes due to their low density and low gravity. The calculation of the radiative atmosphere is intrinsically connected to the cooling of the planet's interior, as the latter sets the lower boundary conditions for the radiative atmosphere's thermal structure. This interplay between the radiative atmosphere and the interior significantly impacts the planet's overall radius evolution. As a result, accurately modeling the radiative atmosphere, alongside interior cooling, is critical for predicting sub-Neptune characteristics more precisely.

A ``boil-off'' phase \citep{Owenwu16} occurs when the rapid decline of confinement pressure from protoplanetary disks causes a planet’s atmosphere to experience hydrostatic disequilibrium, triggering a tremendous hydrodynamic outflow. This happens when the planet is newly formed, still hot, and has an extended atmosphere that is only weakly gravitationally bound. The boil-off phase typically occurs within the first one to two Myr after the inner disk density rapidly diminishes, a process that happens on a timescale of roughly $10^5$ years \citep{Ercolano10}, usually when the disk is a few Myr old \citep{Mamajek09, Gorti2016}. TFM24 demonstrate that boil-off is the dominant mechanism driving mass loss, with time-integrated mass loss often exceeding 90\% of the initial envelope mass. This rapid, energy-intensive process critically shapes the observed sub-Neptune population by significantly affecting the planet's thermal and structural conditions after the boil-off phase. For a more accurate understanding of sub-Neptune evolution, refined boil-off mass loss calculations are essential.

Photoevaporation \citep{Lammer03,Baraffe04}, which TFM24 suggests only removes the last 10\% of the initial H/He mass, is another type of hydrodynamic outflow that critically impacts sub-Neptune evolution over a longer timescale \citep{Lopez12,Owen12,Lopez13,Owen13,Jin14,Chen16,Kubyshkina18,Jin18,Rogers23}. Photoevaporation is powered by stellar ionizing radiation that is absorbed high in a planet's atmosphere. Accurate predictions of planetary radii at lower pressure levels are essential to understanding this process, as they influence the rate of atmospheric mass loss and determine whether sub-Neptunes retain their gas envelopes or evolve into rocky super-Earths.

Therefore, in this work, we develop a state-of-the-art Python-based sub-Neptune evolution model that focuses specifically on the thermal and physical state evolution of the rock/iron core and the structure calculation of the radiative atmosphere. In the framework of \citet{Benneke24}, we investigate ``gas dwarf" (low-metallicity H/He envelopes) and ``miscible envelope'' (H/He with a well-mixed water mass-fraction) scenarios for these planets.  We incorporate a boil-off phase to constrain the initial physical conditions (H/He mass fraction, envelope specific entropy and the rock/iron core's physical states) for late-stage evolution. We focus on the post-boil-off thermal evolution without considering photoevaporation, as photoevaporation is largely dependent on our new radius assessment, which we will study in future work. With this model, we explore the following:
\begin{itemize}
\item How metals in the envelope impact the physical size of the envelope and its thermal evolution in, Section \ref{res:metal}.
\item The solidification and cooling processes of sub-Neptune rock/iron interiors and compare this to those of terrestrial planets, in Section \ref{res:core}.
\item The significance of various physical effects, including the role of the temperature profile, mean molecular weight, and gravity, in the radiative atmosphere, in Section \ref{res:rad}.
\item A direct comparison to previous models that assume simplified rock/iron core and radiative atmosphere, in Section \ref{res:comparison}.
\item A reassessment of boil-off with the improved rock/iron core evolution, in Section \ref{disc:boil-off}.
\item The implications for dynamo operation, in Section \ref{disc:dynamo}.
\end{itemize}

\section{Model description} \label{sec:model}
\subsection{Physical Structure}
\label{subsec:structure}
To model the structure of sub-Neptunes, our new python-based evolution model comprises four one-dimensional spherically symmetric regions, arranged from the interior to the exterior: iron core, silicate mantle, H/He envelope, and radiative H/He atmosphere. Here, the term ``iron core'' exclusively refers to the metallic central interior, analogous to its usage in geophysics. In contrast, we occasionally use the term ``rock/iron core'' to describe the entire high-density interior (encompassing both the iron core and silicate mantle) in comparison to the low-density H/He envelope.

The pressure and radius profiles are calculated with respect to fixed mass shells for the iron core, silicate mantle and H/He envelope (collectively referred to as the interior), under the assumption of hydrostatic equilibrium::
\begin{equation}
\label{masscon}
\frac{dr}{dm} = \frac{1}{4\pi r^2\rho}
\end{equation}

\begin{equation}
\label{hydrostatic}
\frac{dP}{dm} = -\frac{Gm}{4\pi r^4}
\end{equation}
where $r$, $\rho$ and $P$ represent radius, density and pressure at a certain mass shell, respectively, and $m$ denotes the mass enclosed within that mass shell. To improve boundary condition assessments and enhance computational efficiency, we adopt a higher spatial resolution by using smaller mass shells ($dm$) at larger radii. The number of cells in each structural layer is carefully selected to ensure that the physical structures converge across different resolution choices. The governing equations are solved using a relaxation method, where the initial guess is taken from the converged solution of the preceding timestep. The relaxation method follows an iterative approach, repeatedly integrating Eq. \ref{masscon} and Eq. \ref{hydrostatic} until the relative errors are reduced to below 0.1\%.

For the radiative atmosphere, we calculate a hydrostatic structure using fixed discretized pressure grids rather than mass shells, as the majority of the planetary mass is concentrated in the interior, making the self-gravity effect negligible. This is achieved by combining Eq. \ref{masscon} and \ref{hydrostatic} to eliminate $dm$ and substituting the mass $m$ with the total planetary mass $M_{\rm{p}}$. In cases where the atmosphere is unstable and vulnerable to hydrodynamic atmospheric escape, we apply a steady state isothermal Parker wind to improve the assessment of the physical structure. The importance of this treatment is evaluated in Section \ref{res:rad}.

\begin{figure}
\centering
\includegraphics[width=0.5\textwidth]{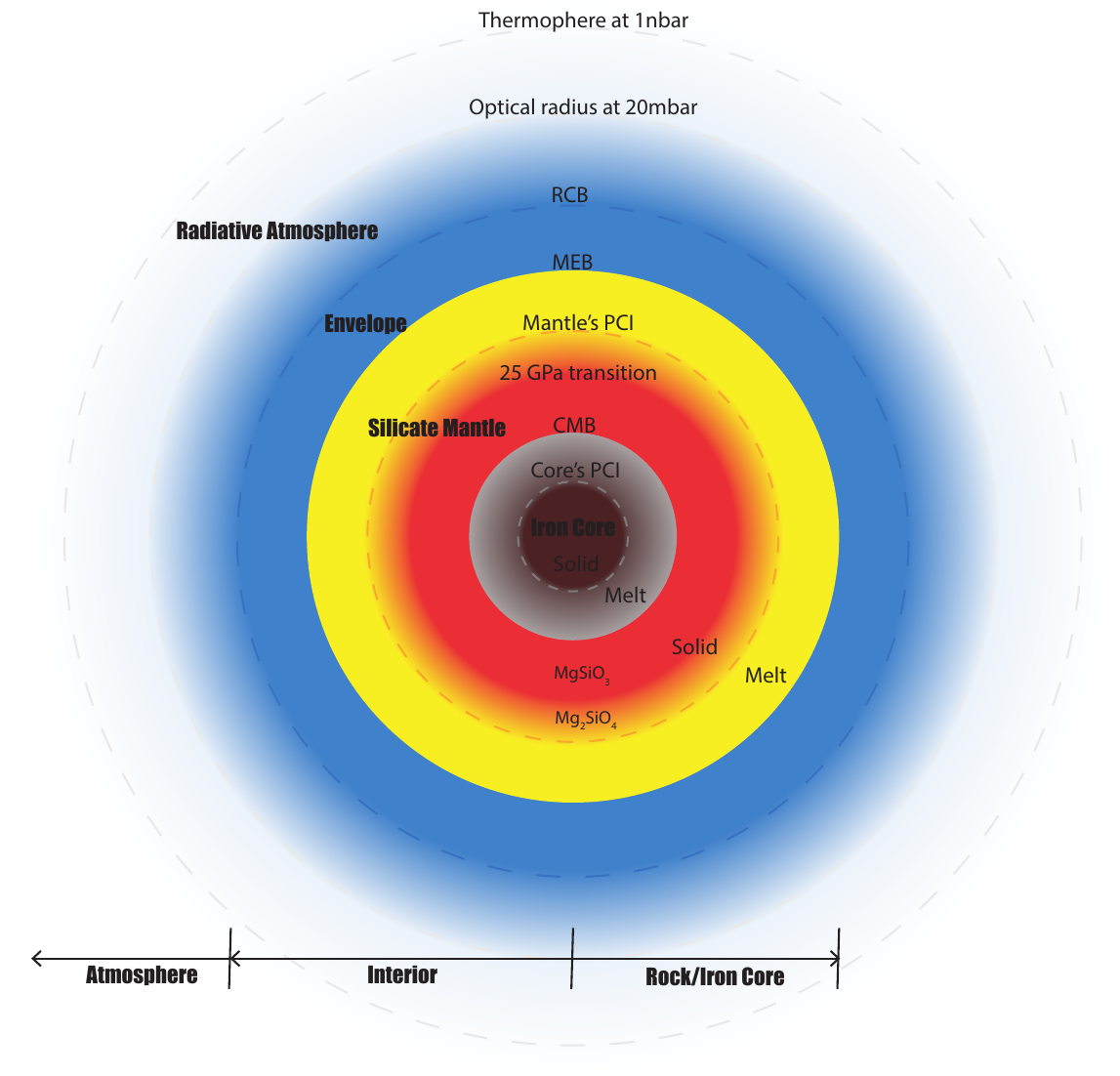} 
\caption{ This schematic is designed to emphasize the essential interfaces, terminologies, and structural components. The relative size of each layer is not to scale. Key locations include the radiative-convective boundary (RCB), the mantle-envelope boundary (MEB), the core-mantle boundary (CMB) and the phase change interfaces (PCI) within both the mantle and core. Additionally, the diagram marks the olivine-to-magnesium perovskite transition at 250 kbar (25 GPa), the optical transit radius at 20 mbar, and the thermosphere, characterized by significant EUV absorption. Our definitions of the rock/iron core, planetary interior, and atmosphere are also provided for clarity, as these may differ from those used in other studies.
}
\label{diagram}
\end{figure}

A schematic representation of the structural layers and key interfaces is provided in Figure \ref{diagram}. In the following subsections, we elaborate on the details of the structure calculation and thermal and physical state evolution for each of these layers.

\subsection{H/He Envelope}
The H/He envelope is assumed to be adiabatic and isentropic, as efficient mixing due to envelope convection is expected even during the most vigorous phases of mass loss (TFM24). This treatment has been widely employed in models of sub-Neptune and giant planet evolution \citep{Fortney07,Lopez14,Owen13}. The thermal state and physical structure of the envelope are primarily characterized by the specific entropy $s$ and the envelope mass. In our default model setup the envelope is assumed to consist of pure H/He, with the Equation of State (EOS) taken from \citet{Chabrier19} with a He solar mass fraction of $Y=0.275$.

The cooling and heating of the H/He envelope determines the amount of gravitational binding energy and internal energy that can be released or gained, influencing whether the H/He envelope contracts or inflates based on the total heat transfer to it. A sub-Neptune is presumed to be born in a hot initial thermal state, due to the primordial heat generated from the planet formation process. The cooling rate $L_{\rm{RCB}}$ at the top interface of the envelope, known as the radiative-convective boundary (RCB), typically dominates over the heating $L_{\rm{MEB}}$ received from the iron/rock core at the bottom, referred to as the mantle-envelope boundary (MEB). As a result, a sub-Neptune generally undergoes thermal contraction throughout most of the evolution. Previous models \citep{Nettelmann11,Lopez14,Rogers24,Tang24} did not consider thermal inflation (our term from TFM24 for the increase in H/He envelope specific entropy), where $L_{\rm{MEB}}>L_{\rm{RCB}}$. However, studies \citep{Ginzburg16,Ginzburg18} argue that the total thermal energy available from the core and mantle can exceed the gravitational binding energy of the envelope, potentially allowing an envelope to thermally inflate, which makes a sub-Neptune more susceptible to Parker wind outflows. A similar phenomenon occurs with XUV-driven escape, which keeps a planet's interior warm and its size large for a long period of time, leading to amplified mass loss \citep{Lopez12}. This phenomenon, which we term core-enhanced mass loss effect, can also affect the late stage boil-off. In this case, a significant decrease in envelope temperature due to adiabatic expansion liberates substantial thermal energy from the rock/iron core, which is expected to be coupled with the mass loss process (TFM24). However, these authors found this effect to be minor for the time-integrated mass loss during boil-off due to its short duration. In this work, we reexamine this physical effect within the context of improved core-mantle evolution modeling. 

The radiative intrinsic luminosity, which is the total energy that a planet's interior radiates per unit time through its radiative atmosphere into space, is given by $L_{\rm{int}}=4\pi R_{\rm{RCB}}^2 \sigma T_{\rm{int}}^4$, where $\sigma$ is Stefan–Boltzmann constant. This luminosity governs the energy transport rate at the RCB, denoted by $L_{\rm{RCB}}$. To assess the intrinsic temperature $T_{\rm{int}}$, and consequently $L_{\rm{RCB}}$, a grid of one-dimensional atmospheric radiative transfer models is employed, by default using solar metallicity for pure H/He envelopes. This evaluation is performed using linear interpolation based on surface gravities, incident bolometric fluxes, and interior temperatures, as described in \citet{Fortney07}.

Between each timestep, we consider a quasi-static evolution, by tracking the heat transfer at the top and bottom of the envelope, which alters the interior thermal states. The isentropic envelope relates the change of specific entropy $\Delta s$ of the envelope adiabat to the total heat transfer $\Delta Q = L_{\rm{env}} \Delta t = \Delta s \int T dm$ over the time interval $\Delta t$, leading to the following expression:
\begin{equation}
\label{thermalcontraction}
\frac{ds}{dt} \int_{M_{{\rm{ric}}}}^{M_{{\rm{ric}}}+M_{\rm{env}}}  T\,dm = - L_{\rm{RCB}} + L_{\rm{MEB}}
\end{equation}
where $M_{\rm{ric}}$ and $M_{\rm{env}}$ are the masses of the rock/iron core and the envelope, respectively, and $T$ is the temperature of each mass shell $dm$. On the right-hand side, the total envelope luminosity $L_{\rm{env}}$ is the sum of the envelope cooling $L_{\rm{RCB}}$ and heating $L_{\rm{MEB}}$. The numerical value of $L_{\rm{MEB}}$ depends on the strength of mantle convection, which is explained in Section \ref{subsec:mantle}.

To evolve the envelope specific entropy, we utilize a fifth-order ODE solver, given the $ds/dt$ calculated from Eq. \ref{thermalcontraction} at each interior structure calculation. In the presence of boil-off, the planetary mass $M_{p}$ is evolved simultaneously in a similar manner to the entropy, given the mass loss rate $\dot{M}$. The envelope mass is given by:
\begin{equation}
\label{envmass}
M_{\rm{env}} = M_p - M_{\rm{ric}} - M_{\rm{atm}}
\end{equation}
where $M_{\rm{atm}}$ is the time-dependent radiative atmospheric mass from the radiative atmosphere calculation. The structure calculation is iterated a few times per timestep until convergence is achieved. The time intervals are chosen to be 1/100 of the smallest timescale among the envelope-mantle-core thermal evolution processes and mass loss. The resulting specific entropy and envelope mass automatically define a new hydrostatic equilibrium of the envelope at the new timestep, based on Eq. \ref{masscon} and \ref{hydrostatic}.

\subsection{Radiative Atmosphere}
\label{subsec:rad}
Above the RCB, the radiative atmosphere is treated as distinct from the interior. In this region, radiation dominates the energy transport and thus controls the temperature-pressure (T-P) profile. Therefore, we assume that its radial structure contracts or inflates passively as the interior undergoes thermal evolution, based on the physical conditions at the RCB. The energy sources for radiation include the visible stellar bolometric energy incident from above and the thermal intrinsic cooling energy emitted from the interior below. The calculation of the radiative atmosphere is essential for evaluating the RCB location, the optical transit radius, and the mass contained within the radiative atmosphere at each structural calculation during a timestep.

Sub-Neptunes' intrinsic flux is usually negligible compared to stellar bolometric flux. Compared to giant planets, sub-Neptunes have cooler interiors due to their smaller energy reservoirs and relatively fast cooling rates, which are facilitated by their lower gravity. This combination of low RCB temperature and rapid cooling leads to a deeper RCB location (at a higher pressure), making the location of the RCB important in the radius calculation. Our numerical model estimates the RCB location by finding the joint point in pressure space between the adiabatic temperature profile of the H/He envelope and the radiative atmospheric profile. 

\begin{figure}
\centering
\includegraphics[width=0.45\textwidth]{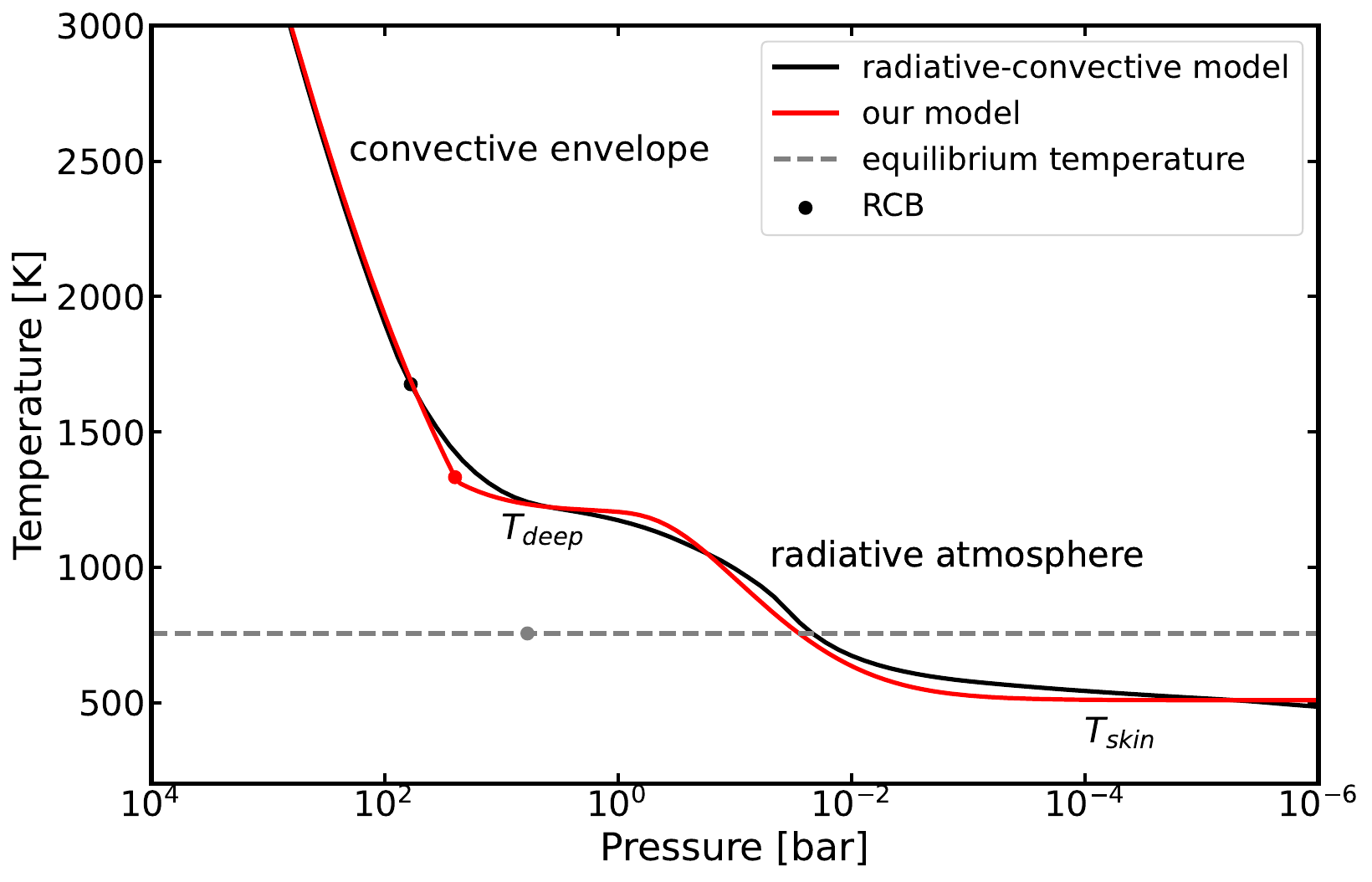} 
\caption{ A close-up view of the top of the convective H/He envelope and radiative atmosphere. For a complete view, see the bottom left panel of Figure \ref{evolution}. We compare the $T-P$ profiles and RCB locations derived from non-gray 1D radiative-convective atmosphere modeling (black) with those from our implementation (red). For reference, we also include the isothermal temperature profile (gray dashed) commonly used in previous sub-Neptune models. This plot is for a planet with $T_{\rm{int}}=150$K, $F_{\rm{bol}}=100F_\oplus$ and gravity of 200 $cm\ s^{-2}$. See text for discussion.
}
\label{rcb_comp}
\end{figure}

This approach results in a deeper RCB location compared to models that assume an isothermal atmosphere. While this depth still remains a slight underestimate relative to the true RCB determined by radiative-convective equilibrium modeling \citep{Ohno&Fortney23}, the agreement is quite good. An example is shown in Figure \ref{rcb_comp}, where we compare the T-P profiles and RCB locations from our model (red) with those from radiative-convective modeling (black) and an isothermal atmosphere model (gray dashed).

\begin{figure*}
\centering
\includegraphics[width=0.9\textwidth]{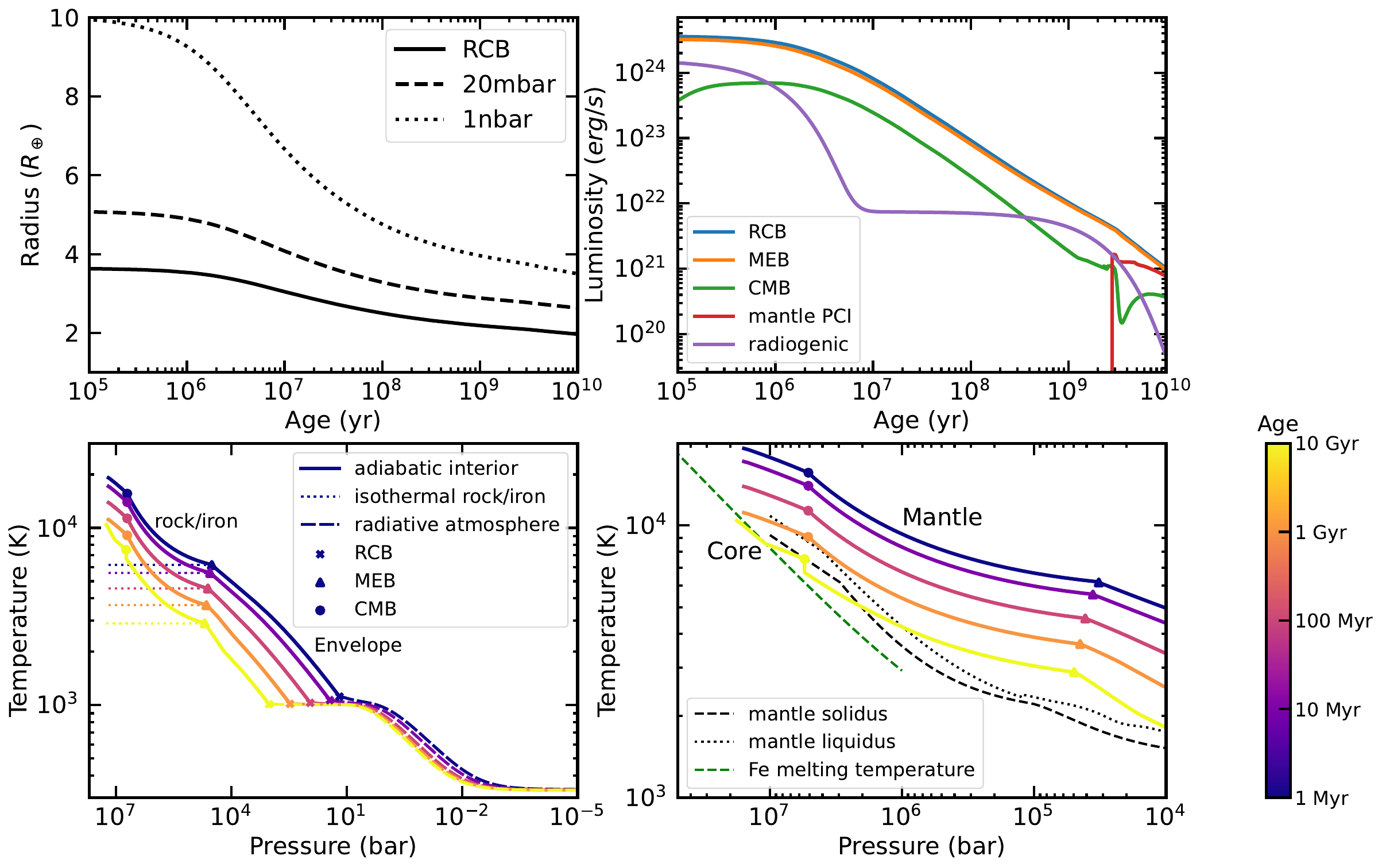} 
\caption{ An example of the evolutionary trajectory for a 4 $M_\oplus$ planet with 2\% H/He mass irradiated with 10 
$F_\oplus$. The top left panel shows the planetary radii at the RCB (solid), 20 mbar (dashed), and 1 nbar (dotted) pressure levels. The top right panel presents the energy budget as the luminosities at each boundary layer, including the mantle’s phase change interface (PCI), and the contribution from the radiogenic heat. Planetary age is defined in the model as the time elapsed since the onset of inner disk dispersal. The bottom panels display the temperature-pressure profile at various ages. In the bottom panels, we mark the radiative-convective boundary (RCB, cross), mantle-envelope boundary (MEB, triangle), and core-mantle boundary (CMB, circle). In the bottom left panel, the radiative atmosphere is shown in dashed colors and the isothermal rock/iron core assumed in earlier models is represented with a dotted line. The bottom right panel zooms in on the rock/iron interior, focusing on its detailed thermal and structural evolution. It highlights phase transitions within the planet's core and mantle. For comparison, we include the mantle solidus (dashed black), liquidus (dotted black), and iron melting temperature (dotted green). 
}
\label{evolution}
\end{figure*}

The RCB pressure of a sub-Neptune undergoes significant changes over time, ranging from around $\sim$ 10 bar in its early stages to 10 kbar in ages exceeding a few Gyrs. As a sub-Neptune ages, its internal temperature decreases, resulting in a reduction of its intrinsic luminosity. Consequently, the disparity between the intense flux emitted by the host star and the diminishing intrinsic flux of the planet widens, causing the RCB to extend deeper into the planet. In Figure \ref{evolution}, we show the locations of the RCBs on the T-P profiles of a sub-Neptune in cross shapes (bottom), and the thermal contraction of the interior in solid curves (top left).

Remarkably, our analysis shows that the radiative atmosphere can constitute a significant portion of the optical radius of low-mass sub-Neptunes due to their weakened gravitational pull, resulting in a decrease in gravity with altitude and consequent increase in scale height outward. This leads to a large planetary radius (top left panel of Figure \ref{evolution}) and low density, which we term the variable gravity effect.  In TFM24 we found this effect to be pivotal for modeling sub-Neptune population studies. Our planetary radius is defined at a pressure level of 20 mbar, typically associated with optical transits of sub-Neptunes. Another significant pressure level for radius assessment occurs at 1 nbar, where the atmosphere becomes optically thick to EUV photons \citep{RMC09}. The radius at 1 nbar determines the amount of high-energy photons a planet receives and thereby influences the intensity of the post-boil-off XUV-driven escape. Moreover, the nbar pressure levels characterize the transit radius for a hazy atmosphere, which is essential for explaining very low-density sub-Neptunes \citep{Gao20}.

To precisely evaluate these radii, we employ a comprehensive temperature profile as a function of thermal optical depth $\tau$, derived from an analytical two-stream radiative transfer model by \citet{Guillot10}:
\begin{equation}
\begin{split}
\label{radiativePT}
T = \Bigl\{\frac{3T_{\rm{int}}^4}{4}\left(\frac{2}{3} +\tau\right)+\frac{3T_{\rm{irr}}^4}{4}f\Bigl[\frac{2}{3}+\frac{1}{\gamma\sqrt{3}} \\ +\left(\frac{\gamma}{\sqrt{3}}-\frac{1}{\gamma\sqrt{3}}\right)e^{-\gamma\tau\sqrt{3}}\Bigr]\Bigr\}^{1/4}
\end{split}
\end{equation}
where $T_{\rm{irr}}=\left(F_{\rm{bol}}/\sigma\right)^{1/4}$ is irradiation temperature and $F_{\rm{bol}}$ is the incident stellar bolometric flux. Time variability in bolometric flux is neglected, as it is generally minor for a main-sequence sun-like star. $f=0.25$ is a constant parameter that controls the angular location where the temperature profile is calculated (our choice corresponds to a global average). 

The temperature profile is primarily influenced by the opacity ratio $\gamma = \kappa_\nu/\kappa_{\rm{th}}$ where $\kappa_\nu$ and $\kappa_{\rm{th}}$ are the opacities to the visible stellar radiation and outgoing thermal radiation, respectively. We assume $\kappa_{th}$ to be a constant value of 0.02 $\rm{cm^2\ g^{-1}}$. To determine $\kappa_\nu$, we compare the analytical model’s $T-P$ profile to full radiative-convective equilibrium models for solar composition atmospheres \citep[e.g.,][]{Marley&McKay99,Fortney+05,Fortney20,Marley&Robinson15,Thorngren19,Ohno&Fortney23}, fitting the best values as a function of the incident bolometric flux. The values show weak dependence on surface gravity and intrinsic luminosity from the interior, as the radiative transfer is primarily driven by re-emitted stellar energy. Note that these $\kappa$ terms do not enter into the thermal evolution calculation, just into assessing shape of the radiative $T-P$ profile. The fit demonstrates very good agreement between the analytical model (red) and the numerical model (black) in Figure \ref{rcb_comp}.

The temperature profiles of the radiative atmosphere typically exhibits two distinct isothermal regions: an outer ``skin temperature'' layer ($T_{\rm{skin}}$), which is optically thin to both incoming optical radiation and outgoing infrared radiation, and a deeper layer ($T_{\rm{deep}}$), characterized by the absorption of incoming optical radiation that is in equilibrium with the optically thick infrared radiation. This is shown in Figure \ref{rcb_comp}.

To transform the temperature profile from optical depth $\tau$ to pressure space, we combine the definition of $\tau$:
\begin{equation}
\label{dtau}
d\tau = -\rho \kappa_{\rm{th}}\, dr
\end{equation}
and Eq. \ref{hydrostatic}, resulting in:
\begin{equation}
\label{tau}
\tau \approx \frac{\kappa_{\rm{th}}}{g(r)} P = \frac{\kappa_{\rm{th}}r^2}{G M_{\rm{p}}} P
\end{equation}
where a radially variable gravity $g(r)$ is considered. We found $\tau$ primarily depends on the local pressure, with minimal contribution from the lower-pressure upper layers, simplifying the expression to a linear form in terms of pressure in Eq. \ref{tau}. The radius structure is then computed by integrating Eq. \ref{hydrostatic} starting from the inner boundary at the RCB. The gas density $\rho$ is evaluated using the ideal gas law $\rho = P\mu m_{\rm{H}}/(RT)$ above 1 bar pressure level, where $\mu$ is the mean molecular weight and $m_{\rm{H}}$ is the mass of atomic hydrogen. Below the 1 bar level, we apply the H/He EOS to account for non-ideal effects. Importantly, molecular dissociation into atomic hydrogen due to photolysis \citep{Kempton12}, occurring at pressure levels between 1 $\mu$bar - 10 nbar, significantly impacts the radius structure of the upper atmosphere. We model the transition of $\mu$ linearly in log pressure from 2.35 to 1.28 for H/He between $10^{-6}-10^{-8}$ bar.

In addition, we calculate the total mass of the atmosphere contained within the hydrostatic radiative layer at each time step by integrating the mass over its respective cells:
\begin{equation}
\label{atmmass}
M_{\rm{atm}} = \int_{R_{\rm{RCB}}}^{R_{\rm{s}}}  4\pi r^2 \rho\,dr
\end{equation}
where $R_{\rm{s}}$ is the sonic radius, defined as:
\begin{equation}
\label{sonicpoint}
R_s = \frac{GM_p}{2c_s^2}
\end{equation}
where $c_s={[kT_{\rm{eq}}/(\mu m_H)]}^{1/2}$ is sound speed and $T_{\rm{eq}}=[F_{\rm{bol}}/(4\sigma)]^{1/4}$ is equilibrium temperature. This radiative atmospheric mass, which does not actively contribute to the thermal evolution of the H/He envelope, is excluded from the assessment of the envelope mass. This exclusion allows for a more self-consistent assessment of the thermal evolution of the planet's interior. 

We find that, for most of the evolutionary history, the proportion of radiative atmospheric mass typically remains minimal, comprising only a few 0.01\% of the total planetary mass. As such, it usually has little impact on the overall thermal evolution. However, a notable exception arises. As a sub-Neptune ages, the RCB penetrates more deeply into the interior, resulting into a larger radiative atmospheric mass fraction. This effect becomes especially pronounced for a highly irradiated planet, which tends to have a deeper RCB location and a more depleted envelope as a result of boil-off. Consequently, the radiative atmospheric mass constitutes most of the total H/He mass. When this occurs, thermal evolution of the envelope greatly slows down, leading the planet to transition into an ``envelope-free'' super-Earth. We terminate thermal evolution when the envelope mass $M_{\rm{env}}$ vanishes, after which the planet is treated as a super-Earth with a stationary radiative atmosphere overlying its core-mantle mixture. This transition is important for low-mass, highly irradiated planets with small envelopes after 1 Gyr.

The radiative atmospheric mass, $M_{\rm{atm}}$, implicitly depends on $M_{\rm{env}}$ because $M_{\rm{env}}$ sets the lower boundary condition at the RCB, which is essential for the calculation of $M_{\rm{atm}}$. As the radiative atmospheric mass varies slowly compared to the thermal contraction, the model assumes for computational efficiency that $M_{\rm{atm}}$ remains constant within the smallest time interval of the ODE solver. The radiative atmospheric mass is then updated only at the end of each timestep, ensuring computational precision without excessive resource expenditure.

\subsection{Metals in the Envelope and Atmosphere}
Beyond a setup for pure H/He envelopes in the previous section, we also investigate the role of metallicity on planetary evolution and structure. Our model is capable of simulating metal-enhanced evolution for up to 100 times solar metallicity. In this study, we default our metal-enhanced models to 50 times solar metallicity.

First, we expand our atmosphere grids to assess the intrinsic luminosity \citep{Lopez14,Ohno&Fortney23} at the metal-rich conditions that may be appropriate from some sub-Neptunes \citep{Fortney13,Benneke24}. The format is similar to that of our default solar metallicity grid. We interpolate within these atmosphere grids, covering [1, 3, 10, 30, 50, 100] times solar metallicity, to evaluate the luminosity at various metallicity values. Moreover, to account for the metals in the envelope, we choose to mix in a $Z$ mass fraction of water in the envelope. We use the additive volume law to obtain the thermodynamic properties, including the specific entropy and density at a certain pressure and temperature. A correction term is incorporated in calculating the entropy of the mixture to recover the ideal mixing case at low pressures and temperatures \citep{Chabrier19}.

We use the water equation of state of \citet{mazevet19}, which covers a wide range of physical states, including precisely calculated liquid, gas, and plasma phases and approximated ice and superionic phases. 

The mass fractions and mean molecular weights at a given metallicity are self-consistently calculated. At 50$\times$ solar, the hydrogen, helium and water mass fractions are X = 0.4317, Y = 0.1638 and Z = 0.4045, respectively. (The hydrogen to helium mass ratio is the same as our pure H/He mixture.) These parameters yield an increase in the mean molecular weight of $\mu=3.58$ (from 2.35) for the molecule-dominated region including the envelope and lower radiative atmosphere and $\mu=1.93$ (from 1.28) for the upper atmosphere where $\rm{H_2O}$ is dissociated into atomic H and OH radicals. To account for the non-ideal effect in the deep atmosphere and photodissociation in the upper atmosphere, we apply the same approaches as used in the pure H/He case. Unless specified, our numerical results are calculated with our default pure H/He setup.

\subsection{Initial States from a Boil-Off Calculation}
\label{subsec:boil-off}
Previous models of planetary evolution typically start with an arbitrary hot initial thermal state, often referred to as a``hot start'' \citep{Marley07,Lopez12}. This approach allows the planet to cool and evolve to thermal state that is insensitive to the initial condition in 10-100 Myr. However, this introduces large uncertainties in determining the planet's thermal behavior during its early evolutionary stages, a period often coupled with XUV-driven mass loss. The initial H/He mass fraction is usually treated as a free parameter. However, this approach lacks strong constraints, making it difficult to accurately predict the thermal history of planets.

To address these uncertainties and to accurately assess the H/He mass fraction for subsequent evolution, a boil-off phase is incorporated as a means to constrain the initial conditions of planetary evolution. In this work, we explore two methods for initializing model planets: one that incorporates a boil-off phase to determine self-consistent initial (post-boil-off) envelope mass fraction for initially gas-rich planets, and one that treats the initial mass fraction as free parameters for initially gas-poor planets. As boil-off may not occur for planets born with minimal H/He mass, the final mass fraction evaluated from a boil-off phase sets an upper limit for the envelope mass a planet may have at the beginning of thermal contraction. In the later section, we call the evolution with boil-off-constrained H/He mass fraction ``post-boil-off evolution''.  

Sustaining the hydrodynamic wind during boil-off requires energy to generate a pressure gradient strong enough to overcome gravity. The primary energy source for boil-off is the bolometric luminosity received from the star, with a small but important portion coming from the radiative cooling of the planet's envelope (TFM24). The adiabatic and quasi-hydrostatic conditions in the interior still hold during this process, due to efficient convection. More details are available in TFM24.

In previous studies, boil-off was modeled using the isothermal Parker wind solution \citep{Owenwu16}, which represents a steady-state hydrodynamic outflow from the planet's atmosphere \citep{Parker1958}. We adopt this approach in our work, incorporating energy considerations into the model. This model is particularly applicable after inner disk dispersal, a process that happens on a very short timescale ($\sim10^5$ years), significantly shorter than both the planet's thermal contraction and the duration of the boil-off phase. Conversely, if the mismatch between the disk's confinement pressure and the wind pressure at the sonic point is insufficient, the outflow can slow down and transition to a non-steady-state condition, which may become relevant during the disk dispersal phase. The disk dispersal phase is neglected here due to its short duration. This choice is discussed in Section \ref{disc:boil-off}. The Parker wind equation:
\begin{equation}
\label{parkerwind}
\frac{1}{2} \left( 1- \frac{c_s^2}{v^2}\right)\frac{d}{dr}(v^2) = -\frac{GM_{\rm{p}}}{r^2} \left(1 - \frac{2c_s^2r}{GM_{\rm{p}}}\right)
\end{equation}
describes the velocity $v$ of the wind as it escapes the planet's gravitational influence, where $c_s$ is the sound speed, $r$ is the radial distance, and $G$ is the gravitational constant. Notably, at the sonic radius $R_s$, the wind speed reaches $c_s$, and the equation simplifies as both sides vanish.
 
The wind advection is significant in the radiative zone, where it dominates both mass and energy transport, while convection takes over below the RCB (TFM24). Consequently, the wind base is positioned at the RCB, where the physical conditions determine the mass loss rate via the steady-state outflow. This rate is quantified by:
\begin{equation}
\label{masslossrate}
\dot{M}_{\rm{parker}} = 4 \pi R_{\rm{RCB}}^2 \rho_{\rm{RCB}} v_{\rm{RCB}} = 4 \pi r^2 \rho v
\end{equation}
Here, $R_{\rm{RCB}}$, $\rho_{\rm{RCB}}$, and $v_{\rm{RCB}}$ represent the planetary radius, density, and wind velocity at the RCB, calculated from the interior structure model and Eq. \ref{parkerwind}. The wind density and pressure profiles as a function of radius are derived using the second equality of Eq. \ref{masslossrate} and applying the ideal gas law.

The Parker wind solution becomes crucial in determining the optical transit radius during the boil-off phase, rather than relying on hydrostatic assumptions (see Section \ref{subsec:rad}). Although the post-boil-off bolometric-driven Parker wind becomes inefficient after the planet's initial stage of mass loss (TFM24), the Parker wind model continues to be relevant. By running the Parker wind calculation in parallel with the radiative atmosphere model after boil-off, we can estimate the wind velocity and evaluate how securely the remaining radiative atmosphere is gravitationally bound to the planet. 

Eq. \ref{masslossrate} represents the maximum possible mass loss rate given sufficient energy input to drive the outflow. However, energy limitations can often create bottlenecks for this process. As shown by TFM24, a bottleneck typically occurs in the deep radiative atmosphere when a critical optical depth to visible radiation $\tau_h$ forms above the RCB. This effectively shields the interior from stellar energy, limiting the available energy to drive mass loss. In this case, the mass loss rate becomes energetically constrained by the interior cooling luminosity $L_{\rm{int}}$, rather than the outflow dynamics alone:
\begin{equation}
\label{mdot_int}
\dot{M}_{\rm{e-lim,\tau_h}} = \frac{L_{\rm{int}}}{GM_p/R_{\rm{RCB}} - GM_p/R_{\tau_h}}
\end{equation}
Here, the location of $\tau_h$, denoted as $R_{\tau_h}$, is computed by identifying the radius at which stellar energy absorption equals the energy dissipation via PdV work in the deep radiative atmosphere.

For very low-mass planets ($< 2M_\oplus$), absorption of bolometric energy can be effectively deposited down to the RCB. However, this energy supply may not be sufficient in the upper atmosphere. In such cases, the mass loss rate is energy-limited by the total amount of stellar heating:
\begin{equation}
\label{mdot_bol}
\dot{M}_{\rm{e-lim,bol}} = \frac{L_{\rm{bol}}}{G M_p/R_{\rm{RCB}}}
\end{equation}
where $L_{\rm{bol}}=4\pi R_{\rm{RCB}}^2 F_{\rm{bol}}$ is the bolometric luminosity evaluated at the RCB. 

The total boil-off mass loss rate is then determined by:
\begin{equation}
\label{mdot_elim}
\dot{M} = \min(\dot{M}_{\rm{parker}},\dot{M}_{\rm{e-lim,bol}},\dot{M}_{\rm{e-lim,\tau_h}})
\end{equation}
This comprehensive calculation is shown to be unnecessary if boil-off mass loss is significantly decoupled from thermal evolution (TFM24), since equations \ref{masslossrate}, \ref{mdot_int}, and \ref{mdot_bol} ultimately converge to the same final mass fraction after boil-off. However, this needs to be reassessed with the improved rock/iron evolution, as the previous models \citep{Owenwu16,Rogers24} underestimate the amount of rock/iron energy release. This may lead to thermal inflation (TFM24), a process potentially coupled with mass loss, impacting post-boil-off mass fractions. Thus, we incorporate Equation \ref{mdot_elim}.

The boil-off mass loss is terminated when the thermal contraction timescale of the planetary envelope, $t_{\rm{cool}}$, becomes comparable to the mass loss timescale, $t_{\dot{M}}$:
\begin{equation}
\label{KHtime}
t_{\rm{cool}} = \frac{U}{ \left|L_{\rm{RCB}}-L_{\rm{MEB}}\right|}
\end{equation}

\begin{equation}
\label{losstime}
t_{\dot{M}} = \frac{M_{\rm{env}}}{\dot{M}}
\end{equation}
where $U$ is the total gravitational binding energy of the envelope. After the transition, the time-integrated mass loss due to the Parker wind is negligible.

The specific envelope entropy at the beginning of boil-off is not well-constrained observationally, but it is chosen such that the thermal contraction timescale at $t=0$ (Eq. \ref{KHtime}) is comparable to the disk's lifetime \citep{Owen13,Tang24}, generally a few Myr. For consistency, a timescale of 5 Myr is used across all models. This parameterization has minimal impact on planets without a boil-off phase, as it corresponds closely to a ``hot start.'' It is important to note that for planets with a boil-off phase, variations in the initial contraction timescale (within 3-10 Myr) influence the post-boil-off mass fraction, as higher initial entropy increases susceptibility to mass loss. Determining the optimal value for this parameter is beyond the scope of this study.

To compute the initial contraction timescale, we assume thermal equilibrium between the envelope and mantle, thus neglecting energy contributions from the rock/iron core and setting $L_{\rm{MEB}}=0$ in Eq. \ref{KHtime}. This assumption holds for most model planets, since the rock/iron core contributes minimally to the total energy with the gravitational binding energy of the envelope dominating. However, for planets with extremely low envelope masses ($\sim$ 0.1\%), the envelope's energy reservoir $E_{\rm{env}}$ can be much smaller than that of the rock/iron core $E_{\rm{ric}}$, making the contraction timescale sensitive to core and mantle heat production. This includes radiogenic heating from 
${}^{26}\mathrm{Al}$ and other heating sources not included in the model (which are likely on the order of $E_{\rm{ric}}$), such as solid mass accretion and rock/iron differentiation, which are beyond the scope of this model. For these cases, our above assumption likely underestimates the interior temperature, leading to partial core solidification at $t=0$. To correct for this, we assume a higher initial envelope specific entropy, ensuring a fully molten interior state at the beginning of thermal evolution. The potential role of core-mantle differentiation is further discussed in Section \ref{disc:inter}.

The initial H/He mass fraction for the boil-off phase is taken from formation models \citep{Ginzburg16}, around 10-20\%. Careful evaluation of the initial conditions reveals that variations in the starting H/He mass fraction within this range minimally affect the final mass fraction (which is also suggested by TFM24 where thermal inflation was not considered). This is because (1) the thermal evolution and mass loss processes are largely decoupled throughout most of the boil-off phase; (2) a more massive envelope experiences a less intensive inflation due to its larger envelope energy reservoir $E_{\rm{ric}}$, which offsets its hotter initial entropy (which encourages more mass loss). 

The initial states of a planet's mantle and core are governed by the envelope's initial conditions, based on the assumption that there are no significant temperature discontinuities at the thermal boundary layers between these regions. This assumption holds because the temperature contrasts at the boundaries are found to be relatively small for a newly born planet, owing to vigorous liquid convection within the rock/iron interior, which ensures an adiabatic profile throughout the whole interior. 


\subsection{Iron Core}
In this model, the iron core is divided into two distinct layers: a solid inner core and a liquid outer core, each governed by their own EOS. Both layers follow an adiabatic temperature gradient, which assumes efficient heat transport through conduction (and convection in the liquid iron) that minimizes the temperature jump across the phase transition boundary between solid and liquid states. The temperature is assumed to be continuous through the solid-liquid interface.

The adiabatic temperature gradient is described by:
\begin{equation}
\label{adiabaticeq}
\left(\frac{\partial T}{\partial P} \right)_{\rm{s}}  = \frac{\alpha T}{\rho c_{\rm{P}}}
\end{equation}
where $c_{\rm{P}}$ is the specific heat capacity at constant pressure, $\alpha$ is the thermal expansivity. The specific heat capacity $c_{\rm{P}}$ is considered constant across the core (with the value and reference summarized in Table \ref{table:val}), while $\alpha$ varies with radius and is locally evaluated using the relation:
\begin{equation}
\label{alpha}
\alpha = \frac{\gamma\rho c_{\rm{V}}}{K_{\rm{T}}}
\end{equation}
In this equation, $\gamma$ is the Grüneisen parameter, $c_{\rm{V}}$ is the specific heat capacity at constant volume, and 
$K_{\rm{T}}$ is the isothermal bulk modulus, which is calculated from:
\begin{equation}
\label{KT}
K_{\rm{T}} = \rho\frac{\partial P}{\partial \rho} \bigg|_{\rm{T}}
\end{equation}
Additionally, the relation between $c_{\rm{V}}$ and 
$c_{\rm{P}}$ is given by:
\begin{equation}
\label{cv}
c_{\rm{V}} = \frac{c_{\rm{P}}}{1+\gamma\alpha T}
\end{equation}
The thermal expansivity $\alpha$ plays a crucial role in determining the slope of the T-P profile and thus influences the timing of the core's solidification. $\alpha$ typically increases by a factor of 2-3 from the deepest location in the iron core to the core-mantle boundary (CMB). We find that for an Earth-mass planet $\alpha$ is around $1\times10^{-5}$, while for more massive planets like a Neptune-sized one ($\sim 20 M_\oplus$), $\alpha$ can drop to less than $1\times10^{-6}$. This variability becomes especially important for the mantle, where even small changes in 
$\alpha$ can significantly affect the solidification process and shift the cooling timescales.

The Grüneisen parameter $\gamma=-(\partial \ln{T}/\partial \ln{V})_s$ is a crucial dimensionless thermodynamic quantity that describes the efficiency of thermal expansion in the iron core and mantle, thereby influencing the thermal profile of these layers. Several empirical formulations of $\gamma$ \citep{Mosenfelder09} have been proposed to align with experimental data. The simplest approach is a power-law function of density $\rho$, but more sophisticated models, such as the Al'tshuler form \citep{Altshuler}, capture the gradual changes in $\gamma$ under high compression conditions. In this work, we utilize the Al'tshuler formulation given by:
\begin{equation}
\label{gamma}
\gamma = \gamma_\infty + (\gamma_0-\gamma_\infty)\left(\frac{\rho}{\rho_0}\right)^{-q}
\end{equation}
where $\gamma_0$ and $\gamma_\infty$ represent the Grüneisen parameter at zero and infinite compression, respectively, while $\rho_0$ is a reference density. The parameter $q$ is empirically determined and governs how $\gamma$ changes with increasing density. The values of these parameters are sourced from various literature for each structural component and are detailed in Table \ref{table:EOS}.  

To determine the physical state of the iron core, we compare the calculated temperature profile with a melting curve, designating regions of the T-P profile that exceed the melting temperature as liquid and \textit{vice versa}. We use Simon's equation to parameterize the melting temperature of the core:
\begin{equation}
\label{melting}
T_{\rm{melt}} = T_{\rm{ref}}\left(\frac{P}{a} + 1\right)^b
\end{equation}
where the reference temperature $T_{\rm{ref}}$ and the coefficients $a$ and $b$ are summarized in Table \ref{table:melt}. To incorporate the effects of impurities, primarily sulfur, we adjust the melting temperature of pure iron by a specific fraction. Previous studies \citep{Stevenson83, Stixrude14, Zhang22} on Earth's interior evolution suggest using 80\% of the pure iron melting temperature to accurately match the size of Earth's inner liquid iron, which we adopt as our default value for super-Earths and vary to examine its influence on the solidification timescale in Section \ref{res:core}. 

Given the interdependence of the physical states, temperature profiles and physical structures, we employ the iterative relaxation method described in Section \ref{subsec:structure} for self-consistent calculations of all these variables. This relaxation approach is crucial due to the sensitive nature of the solidification process.

Density at each mass shell in each compositional layer is derived from the EOS, which relates density to pressure at a reference temperature $T_0$. The density calculation also depends weakly on the temperature profile due to thermal perturbations. Thus, the full EOS is expressed as:
\begin{equation}
\label{perturbation}
P(\rho,T) = P(\rho,T_0) + P_{\rm{th}}(\rho,T) - P_{\rm{th}}(\rho,T_0)
\end{equation}
where $P_{\rm{th}}$ represents the temperature-dependent thermal pressure. The thermal pressure terms are calculated using the Mie-Grüneisen-Debye model, expressed as:
\begin{equation}
P_{\rm{th}}(\rho,T) = \gamma \rho E_{\rm{th}}(\rho, T)
\end{equation}
where the parameters for $E_{\rm{th}}$ are sourced from the references listed in Table \ref{table:EOS}. The density $\rho(P,T)$ is determined through linear interpolation of Eq. \ref{perturbation}. By integrating Eq. \ref{masscon} and \ref{hydrostatic}, we obtain the radius and pressure distributions.

For computational efficiency, we have switched off the temperature-dependent thermal pressure when generating the model results. This simplification leads to an overestimation of density and an underestimation of temperature at low pressure levels and high temperatures by less than 5\%. The resulting impact on the planetary radius is minimal, typically on the order of 0.1\%, which is insignificant since the overall radius of most sub-Neptunes is primarily determined by the H/He envelope and radiative atmosphere, rather than the rock/iron core.

For solid iron, we utilize the SESAME2140 EOS \citep{Lyon1992}, while for liquid iron, we adopt the third-order Birch-Murnaghan (BM3) formulation \citep{Birch1952}:
\begin{equation}
\label{bm3}
\begin{split}
P(\rho) = \frac{3K_0}{2}\left[\left(\frac{\rho}{\rho_0}\right)^{\frac{7}{3}}-\left(\frac{\rho}{\rho_0}\right)^{\frac{5}{3}}\right] \Biggl\{1+\\\frac{3}{4}(K^\prime_0-4)\left[\left(\frac{\rho}{\rho_0}\right)^{\frac{2}{3}}-1\right]\Biggr\}
\end{split}
\end{equation}
This choice of EOS allows for accurate modeling of the pressure-density relationship in both solid and liquid states of iron, essential for understanding the thermal and structural properties of the core.

The thermal evolution of the core adiabat is closely tied to both cooling processes and heat production. Between two hydrostatic timesteps, we solve the energy budget equation to assess the cooling rate:
\begin{equation}
\label{ironevolution}
\begin{split}
\int_{0}^{M_{\rm{c}}}  c_{\rm{Pc}}\dot{T}\, dm -  \int_{0}^{M_{\rm{c}}}  \frac{\alpha T}{\rho}\dot{P}\, dm \\ = c_{\rm{Pc}} \dot{T}_{\rm{c}} \int_{0}^{M_{\rm{c}}} \frac{\partial T}{\partial T_{\rm{c}}} \bigg|_{P}\, dm = - L_{\rm{CMB}} + L_{\rm{Fe}}
\end{split}
\end{equation}
where $M_{\rm{c}}= f_{\rm{Fe}} * M_{\rm{ric}}$ represents the iron core mass, which is a function of the iron core mass fraction, $f_{\rm{Fe}}$, and the total mass of the rock/iron core $M_{\rm{ric}}$. The term $c_{\rm{Pc}}$ refers to the specific heat capacity at constant pressure for iron.

The left-hand side of the equation consists of two terms: the first term represents the change in total enthalpy, while the second accounts for the energy contribution from decompression. Here, $\dot{T}$ and $\dot{P}$ indicate the rates of change in temperature and pressure within each mass shell, respectively. On the right-hand side, we include the cooling at the CMB, denoted by  $L_{\rm{CMB}}$, which is constrained by mantle convection rather than heat transport within the iron core. Additionally, we account for the heating produced during iron solidification, represented as $L_{\rm{Fe}}=l_{\rm{Fe}} \dot{M}_{\rm{Fe,l}}$, where $l_{\rm{Fe}}$ is the specific latent heat of iron and $\dot{M}_{\rm{Fe,l}}$ denotes the rate of change of liquid iron mass. 

It is important to note that we have neglected the contribution of gravitational energy release due to the contraction of the iron core. This treatment is based on our finding that the radius contraction rate is minimal over Gyr timescales ($\sim$1\%), resulting in gravitational energy release contributing at most 10\% of the energy flux at the CMB. This energy release is negligible compared to the cooling energy $L_{\rm{int}}$ that drives envelope contraction.

It is convenient to project the temperatures of each mass shell to a specific pressure level along the adiabatic temperature profile, which simplifies the left-hand side of the equation to focus solely on the potential temperature, denoted as $T_c$, for the iron core. This leads to the middle equality in the equation. 

The $\partial T/\partial T_{\rm{c}}$ term can be readily derived from Eq. \ref{adiabaticeq}. The rate of change in the iron potential temperature,
$\dot{T}_{\rm{c}}$, is closely linked to the time derivatives of physical properties at the CMB:
\begin{eqnarray}
\label{temperaturedot}
\dot{T}_{\rm{c}} &=& \frac{\partial T_{\rm{c}}}{\partial T_{\rm{CMB,c}}}\bigg|_{P_{\rm{CMB}}} \dot{T}_{\rm{CMB,c}}+ \frac{\partial T_{\rm{c}}}{\partial P_{\rm{CMB}}}\bigg|_{T_{\rm{CMB,c}}} \dot{P}_{\rm{CMB}} \nonumber \\
&=& \dot{T}_{\rm{CMB,c}} - \frac{\alpha_{\rm{CMB,c}} T_{\rm{CMB,c}}}{\rho_{\rm{CMB,c}} c_{\rm{Pc}}} \dot{P}_{\rm{CMB}}
\end{eqnarray}
where $T_{\rm{CMB,c}}$, $\alpha_{\rm{CMB,c}}$, $\rho_{\rm{CMB,c}}$ are the temperature, thermal expansivity and density of the iron core at the CMB. The second equality holds when the potential temperature $T_c$ is defined at the CMB pressure for each timestep for computational convenience. The choice of this reference pressure does not influence the final result. By tracking $T_{\rm{CMB,c}}$, we can accurately evolve the iron core's thermal state along the adiabat. The second term in the middle, associated with the pressure change at the CMB, $\dot{P}_{\rm{CMB}}$, is updated during each hydrostatic calculation. Although this pressure change typically evolves over a long timescale and contributes minimally to the energy budget during the thermal contraction phase, it becomes significantly impactful during periods of boil-off. Rapid depressurization due to the removal of envelope mass can substantially influence the interior energy budget, highlighting the necessity of this decompression term.

\subsection{Silicate Mantle}
\label{subsec:mantle}
To calculate the physical and thermal structure of the silicate mantle, we divide this region into up to three adiabatic layers depending on the physical state: the magma ocean, the upper solid mantle and the lower solid mantle. Unlike terrestrial planet evolution models, which require a non-adiabatic crust layer to match the hot interior adiabat to the cold surface conditions, sub-Neptunes likely lack such a substantial, complex surface layer due to the blanketing effect of the H/He envelope. This envelope sets the mantle surface temperature above 1000 K, rendering a crust layer unnecessary. In the current model we do not treat the possibility of a composition gradient from silicate vapor escaping the magma ocean, which may influence the envelope's temperature profile \citep{Misener22}. 

The adiabatic temperature gradients for the silicate mantle are calculated in the same manner as for the iron core, using Eq. \ref{adiabaticeq}-\ref{gamma}. The parameters for the EOS of each layer are sourced from analytical fits to molecular dynamics simulations or shock wave experiments, as shown in Table \ref{table:EOS}. The available data for the EOS fits typically extend up to 250 GPa. However, the pressure in a sub-Neptune's mantle often exceeds this, reaching up to 1000 GPa at the CMB. To account for these higher pressures, we use the EOS fits for the liquid and solid phases of $\rm{MgSiO_3}$ that have shown good extrapolation behavior because they are consistent with thermodynamic principles even in these high-pressure regimes \citep{Cohen00,Stamenkovic11,WolfBower2018,Zhang22}. Although using these extrapolated EOS fits introduces some uncertainty in the precise timing of solidification, it does not significantly affect the overall trends and conclusions of our model.

The EOS for the magma ocean in the upper mantle is taken from \citet{WolfBower2018}, using the Vinet form \citep{Vinet89}:
\begin{equation}
\begin{split}
\label{vinet}
P(\rho) = 3K_0 \left(\frac{\rho}{\rho_0}\right)^{\frac{2}{3}} \left[1-\left(\frac{\rho}{\rho_0}\right)^{-\frac{1}{3}}\right] \exp\Biggl\{\frac{3}{2}(K^\prime_0-1) \\ \left[1-\left(\frac{\rho}{\rho_0}\right)^{-\frac{1}{3}}\right]\Biggr\}
\end{split}
\end{equation}
where $K_0$, $K^\prime_0$ and $\rho_0$ are the model parameters. 

For the solid mantle, the dominant mineral types vary with pressure. Below 25 GPa, the lower mantle consists primarily of Mg-perovskite ($\rm{MgSiO_3}$), while the upper solid mantle is composed of olivine ($\rm{Mg_2SiO_4}$). For simplicity, we divide the solid-phase mantle into two distinct layers to account for the pressure-dependent variation in the Mg:Si ratio, each with a fixed composition. However, it is important to acknowledge that a more complex composition, including a variable Mg:Si ratio and the presence of chemical impurities, could significantly influence interior dynamics, resulting in varied thermal evolution. Addressing these complexities will be an important focus for future studies. 

The EOS for Mg-perovskite is based on Keane's formulation \citep{Keane54}, which reduces uncertainties at high pressures compared to the BM3 model:
\begin{equation}
\label{keane}
P(\rho) = K_0\left\{\frac{K^\prime_0}{K_0^{\prime\prime2}}\left[\left(\frac{\rho}{\rho_0}\right)^{K_0^{\prime\prime}}-1\right]-\left(\frac{K^\prime_0}{K_0^{\prime\prime}}-1\right)\ln{\left(\frac{\rho}{\rho_0}\right)}\right\}
\end{equation}
where $K_0$, $K^\prime_0$, $K_0^{\prime\prime}$ and $\rho_0$ are the model parameters. For the upper mantle, the EOS follows the BM3 formulation (Eq. \ref{bm3}).

At the beginning of the evolution, the silicate mantle is fully molten. As the planet cools, solidification initiates from the bottom, and the phase change interface (PCI) progressively moves outward. The PCI is defined as the location where the local melt fraction $\phi$ approaches 0.5, marking the transition from a predominantly solid to a molten state.

The local melt fraction characterizes the physical state of each mass shell, representing the mass fraction of molten minerals when impurities are present. This melt fraction is solved simultaneously with the hydrostatic equations during each relaxation step, along with thermal and physical quantities such as temperature $T$, density $\rho$, pressure $P$, and radius $r$. The mantle material is fully solid below the solidus temperature $T_{\rm{sd}}$ and fully liquid above the liquidus temperature $T_{\rm{ld}}$. Between the solidus and liquidus, the mantle exists in a mushy phase, where the melt fraction is approximately proportional to the dimensionless homologous temperature $T_{\rm{H}}$ \citep{Iwamori95}, given by:
\begin{equation}
\label{meltfraction}
\phi = 
\begin{dcases}
    T_{\rm{H}}(p) \equiv \frac{T-T_{\rm{sd}}(p)}{T_{\rm{ld}}(p)-T_{\rm{sd}}(p)},& \text{if } T_{\rm{sd}}(p) < T < T_{\rm{ld}}(p)\\
    0,              & \text{if } T \leq T_{\rm{sd}}(p)\\
    1,              & \text{if } T \geq T_{\rm{ld}}(p)
\end{dcases}
\end{equation}

The solidus for the upper mantle (pressures $<$ 10 GPa) is based on experiments on peridotite \citep{Hirschmann2000}. Since the liquidus data from the same experiments carry significant uncertainty, we assume the liquidus is 200 K above the solidus. At higher pressures, the solidus and liquidus are derived from experimental data on chondrites \citep{Andrault11}. For pressures exceeding 300 GPa, where experimental data is unavailable, the liquidus is taken from the \textit{ab initio} melting curve of pure Mg-perovskite \citep{Belonoshko05}, with the solidus assumed to be 85\% of the liquidus temperature. The T-P relationship of the solidus and liquidus are shown in the bottom right panel of \ref{evolution}. We integrate the local melt fraction $\phi$ over each mass shell to determine the total masses of the solid and molten silicates, denoted as $M_s$ and $M_m$, respectively. 

The temperature difference across the interface between the two solid layers (the upper and lower solid mantle) is assumed to be zero, meaning they evolve together as a single thermodynamic entity. We evolve the magma ocean adiabat and solid mantle adiabat separately due to their contrasting rates of heat transport. The temperature jump at the PCI governs the energy transfer rate between the solid and liquid layers. Nevertheless, our results indicate that this temperature jump is relatively small ($\lesssim$ 100K) compared to that at the CMB (exceeding 1000K). Consequently, the mantle bottom, rather than the PCI, acts as the rate-limiting layer for energy transport in the mantle.

Similar to the treatment of the iron core, the energy balance equations for the magma ocean and solid mantle are given by:
\begin{equation}
\begin{split}
\label{magmaevolution}
\int_{M_{\rm{Fe}}+M_{\rm{s}}}^{M_{\rm{Fe}}+M_{\rm{s}}+M_{\rm{m}}}  c_{\rm{Pm}} \frac{\partial T}{\partial T_{\rm{m}}} \dot{T}_{\rm{m}}\, dm \\ = - L_{\rm{MEB}} + L_{\rm{PCI}} + \frac{M_{\rm{m}}}{M_{\rm{Si}}}\left( L_{\rm{radio}} + L_{\rm{Si}}\right)
\end{split}
\end{equation}

\begin{equation}
\label{solidevolution}
\int_{M_{\rm{Fe}}}^{M_{\rm{Fe}}+M_{\rm{s}}}  c_{\rm{Pm}} \frac{\partial T}{\partial T_{\rm{s}}} \dot{T}_{\rm{s}}\, dm = - L_{\rm{PCI}} + L_{\rm{CMB}} + \frac{M_{\rm{s}}}{M_{\rm{Si}}}\left( L_{\rm{radio}} + L_{\rm{Si}}\right)
\end{equation}
where $M_{\rm{Si}}=M_s+M_m$ represents the total silicate mass. In these equations, the rate of change in the potential temperatures of the magma ocean, $T_m$, and solid mantle, $T_s$, are evaluated similarly to Eq. \ref{temperaturedot}. $T_c$, $T_m$ and $T_s$ are evolved simultaneously with $s$ and $M_{\rm{env}}$ using the ODE solver, based on Eq. \ref{ironevolution}, \ref{magmaevolution} and \ref{solidevolution}. On the right hand sides, the energy transport rates ($L_{\rm{CMB}}$, $L_{\rm{PCI}}$ and $L_{\rm{MEB}}$) at each boundary layers are included, alongside heat production from silicate solidification $L_{\rm{Si}} = l_{\rm{Si}} \dot{M}_{\rm{s}}$ and radioactive decay $L_{\rm{radio}}$. Similar to the iron core, we find that the gravitational energy release from mantle contraction contributes minimally to heating. An example of the evolutionary trajectory of these luminosities is shown in the top right panel of Figure \ref{evolution}. $L_{\rm{radio}}$ is evaluated with:
\begin{equation}
\label{radiogenic}
L_{\rm{radio}} = M_{\rm{Si}}\sum q_i\,  2^{-(t+\tau_{\rm{CAI}})/\tau_i}
\end{equation}
where $q_i$ and $\tau_i$ are the heat production rate and half-life of each radiogenic isotope (${}^{26}\mathrm{Al}$, ${}^{40}\mathrm{K}$, ${}^{232}\mathrm{Th}$, ${}^{235}\mathrm{U}$ and ${}^{238}\mathrm{U}$), with their values summarized in Table \ref{table:iso}. The heat production is assumed to be uniformly distributed in the mantle, proportional to the mass of each layer.

Though the short-lived isotope ${}^{26}\mathrm{Al}$ has little impact on long-term thermal evolution, it can significantly influence early planet structure, particularly enhancing thermal inflation of the H/He envelope for low-mass planets experiencing vigorous boil-off mass loss, thereby affecting time-integrated mass loss. $\tau_{\rm{CAI}}$ sets the the beginning time of our evolution $t_0$, following the formation of calcium-aluminum inclusions (CAIs), and is roughly comparable to the disk lifetime. It determines the abundance of ${}^{26}\mathrm{Al}$ at $t_0$ and, consequently, the intensity of the radiogenic heat production at early times ($\sim$ Myrs).

The heat transport process in the mantle is dominated by mantle convection. To evaluate the luminosity at each boundary layer, we employ parameterized convection, which models the process using scaling laws derived from fluid dynamics instead of solving the full set of fluid dynamic equations. This approach has been widely used for Earth and other solar system bodies \citep[e.g.,][]{Turcotte79, McKenRicht:1981,Nimmo2000,Nimmo2004}.

The most common approach to parameterized convection focuses on the thermal boundary layers at the top and bottom of the convective region. In these layers, conduction dominates over convection, making them the rate-limiting factors for heat transfer. These boundary layers are well-described by simple scaling laws \citep{Howar:1964,Solom:1995}, allowing us to bypass the need for resolving the entropy gradient of the interior and instead solve for the superadiabatic part of the temperature gradient. Comparisons between parameterized convection models and full numerical simulations show good agreement \citep{VanKe:2001}.  

Our implementation assumes that the superadiabatic temperature gradient concentrates in a thin thermal boundary layer, which leads to a temperature jump between structural layers, and thus the conduction equation gives the heat transport rate:
\begin{equation}
\label{flux}
L_{\rm{CMB}} = 4\pi R_{\rm{CMB}}^2 F_{\rm{CMB}}= 4\pi R_{\rm{CMB}}^2 \frac{k \Delta T_{\rm{CMB}}}{\delta}
\end{equation}
where $k$ is thermal conductivity of the mantle, whose value varies depending on the physical state and the pressure level. $\Delta T_{\rm{CMB}}=T_{\rm{CMB},c}-T_{\rm{CMB},m}$ sets the temperature gradient at the boundary.  The thickness of the thermal boundary layer $\delta$ is parameterized using scaling laws. Two regimes are considered here. The mantle convects in the so-called ``hard turbulence" regime \citep{ShraiSiggi:1990} in the liquid state ($\phi>\phi_{\rm{crit}}$), because of the low viscosity and large thickness of the magma ocean. For the solid convection ($\phi \leq \phi_{\rm{crit}}$), $\delta$ is evaluated using the scaling law for the local Rayleigh number $Ra$, which approaches the critical Rayleigh number $Ra_c$ at the thermal boundary \citep{Nimmo2000,Nimmo2004}. These give:
\begin{equation}
\label{delta}
\delta = 
\begin{dcases}
    d/Nu = d/(0.3 Ra^{2/7}),     & \text{if } \phi >\phi_{\rm{crit}} \\
    \left[\frac{Ra_c\kappa_m \eta}{\rho g\alpha\Delta T_{\rm{CMB}}}\right]^{1/3},     & \text{if } \phi \leq \phi_{\rm{crit}}\\
\end{dcases}
\end{equation}

\begin{equation}
\label{ra}
Ra = \frac{\rho g \alpha \Delta T_{\rm{CMB}} d^3}{\kappa_m \eta}
\end{equation}
where $\rho$ is local density, $g$ is local gravity, $\alpha$ is local thermal expansivity, $\kappa_m$ is the thermal diffusivity of the mantle, $\eta$ is the dynamic viscosity of the mantle, and $d$ is the thickness of the magma ocean. The critical melt fraction $\phi_{\rm{crit}}$ is chosen to be 0.5, inspired by \citet{Costa09,Bower18}. The Nusselt number $Nu$ parameterizes the increased heat transfer due to convection and depends on $Ra$.

Adopting the \citet{ShraiSiggi:1990} parameterization yields the second equality in the first regime. In the above equations, we take the CMB as an example and similar implementation is also applied to the energy transport through other thermal boundaries including the MEB and mantle PCI. We find that the temperature jump at the boundary is usually very small, on the order of 1 K for the liquid convection, and is much more substantial, up to a few thousand K, for the solid convection.

The temperature jump at the thermal boundary and the vigor of the convective heat transport are greatly influenced by the mantle viscosity $\eta$. Given the high efficiency of magma convection, resulting in a negligible temperature gradient, we find that the choice of magma viscosity $\eta_m$ does not significantly affect the thermal evolution and temperature profiles. Therefore, we treat $\eta_m$ as a constant. 

The solid viscosity $\eta_s$ varies significantly with temperature and pressure. We model it using the Arrhenius form for post-perovskite from \citet{Tackley13}:
\begin{equation}
\label{solidviscosity}
\eta_s = \eta_0 \exp{\left[ \frac{H_a(p)}{R T}-\frac{H_a(0)}{R T_{0}}\right]}
\end{equation}

\begin{equation}
\label{enthalpy}
H_a(p) = E_a + pV_a(p)
\end{equation}

\begin{equation}
\label{volume}
V_a(p) = V_{a0} \exp{\left(-\frac{p}{p_{\rm{decay}}}\right)}
\end{equation}
The parameters $\eta_0$, $E_a$, $V_{a0}$ and $T_{0}$ are provided in Table \ref{table:val}. The transition from the low-viscosity magma (with $100\ \rm{Pa\ s}$) to the highly viscous solid phase (with $10^{22}\ \rm{Pa\ s}$) is quite abrupt. To capture this behavior, we adopt a hyperbolic switch function  \citep{Bower18} $w$ with a transition width $\Delta \phi=0.1$ and a critical melt fraction $\phi_{\rm{crit}}=0.5$:
\begin{equation}
\label{weight}
w = \frac{1}{2} + \frac{1}{2} \tanh{\left(\frac{\phi-\phi_{\rm{crit}}}{\Delta \phi}\right)}
\end{equation}
Finally, we combine the contributions from both the solid and magma viscosities to express the effective viscosity as:
\begin{equation}
\label{viscosity}
\log{\eta} = w \log{\eta_m} + (1-w) \log{\eta_s}
\end{equation}

With the above description, the physical structure and thermal/compositional evolution of the silicate mantle and iron core are completely defined. These processes are largely influenced by and coupled with the thermal evolution of the H/He envelope, specifically through the envelope temperature at the MEB. The thermal evolution and mass loss of the H/He envelope are, in turn, affected by the rock/iron interior, due to the substantial heat reservoir of the mantle and core.

\begin{deluxetable*}{c c c c c c}
\tabletypesize{\scriptsize}
\tablewidth{0pt} 
\tablecaption{Values of physical constants assumed}\label{table:val}
\tablehead{
 \colhead{Symbol} & \colhead{Value} & \colhead{Units}  & \colhead{Quantity} & \colhead{Eq.}  & \colhead{Reference}
} 
\startdata 
    $c_{\rm{Pm}}$ & 1200 & $\rm{J\ kg^{-1}\ K^{-1}}$ & Specific heat capacity of silicate & \ref{magmaevolution}, \ref{solidevolution} & 1 \\ [0.5 ex]
    $c_{\rm{Pc}}$ & 840 & $\rm{J\ kg^{-1}\ K^{-1}}$ & Specific heat capacity of Fe & \ref{ironevolution} & 2\\[0.5 ex]
    $l_{\rm{Si}}$ & 732.2 & $\rm{kJ\ kg^{-1}}$ & Specific latent heat of silicate & \ref{magmaevolution}, \ref{solidevolution} & 3\\[0.5 ex]
    $l_{\rm{Fe}}$ & 1200 & $\rm{kJ\ kg^{-1}}$ & Specific latent heat of Fe & \ref{ironevolution} & 4\\[0.5 ex]
    $k_{\rm{um}}$ & 5 & $\rm{W\ m^{-1}\ K^{-1}}$ & Thermal conductivity of upper mantle & \ref{flux} & 1 \\[0.5 ex]
    $k_{\rm{lm}}$ & 10 & $\rm{W\ m^{-1}\ K^{-1}}$ & Thermal conductivity of lower mantle & \ref{flux} & 1\\[0.5 ex]
    $k_{\rm{m}}$ & 4 & $\rm{W\ m^{-1}\ K^{-1}}$ & Thermal conductivity of magma & \ref{flux} & 5\\[0.5 ex]
    $k_{\rm{c}}$ & 40 & $\rm{W\ m^{-1}\ K^{-1}}$ &  Thermal conductivity of core & \ref{fluxconv} & 6\\[0.5 ex]
    $\kappa_{\rm{m}}$ & $\rm{1\times10^{-6}}$ & $\rm{m^2\ s^{-2}}$ & Thermal diffusivity of mantle & \ref{flux} & 7\\[0.5 ex]
    $\eta_{\rm{m}}$ & 100 & $\rm{Pa\ s}$ & Viscosity of magma & \ref{viscosity} & 8\\[0.5 ex]
    $Ra_{\rm{c}}$ & 600 & - & Critical Rayleigh number & \ref{delta} & 7 \\[0.5 ex]
    $\eta_0$ & $1.9\times 10^{21}$ & $\rm{Pa\ s}$ & Reference viscosity for solid mantle & \ref{solidviscosity} & 9\\[0.5 ex]
    $E_a$ & 162 & $\rm{kJ\ mol^{-1}}$ & Activation energy & \ref{enthalpy} & 9\\[0.5 ex]
    $V_{a0}$ & 1.4 & $\rm{cm^3\ mol^{-1}}$ & Reference activation volume & \ref{volume} & 9\\[0.5 ex]
    $T_{0}$ & 1600 & $\rm{K}$ & Reference temperature & \ref{solidviscosity} & 9\\[0.5 ex]
    $p_{\rm{decay}}$ & 1610 & $\rm{GPa}$ & pressure decay & \ref{volume} & 9\\[0.5 ex]
    $\tau_{\rm{CAI}}$ & 3 & Myr & Time after CAI & \ref{radiogenic} & 10\\[0.5 ex]
    $\phi_{\rm{crit}}$ & 0.5 & - & critical melt fraction & \ref{weight}\\[0.5 ex]
    $\Delta \phi$ & 0.1 & - & transition width & \ref{weight}\\[0.5 ex]
    $f_{\rm{Fe}}$ & 0.325 & - & iron core mass fraction & - & 11\\ [0.5 ex]
    $\sigma$ & $1\times{10^{6}}$ & $\rm{S\ m^{-1}}$ &  Electrical conductivity of core & \ref{rm} & 12\\[0.5 ex]
\enddata
\tablerefs{(1)\citet{Stacey92};(2)\citet{Anderson95};(3)\citet{Hess90};(4)\citet{Anderson97};(5)\citet{Bower18};(6)\citet{Konopkov16};(7)\citet{Nimmo2004};(8)\citet{Abe97};(9)\citet{Tackley13};(10)\citet{Rubie07};(11)\citet{Anderson89,Allegre95};
}
\end{deluxetable*}

\begin{deluxetable*}{c | c c c c c c}
\tabletypesize{\scriptsize}
\tablewidth{0pt} 
\tablecaption{EOS parameters for the core and mantle}
\label{table:EOS}
\tablehead{
 \colhead{Composition} & \colhead{Olivine} & \colhead{Perovskite}  & \colhead{$\rm{MgSiO_3}$ melt}  & \colhead{Liquid iron}  & \colhead{Solid iron}
} 
\startdata  
    Model & BM3 & Keane & Vinet & BM3 & SESAME2140\\ [0.5 ex]
    $\rho_0\ (\rm{g\ cm^{-3}})$ & 3.2137 & 4.1059 & 2.5844 & 7.0378 & 8.2694\\ [0.5 ex]
    $K_0\ (\rm{GPa})$ & 127.4 & 267.7 & 13.2 & 83.7 & -\\ [0.5 ex]
    $K_0^{\prime}$ & 4.2 & 4.04 & 8.238 & 5.97 & -\\ [0.5 ex]
    $K_0^{\prime\prime}$ & - & 2.63 & - & - & -\\ [0.5 ex]
    $\gamma_0$ & 1.31 & 1.506 & 0.46 & 2.033 & 1.875\\ [0.5 ex] 
    $\gamma_\infty$ & 0 & 1.148 & 0 & 0 & 1.305\\ [0.5 ex]
    $q$ & 3.2 & 7.025 & -1.35 & 1.168 & 3.289\\ [0.5 ex]
    ref. & 1 & 2 & 3 & 4 & 5
\enddata
\tablerefs{(1)\citet{Katsura09};(2)\citet{Oganov01}; \citet{Zhang22}; (3)\citet{WolfBower2018,Mosenfelder09};(4)\citet{Dorogokupets17}; (5) \citet{Lyon1992,Dewaele06}}
\end{deluxetable*}

\begin{deluxetable*}{c | c c c c}
\tabletypesize{\scriptsize}
\tablewidth{0pt} 
\tablecaption{Melting curve parameters}
\label{table:melt}
\tablehead{
 \colhead{Composition} & \colhead{Chondrite solidus} & \colhead{Chondrite liquidus}  & \colhead{Perovskite}  & \colhead{Iron} 
} 
\startdata  
    $T_{\rm{ref}}$ (K) & 2045 & 1940 & 1831 & 1900 \\[0.5 ex]
    a (GPa) & 92 & 29 & 4.6 & 31.3\\ [0.5 ex]
    b & 1.3 & 1.9 & 3 & 1.99\\ [0.5 ex]
    ref. & 1 & 1 & 2 & 3
\enddata
\tablerefs{(1)\citet{Andrault11};(2)\citet{Belonoshko05};(3)\citet{Zhang15};
}
\end{deluxetable*}
\begin{deluxetable*}{c c c c }
\tabletypesize{\scriptsize}
\tablewidth{0pt} 
\tablecaption{Parameters of radioactive isotopes}
\label{table:iso}
\tablehead{
 \colhead{Isotope} & \colhead{Power $q_i$ ($\rm{erg\ s^{-1} g^{-1}}$)} & \colhead{Half life $\tau_i$}  & \colhead{ref.} 
} 
\startdata  
    ${}^{26}\mathrm{Al}$ & $2.27\times 10^{-3}$ & 0.717 Myr & 1\\ [0.5 ex] 
    ${}^{40}\mathrm{K}$ & $4.59 \times 10^{-7}$ & 1.27 Gyr & 2\\[0.5 ex] 
    ${}^{232}\mathrm{Th}$ & $9.76 \times 10^{-10}$ & 14.05 Gyr & 2\\[0.5 ex]
    ${}^{235}\mathrm{U}$ & $3.05 \times 10^{-9}$ & 0.704 Gyr & 2\\[0.5 ex]
    ${}^{238}\mathrm{U}$ & $1.38 \times 10^{-9}$ & 4.47 Gyr & 2\\[0.5 ex]
\enddata
\tablerefs{(1)\citet{Rubie07};(2)\citet{Anders89}
}
\end{deluxetable*}

\section{Model results} \label{sec:results}
\subsection{Contraction Due to Thermal Evolution}
\label{res:contraction}
The thermal evolution history of the H/He envelope significantly influences the radii distribution of the sub-Neptune population, as planets with different physical properties (e.g., rock/iron mass, incident flux, and envelope mass fraction) contract at varying rates. The energy contribution from the rock/iron interior substantially delays thermal contraction by injecting heat into the envelope, making the evolutionary trajectory highly sensitive to the luminosity of this part of the planet. The modeling details in previous studies vary greatly, with some models either simplifying or ignoring this deep energy flux. In this section, we discuss thermal evolution and associated radius contraction with the improved rock/iron interior evolution.

\begin{figure}
\centering
\includegraphics[width=0.5\textwidth]{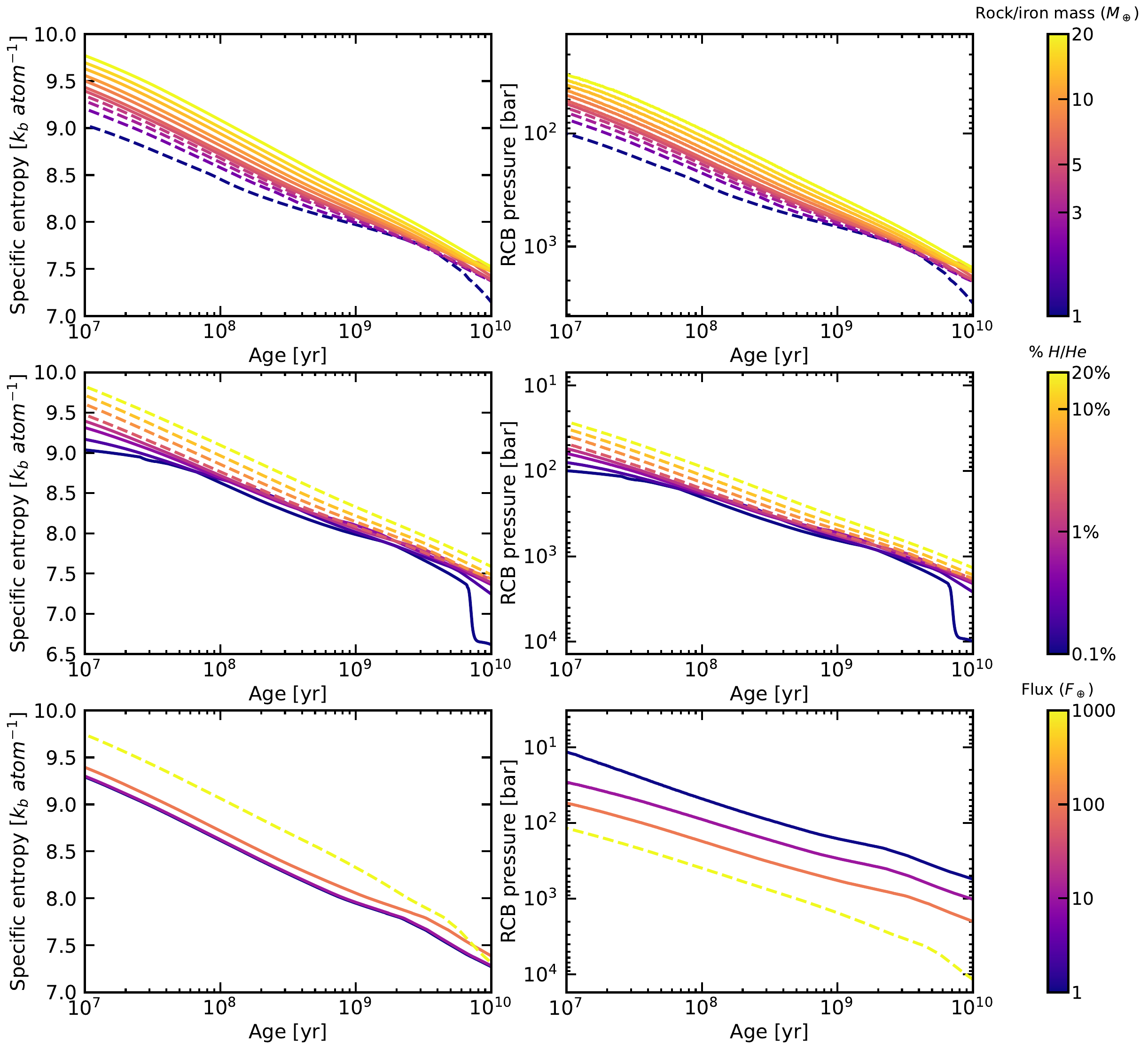} 
\caption{ The evolution of the specific entropy of the envelop (left panels) and the RCB pressure (right) as a function of rock/iron mass (top), H/He mass fraction (middle) and stellar heating (bottom). Our default model has 5$M_\oplus$, 1\% and 100$F_\oplus$. Since the boil-off phase is not included in this calculation, some planets with low masses, high insolation, and substantial envelopes are indicated as being in boil-off conditions (meaning having a H/He mass fraction higher than the values in Table \ref{table:1x}), marked by dashed lines.
}
\label{entropy-rcb}
\end{figure}

In the left panels of Figure \ref{entropy-rcb}, we show the evolution tracks of envelope specific entropy as a function of iron/rock mass, H/He mass fraction and incident bolometric flux. Our findings indicate that a sub-Neptune consistently contracts over time outside the boil-off phase, due to a decline in envelope specific entropy. We find the planets with either a more massive rock/iron core, more substantial envelope mass or a higher stellar insolation possess a warmer interior at a given age. This occurs because they retain a larger interior energy reservoir, even though their cooling luminosity is slightly higher. The difference is found to be small between planets with cold surfaces (see the $1 F_\oplus$ and $10 F_\oplus$ models in the bottom left).

The envelope specific entropy of sub-Neptunes declines steadily for most of their evolution. During this phase, the upper mantle remains in a liquid state, vigorously convecting and efficiently releasing thermal energy from the rock/iron interior. This results in a negligible temperature difference between the mantle surface and the bottom of the H/He envelope, aligning with earlier models that assume the rock/iron interior temperature is equal to the temperature at the base of the envelope. However, our improved rock/iron evolution model reveals a distinct second phase with contrasting behavior. This occurs when the surface of the mantle begins to solidify, significantly reducing the efficiency of heat transport due to the high viscosity of the solid mantle. Without adequate heat from the deep interior, the envelope contracts more rapidly (blue of Figure \ref{entropy-rcb}). This transition is abrupt, reflecting the vast difference in the efficiency of magma convection versus solid mantle convection. Importantly, though, the solid mantle convection phase is rare, appearing only in the 0.1\% H/He mass model by 10 Gyr shown in the middle left panel. 

\begin{figure}
\centering
\includegraphics[width=0.5\textwidth]{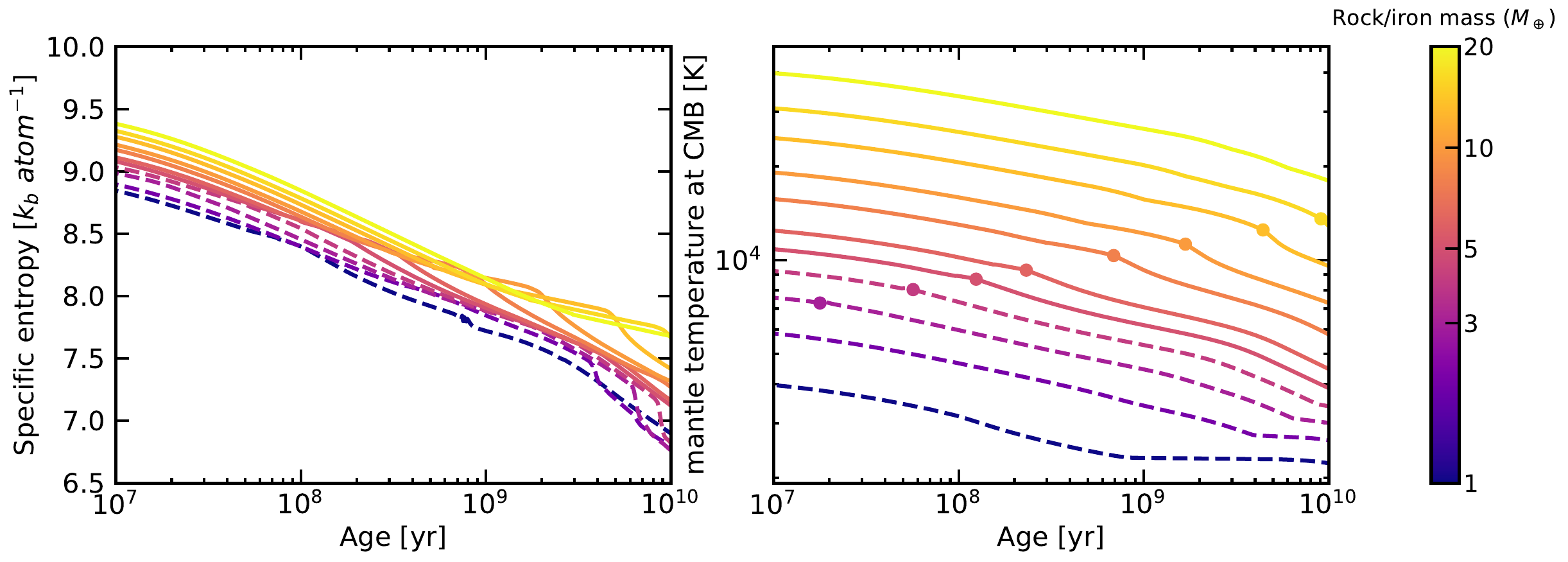} 
\caption{ The evolution of the envelope specific entropy (left) and the temperature at the bottom of the mantle (right) for planets with 0.2\% and 10$F_\oplus$. Note that the planets cool off significantly faster after solidification at the bottom of the mantle (circles), which reduces the cooling rate of the rock/iron interior at the MEB. Consequently, the envelope contraction speeds up. This effect is also observed in the 0.1\% model shown in Figure \ref{entropy-rcb}.
}
\label{entropy-temperature}
\end{figure}

In most cases, the thermal evolution of a sub-Neptune's envelope is primarily controlled by its own energy budget (as in Eq. \ref{thermalcontraction}), making it relatively insensitive to the solidification processes occurring deep within the rock/iron interior. However, exceptions arise for planets with both a substantial rock/iron interior($\geq 8 M_\oplus$) and a small envelope mass (H/He mass fraction $\leq$ 0.2\%). In such cases, the planet's energy reservoir is predominantly governed by the rock/iron core, causing the envelope's contraction rate to be directly influenced by both the cooling rate and the physical state of the mantle. As the bottom of the mantle solidifies, the mantle's cooling rate accelerates, due to the reduced heat input from the iron core. This leads to a sudden increase in the envelope contraction rate. Figure \ref{entropy-temperature} illustrates these shifts in both the envelope's contraction rate (left panel) and the mantle's cooling rate (right panel), which coincide with the phase transition at the CMB (circles). After the phase transition, the iron core begins to decouple from the thermal evolution of the mantle and envelope. These model planets have a H/He mass fraction of 0.2\%.

Radiogenic heating typically has minimal impact on thermal contraction, particularly during the early stages of a planet's evolution (ages $<1Gyr$) and after the mantle surface has solidified. Its influence becomes significant primarily for planets with both a low rock/iron core mass and a low envelope mass, as these conditions result in a limited envelope energy reservoir.

In the right panels of Figure \ref{entropy-rcb}, we illustrate the evolution of the RCB pressure over time. We examine planets with 1 to 20 $M_\oplus$, H/He mass fractions from 0.1\% to 20\%, and incident bolometric flux from 1 to 1000 $F_\oplus$. As a planet cools, the RCB penetrates to deeper levels, indicating that the radiative atmosphere becomes a larger portion of the gaseous H/He layer. This behavior is anticipated, as the approximately constant RCB temperature ($\sim T_{\rm{eq}}$) must correspond to a higher pressure level of a colder adiabat. With an increase in incident bolometric flux, the disparity between the interior heat transport (which establishes the envelope adiabat) and the stellar heating (which determines the radiative atmospheric temperature) widens, causing the RCB to shift inward to depths exceeding several kilobars (bottom right). Additionally, planets with more massive rock/iron cores or a more massive envelope generally exhibit a shallower RCB (top and middle panels on the right) due to their warmer interiors. Among the parameters chosen and within the ranges considered in this study, incident bolometric flux exerts the most significant effect on the RCB location.

Our results indicate that in extreme cases (bolometric flux $\gtrsim 100 F_\oplus$, H/He mass fraction $\lesssim 0.2\%$, and core mass $\lesssim 3 M_\oplus$), the incident bolometric radiation penetrates to the MEB after 1 Gyr, leading to an envelope-free, thick-atmosphere super-Earth, with the radiative atmosphere contributing more than 40\% of its total radius.

Note that the contribution of the radiative atmospheric mass has a significant evolutionary influence on the most highly irradiated planets (see Section \ref{subsec:rad}) but not for the others. In the case of the $1000 F_\oplus$ model (yellow), the radiative zone occupies a substantial portion of the H/He mass at older ages due to the deep RCB (bottom right). Consequently, the reduced envelope mass corresponds to a smaller energy reservoir for cooling, leading to a faster contraction rate compared to the planets with negligible radiative atmospheric mass (those experiencing lower stellar insolation). This behavior is evident in the yellow curve in the bottom left panel, where the contraction rate accelerates significantly after a few Gyrs. This effect is particularly pronounced in planets with lower rock/iron mass and/or lower H/He mass.

\subsection{Role of Enhanced Metallicity}
\label{res:metal}
Heavy elements (usually water and silicates) in a sub-Neptune's envelope and radiative atmosphere influences its evolution and physical structure in three key ways. First, the denser metal species make the envelope more compact, rendering it less susceptible to boil-off compared to a pure H/He envelope of the same planetary mass. Second, we find that a metal-rich radiative atmosphere exhibits a lower intrinsic luminosity at a given surface gravity, incident bolometric flux and envelope specific entropy. Consequently, our treatment in Eq. \ref{KHtime} predicts a higher initial entropy and a hotter interior at a certain age compared to a pure H/He envelope. (The higher opacity due to increased metallicity does not allow the planet to cool as quickly.)  Finally, a metal-rich radiative atmosphere has a higher mean molecular weight, leading to a reduced scale height and a less extensive radiative atmosphere.

\begin{figure}
\centering
\includegraphics[width=0.5\textwidth]{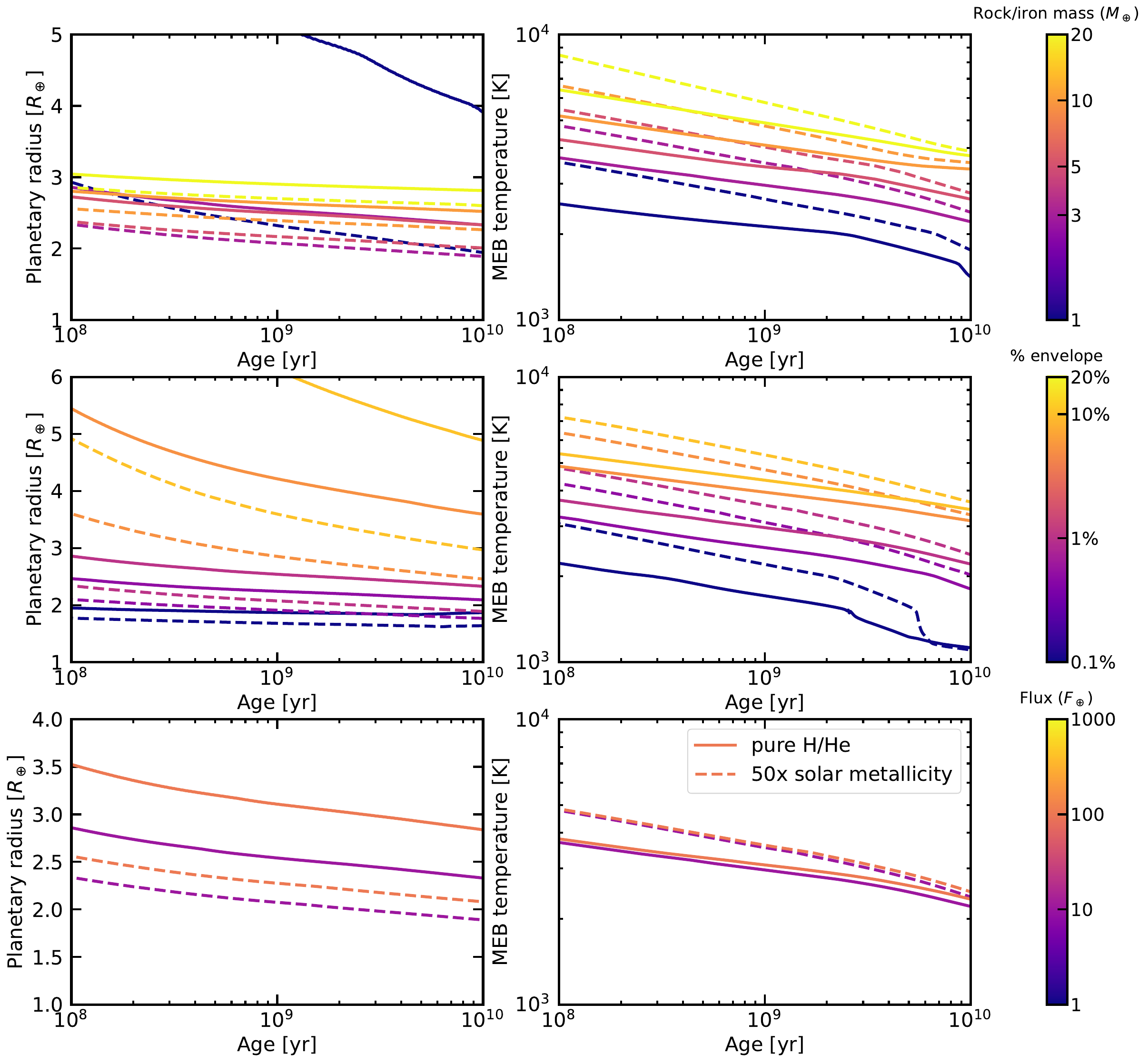} 
\caption{We show the evolution of optical radii (left panels) and temperatures at the bottom of the envelope (bottom) for planets with varying rock/iron mass (top panels), H/He mass fraction (middle) and bolometric flux (bottom). Our standard model has 3$M_\oplus$, 1\% and 10$F_\oplus$. For planets with the same parameters, a metal-rich planet (dashed curves) is more compact but has a hotter interior at a given age compared to the pure H/He planet (solid). Note that for comparison purposes, we did not incorporate a boil-off phase to constrain the initial envelope mass fraction, so some model planets are not realistic. Those planets hold a substantial radiative atmosphere (e.g. blue solid in top left). Refer to Table \ref{table:1x} and \ref{table:50x} for reasonable parameters.
}
\label{metallicity}
\end{figure}

In Figure \ref{metallicity}, we present the evolution of the planetary optical radius (left panels) and the MEB temperature of the envelope (right panels) for planets with both 50$\times$ solar metallicity (dashed) and pure H/He composition (solid). Overall, we find that these three effects of metallicity reduce the optical radius (left) despite the hotter interior thermal state (right). From top to bottom panels, we observe that the difference in radius between a metal-rich planet and a pure H/He planet increases with either a lowered rock/iron mass, an increased envelope mass fraction, or an increased incident flux. In contrast, there is no significant variation in the difference in interior temperature across these parameters.

The reduced optical radius in metal-rich planets is primarily due to the higher mean molecular weight in the radiative atmosphere (the third effect mentioned above). This mechanism can effectively explain the trends observed earlier in left panels of Figure \ref{metallicity}: (1) a lower rock/iron mass (top) corresponds to lower surface gravity, leading to a greater scale height contrast between metal-rich and metal-free envelopes, making the optical radius more sensitive to mean molecular weight; (2) an increased envelope mass fraction (middle) also reduces surface gravity by increasing planetary radius, similarly affecting the optical radius; and (3) a higher incident flux also increases the scale height contrast, producing a similar trend.

\begin{figure}
\centering
\includegraphics[width=0.5\textwidth]{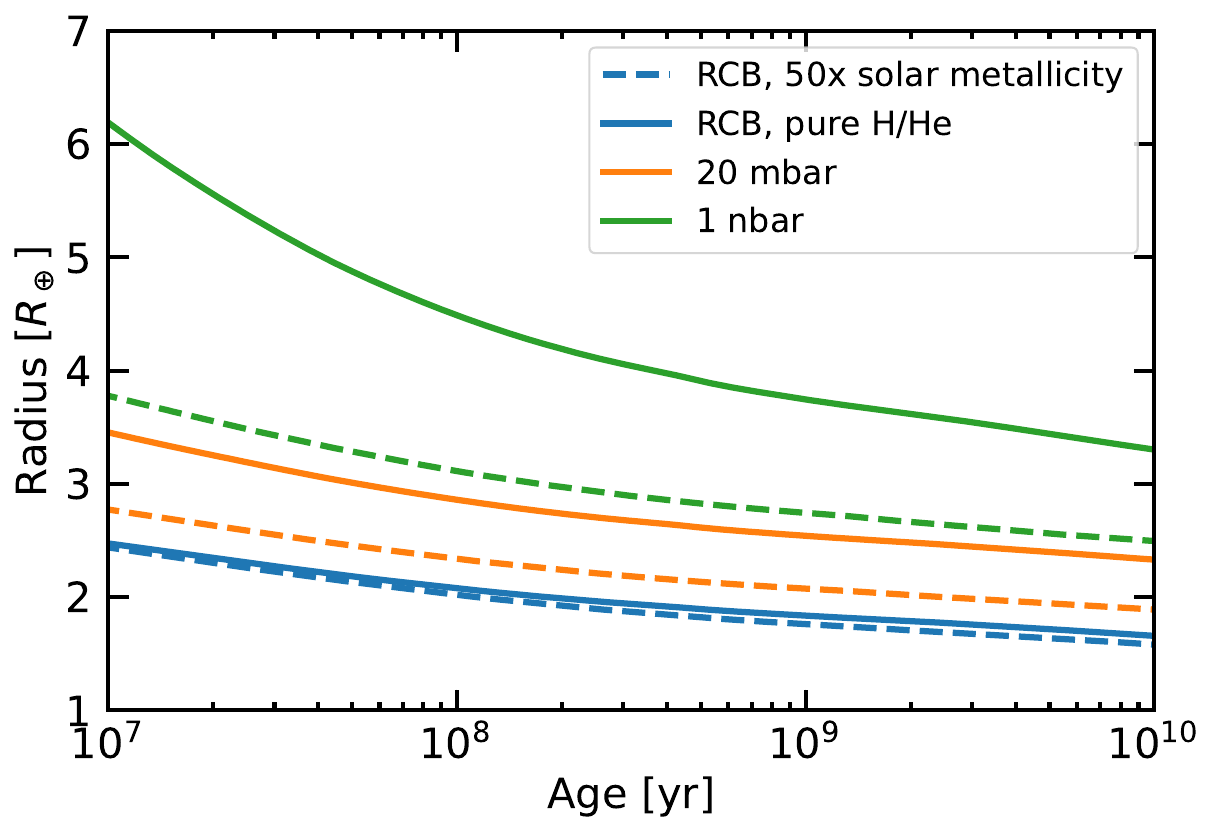} 
\caption{ We show the planetary radii at the RCB (blue), 20 mbar (orange) and 1 nbar (green) pressure levels for planets with pure H/He envelopes (solid) and 50$\times$ solar metallicity (dashed). The planetary parameters are consistent with the standard model in Figure \ref{metallicity}. The smaller optical radius for metal-rich planets results from the smaller scale height of the radiative atmosphere due to the heavier mean molecular weight. Note that the RCB radii are very similar for both metal-rich and pure H/He envelopes. 
}
\label{radius-metal}
\end{figure}

To validate this, in Figure \ref{radius-metal} shows planetary radii defined at the RCB (blue), 20 mbar (orange), and 1 nbar (green) pressure levels. The RCB radii are comparable for both the 50$\times$ solar model (dashed) and the pure H/He model (solid). This trend holds across all modeled planets, with the difference in radius between these two envelope compositions being negligible for planets with small H/He mass fractions ($\leq 2\%$) and greater (up to 15\%) for planets with higher H/He mass fractions. This suggests that the density and entropy effects discussed earlier largely cancel each other out, leaving the difference in radius primarily from the radiative atmosphere. As the pressure level decreases in the radiative atmosphere, we observe a larger relative difference between compositions, ranging from 20-30\% at 20 mbar to a factor of 2 at 1 nbar, due to the variable gravity effect (the radius decreases faster with altitude for planets with larger scale heights).

\subsection{Post-Boil-Off Mass Fraction}
\label{post-boil-off}
In this section, we present the mass fractions at the end of boil-off, which determine the maximum initial mass fraction permitted for subsequent evolution. Tables \ref{table:1x} and \ref{table:50x} show the mass fraction grids as functions of rock/iron mass and bolometric flux for pure H/He envelopes and envelopes with 50$\times$ solar metallicity, respectively. For low-mass and highly irradiated planets (bottom left), the envelopes are completely stripped due to boil-off mass loss, denoted by zeroes. Note that they may still hold a radiative atmosphere with a negligible H/He mass. Conversely, more massive and cooler planets (top right) retain all of their initial envelope mass post-boil-off and, therefore, could potentially start the later evolution with any amount of envelope mass. Constraining the initial envelope mass for these planets requires a planet formation model, which is beyond the scope of this study.

At the same rock/iron mass and bolometric flux, an envelope with 50$\times$ solar metallicity retains a higher final mass fraction at the end of boil-off, typically by a factor of 2 to 3, compared to a pure H/He envelope. This is because an atmosphere with a higher mean molecular weight ($\mu=3.58$) is more strongly bound to the planet and thus harder to escape. It is important to note that our model predicts that the metal-rich atmosphere would start at a higher initial temperature and therefore lose more mass than the pure H/He envelope if they had the same mean molecular weight. However, this initial entropy effect is relatively minor compared to the impact of the mean molecular weight in this comparison.

To fit the post-boil-off mass fraction, we use a power-law analytical formula:
\begin{equation}
f = a*(M_{\rm{ric}}/M_\oplus)^b*(F_{\rm{bol}}/F_\oplus)^c
\end{equation}
where $f$ is the post-boil-off mass fraction in percentage. The best fit parameters are $a=0.1178, b=3.801, c=-0.8830$ for the pure H/He grid and $a=0.4458, b=4.713, c=-1.188$ for the 50$\times$ solar metallicity grid. Due to the complex nature of boil-off, this fit introduces an uncertainty of up to a factor of 5, though it typically does not exceed a factor of 2.

\begin{deluxetable*}{c | c c c c c c c c c c c}
\tabletypesize{\scriptsize}
\tablewidth{0pt} 
\tablecaption{Maximum H/He mass fraction allowed after boil-off (pure H/He) \label{table:1x}}
\tablehead{
\colhead{Flux ($F_\oplus$)} & \colhead{1 $M_\oplus$} & \colhead{2 $M_\oplus$} & \colhead{3 $M_\oplus$} & \colhead{4 $M_\oplus$} & \colhead{5 $M_\oplus$} & \colhead{6 $M_\oplus$} & \colhead{8 $M_\oplus$} & \colhead{10 $M_\oplus$} & \colhead{13 $M_\oplus$} & \colhead{16 $M_\oplus$} & \colhead{20 $M_\oplus$} 
} 
\startdata 
 1 & 0.18\%	& 1.84\%	& 6.49\% & 14.25\% & - & - & - & - & - & - & -\\[0.5 ex]
 3 & 0.05\%	& 0.70\% & 2.97\% & 7.16\% & 12.55\% & - & - & - & - & - & - \\        [0.5 ex]
 10 & 0 & 0.22\% & 1.19\% &	3.29\%	& 7.03\% & 14.53\% & - & - & - & -& - \\        [0.5 ex]
 30 & 0 & 0.04\% & 0.36\% & 1.35\%	& 3.19\% & 6.29\% & 17.58\% & - & - & - & - \\        [0.5 ex]
 100 & 0 & 0 & 0.10\% & 0.27\%	& 0.86\% & 1.92\% & 6.32\% & 11.12\% & - & - & - \\[0.5 ex]
 300 & 0 & 0 & 0 & 0.07\% & 0.14\%	& 0.46\% & 2.18\% & 5.84\% & 13.54\% & - & - \\[0.5 ex]
 1000 & 0 & 0 & 0 & 0 & 0 & 0.03\% & 0.25\%	& 1.02\% & 3.91\% & 8.84\%	& 17.27\% \\[0.5 ex]
\enddata
\tablecomments{The model planets with low masses and high insolation (bottom left) lose their entire envelopes due to boil-off mass loss, becoming terrestrial planets, which are marked with zeros. Note that they can still retain a H/He radiative atmosphere, with an atmospheric mass fraction estimated to be less than 0.01\% depending on the surface pressure. The dash symbols denote planets that are not susceptible to boil-off and can retain any amount of H/He mass.
}
\end{deluxetable*}

\begin{deluxetable*}{c | c c c c c c c c c c c}
\tabletypesize{\scriptsize}
\tablewidth{0pt} 
\tablecaption{Maximum H/He mass fraction allowed after boil-off (50$\times$ solar metallicity) \label{table:50x}}
\tablehead{
\colhead{Flux ($F_\oplus$)} & \colhead{1 $M_\oplus$} & \colhead{2 $M_\oplus$} & \colhead{3 $M_\oplus$} & \colhead{4 $M_\oplus$} & \colhead{5 $M_\oplus$} & \colhead{6 $M_\oplus$} & \colhead{8 $M_\oplus$} & \colhead{10 $M_\oplus$} & \colhead{13 $M_\oplus$} & \colhead{16 $M_\oplus$} & \colhead{20 $M_\oplus$} 
} 
\startdata 
 10 & 0 & 0.40\% & 4.67\% &	11.10\%	& - & - & - & - & - & -& - \\        [0.5 ex]
 30 & 0 & 0.06\% & 1.21\% & 5.72\%	& 10.78\% & - & - & - & - & - & - \\        [0.5 ex]
 100 & 0 & 0 & 0.10\% & 1.25\%	& 4.62\% & 8.92\% & - & - & - & - & - \\[0.5 ex]
 300 & 0 & 0 & 0.02\% & 0.10\% & 0.69\%	& 2.67\% & 8.81\% & 16.49\% & - & - & - \\[0.5 ex]
\enddata
\tablecomments{Due to the limited $T_{\rm{int}}$ data available for atmospheres with 50 times solar metallicity, we model planets with incident bolometric fluxes ranging between $10-300 F_\oplus$.
}
\end{deluxetable*}

\subsection{Evolution of the Core and Mantle}
\label{res:core}
The envelope largely governs the evolution of the rock/iron interior. The cooling rates of the iron core and silicate mantle are primarily dictated by envelope cooling; a rapidly cooling envelope induces a larger temperature contrast at the CMB, thereby encouraging more vigorous energy transport in the rock/iron core. The adiabatic envelope maintains warm conditions at the surface of the silicate mantle, keeping it molten for an extended period. This blanketing effect leads to a divergent thermal and physical state evolution of the core and mantle compared to terrestrial planets.

\begin{figure*}
\centering
\includegraphics[width=0.9\textwidth]{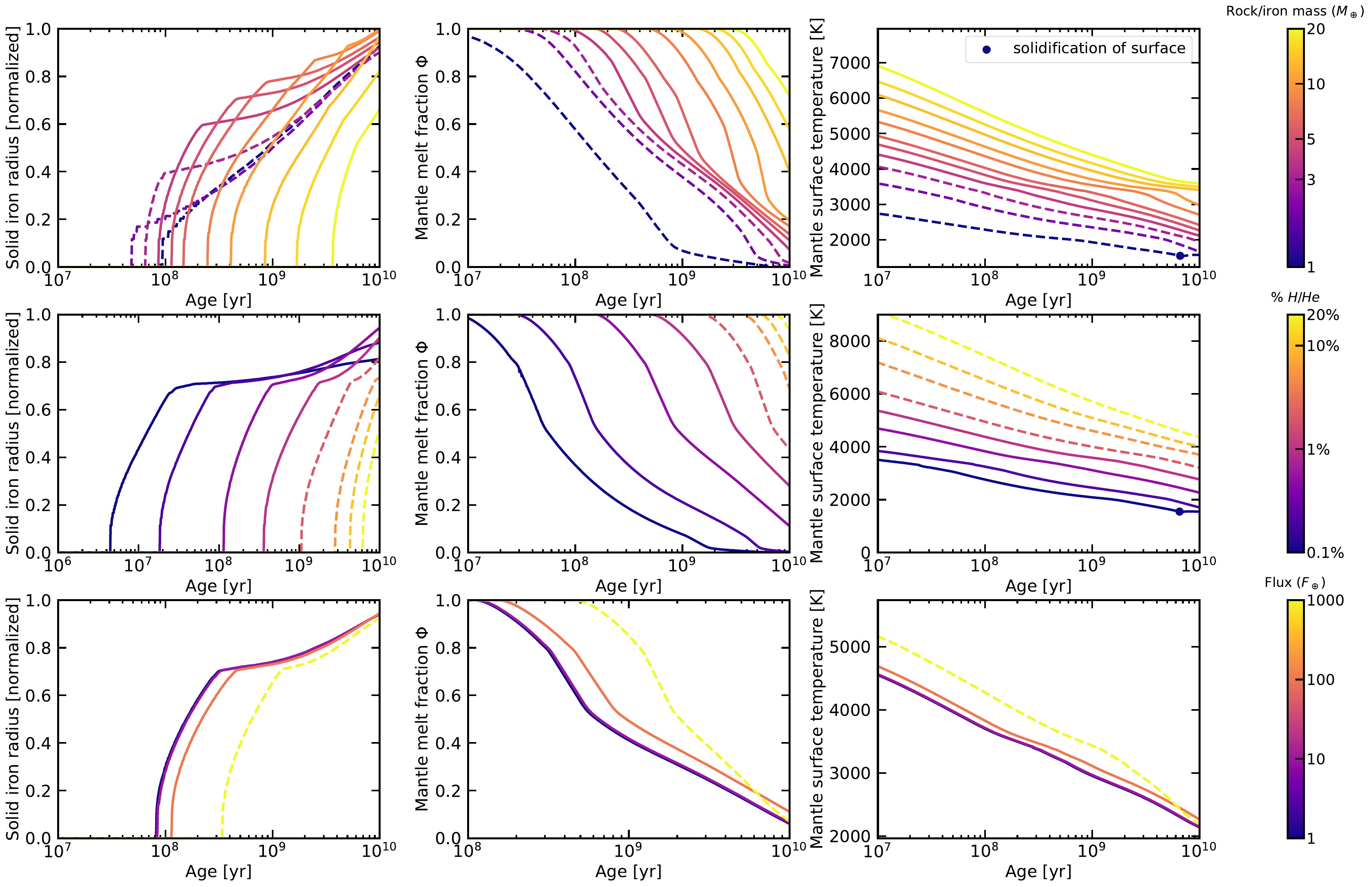} 
\caption{ We show the solidification timescales of the silicate mantle and iron core as a function of rock/iron mass (top panels), H/He mass fraction (middle panels), and planetary bolometric flux (bottom panels). Our default model assumes a planet with a rock/iron core mass of 5$M_\oplus$, a H/He mass fraction of 0.5\% and a bolometric flux of 100$F_\oplus$. The left panels depict the physical size of the inner solid iron core, normalized by the total iron core radius. The kinks in those curves (except for the one corresponding to the lowest rock/iron mass) occur when the mantle starts to solidify at its bottom, which slows the growth of the inner solid core. See text for detailed discussions. The middle panels display the bulk melt fraction of the silicate mantle. In the right panels, we show the temperature of the mantle surface, which is significantly higher than that of terrestrial planets due to the blanketing effect of the H/He envelope. The solidification of the mantle surface is indicated by circle marks. Similar to Figures \ref{entropy-rcb} and \ref{entropy-temperature}, model planets in boil-off conditions (which technically should not exist in the post-boil-off phase) are represented with dashed lines.
}
\label{solidification}
\end{figure*}

In the right panels of Figure \ref{solidification}, we show the evolution of the mantle surface temperature as a function of rock/iron mass (top), H/He mass fraction (middle), and stellar heating (bottom). The surface temperature can exceed 2000 K by 5 Gyr, even for planets with low stellar irradiation, contrasting sharply with the much lower surface temperature of a terrestrial planet like Earth, which is around 300 K. Complete solidification of the mantle is indicated by the solidification of the mantle surface and occurs only in planets with very low mass ($\leq 1 M_\oplus$, top) and those with minimal envelope masses (H/He mass fraction $\leq$ 0.1\%, middle), as marked by the circles. This solidification is found after 5 Gyr. This finding implies that for these thin-envelope planets, other physical effects, through the interaction between the magma ocean and the envelope, could potentially influence the solidification timescales. This is discussed in Section \ref{disc:inter}.

The thermal states of the core and mantle determine their physical states. As cooling progresses, the interior temperature decreases, leading to solidification. The internal physical state of the silicate mantle, in turn, influences energy transport (cooling) rates, resulting in a coupling between these processes. We find that the silicate mantle solidifies in a \emph{bottom-up order for low-mass planets ($\leq 3 M_\oplus$)}, where the local melt fraction monotonically increases with radius. This behavior is illustrated in the top panel of Figure \ref{meltfraction}, which shows the melt fraction as a function of planetary radius in the mantle. 

However, for intermediate and high-mass planets (bottom panel), the melt fraction drops below 1 (indicating a transition out of the pure liquid state) starting from the \emph{middle of the lower mantle} rather than from the bottom. Consequently, the crystallization region quickly expands and propagates both upward and downward, leading to a large portion of the lower mantle entering a partially molten phase and solidifying nearly simultaneously. Notably, the density gradient remains negative regardless of the reversed gradient of the melt fraction at the mantle bottom, ensuring that the solid/liquid mixture is gravitationally stable. We observe that complete solidification always occurs first at the bottom rather than at the surface. 

A middle-out solidification process could theoretically form a solid layer that inhibits cooling at the mantle bottom due to high viscosity, potentially leading to a persistent basal magma ocean \citep{Lherm23}. However, the intermediate- and high-mass planets modeled in this study are \emph{likely} not in this case, as nearly synchronized solidification appears to occur in the lower mantle once solid convection begins ($\phi < 0.5$). Since our model does not explicitly capture energy transport within individual mantle mass shells, this behavior requires further investigation in future studies.

In the middle panels of Figure \ref{solidification}, we present the bulk melt fraction of the mantle as a function of age. The onset of mantle solidification is found to increase with both rock/iron mass and envelope mass fraction, ranging from 10 Myr to several Gyr. This trend arises because massive, gas-rich planets experience a stronger insulation effect from the envelope, resulting in a hotter rock/iron interior. By 5 Gyr, most planets still possess a substantial magma ocean, except for those with low rock/iron mass and thin envelopes. This results in a typical solidification timescale of Gyr, contrasting with the shorter 1-100 Myr year timescales observed for terrestrial planets.

For all modeled planets, we find that iron solidification begins from the center, gradually forming a solid inner core over time. The solidification onset timescale varies greatly among planets, ranging from a few Myr to a few Gyr. In the left panels, we show the fractional solid core radius normalized by the iron core radius. As expected, planets with either a higher rock/iron mass or more substantial envelope mass begin solidification at a later age due to their hotter mantles (which results from the hotter envelopes) that keep the iron core warm. An exception is observed for the $1 M_\oplus$ planet (dashed curve with the latest onset of iron solidification in the top left panel), as explained below. None of the modeled planets' cores are completely solidified by 10 Gyr. For planets with low and intermediate rock/iron mass (top), their fractional solid core radius tends to converge after a few Gyr, after which the solidification rate (the slope) becomes similar. In contrast, planets with varying envelope masses exhibit divergent solidification rates at later ages, with those having more massive envelopes solidifying at a faster pace, resulting in a wide dispersion in fractional solid core radii.

We argue that the physical state of the mantle significantly impacts the iron solidification timescale. If the bottom of the mantle remains in the liquid phase, iron solidification proceeds quickly due to vigorous mantle convection that extracts heat from the core. Once the bottom of the mantle becomes solid, the solidification rate sharply decreases, leading to changes in the slopes of the solidification curves. It is important to note that the solidification of the $1 M_\oplus$ planet (top left) is delayed, indicating a warmer core condition compared to the $2-4 M_\oplus$ planets, despite having a colder mantle (middle). This is because its mantle solidifies at a very young age from the bottom, slowing down the cooling of the iron core. Consequently, a positive feedback is triggered: the mantle experiences a reduced heat flux at the CMB due to the iron's latent heat, which accelerates the mantle solidification process while delaying the solidification of the iron core. This effect becomes more pronounced in planets with less substantial envelopes and higher iron melting temperatures (see below) and may extend to more massive planets.

\begin{figure}
\centering
\includegraphics[width=0.45\textwidth]{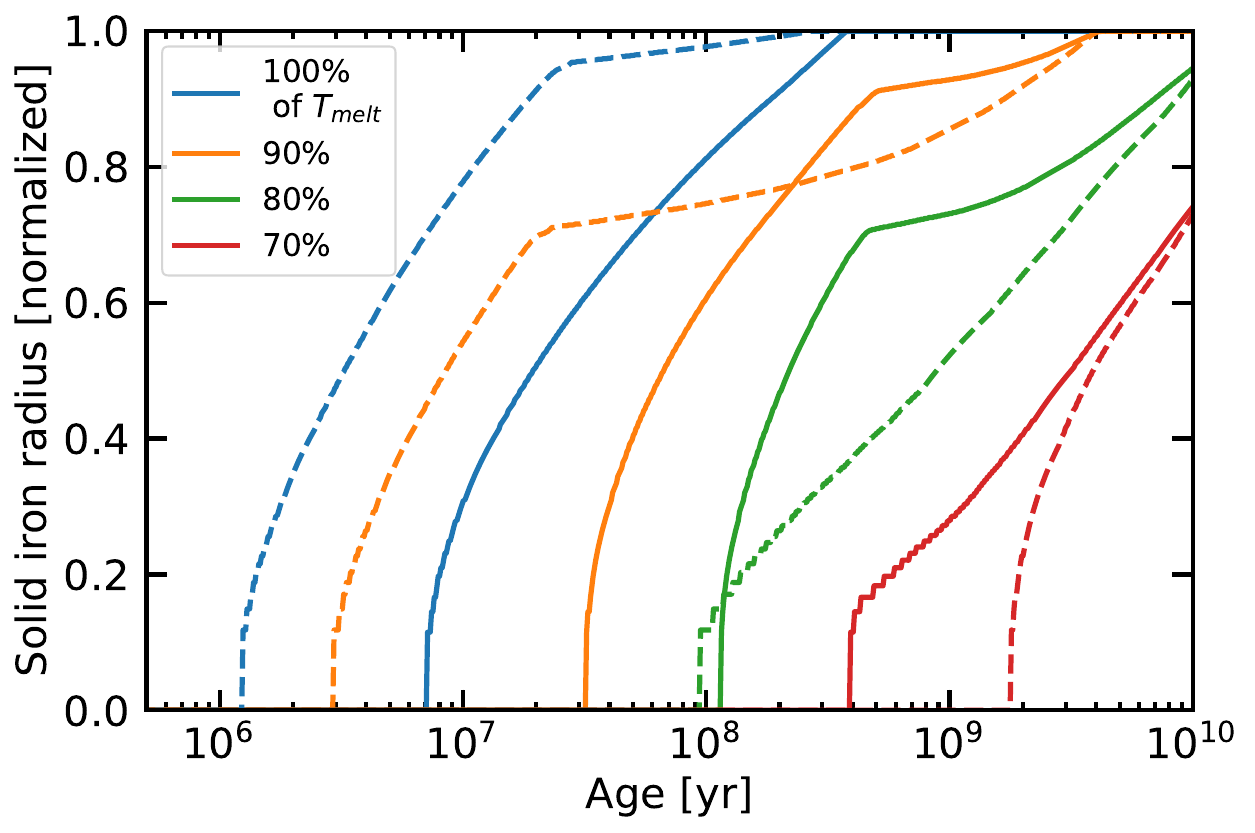} 
\caption{ We present the iron solidification timescales using different melting temperature estimates, taken as a fraction of the melting curve for pure iron $T_{\rm{melt}}$ to account for chemical impurities. These estimates are shown in different colors. The solidification timescale is highly sensitive to the melting temperature, leading to significant uncertainty in the evaluation. Low-mass planets (dashed, 1$M_\oplus$) are notably more sensitive than heavier planets (solid, 5$M_\oplus$).
}
\label{iron}
\end{figure}

While our model effectively captures the relative solidification timescales among planets with different parameters, the quantitative details are expected to carry some uncertainty. We find that the iron solidification timescale is sensitive to the melting temperature, which is expected to be significantly influenced by impurities. In Figure \ref{iron}, we explore various melting curves that represent fractions of the melting temperature for a pure composition to account for different levels of impurity. We find that the onset of iron solidification for a $5 M_\oplus$ planet (solid line) ranges from 10 Myr to a few 100 Myrs, given a 30\% difference in melting temperature. For a lower-mass planet (dashed, 1 $M_\oplus$), the difference in solidification onset timescale is even more pronounced, varying from just 1 Myr to over 1 Gyr. This large dispersion (note the reversed trend between the 1$M_\oplus$ and 5$M_\oplus$ models in red) arises from early solidification at the bottom of the mantle, which greatly slows down core cooling, resulting in this sensitive behavior.

For a comparison between our results, which incorporates the improved rock/iron core modeling, and previous models that assume isothermal cores, see Section \ref{res:comparison}.
 
\begin{figure}
\centering
\includegraphics[width=0.45\textwidth]{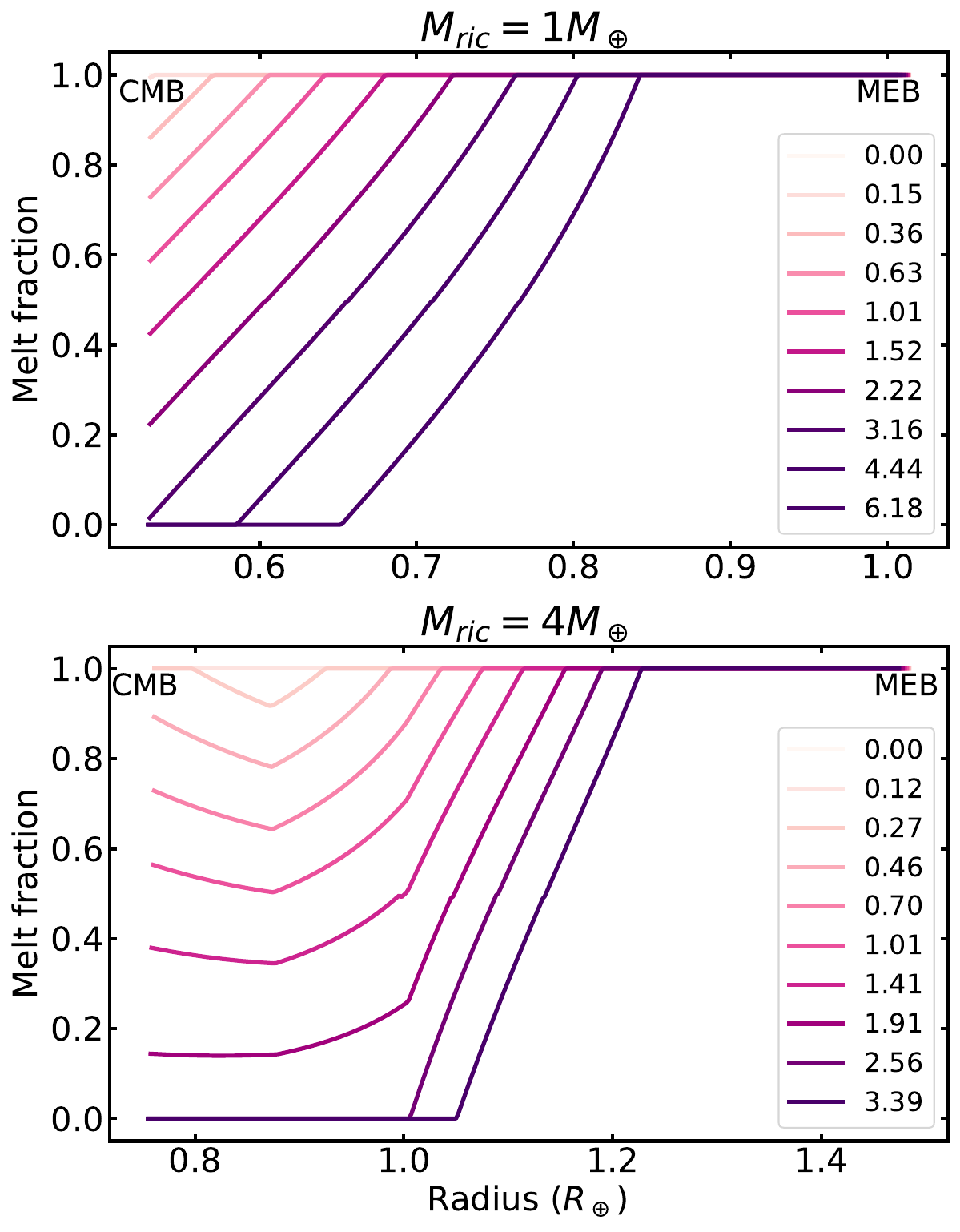} 
\caption{ Local melt fraction as a function of planetary radius in the silicate mantle. From top to bottom, the curves represent increasing time since the onset of mantle solidification (in units of Gyr). The top panel depicts a low-mass planet ($1 M_\oplus$, 5\%) that solidifies in a bottom-up manner, while the bottom panel illustrates a more massive planet ($4 M_\oplus$, 1\%) that begins solidification in the middle of the lower mantle. Both planets have stellar insolation of $100 F_\oplus$. See text for further discussion.
}
\label{melt}
\end{figure}

\subsection{Importance of the Radiative Atmosphere}
\label{res:rad}
In this section, we review the common assumptions made in previous models \citep{Nettelmann11,Lopez12,Owen13,Howe15,Chen16,Ginzburg18} for the radiative atmosphere and quantitatively discuss how they impact the planetary radii calculation.

\subsubsection{Traditional Treatments}
The temperature profile of the radiative atmosphere largely controls its vertical structure through Eq. \ref{masscon} and \ref{hydrostatic}. In these calculations, the ideal gas law $P\mu = \rho k T$ is commonly employed to relate density ($\rho$), pressure ($P$), and temperature ($T$), where $\mu$ is the mean molecular weight of the radiative atmosphere and $k$ is the Boltzmann constant. For convenience, an isothermal temperature profile with $T_{\rm{eq}}$ is often used, allowing for a straightforward analytical solution for the physical structure in exponential form: 
\begin{equation}
\label{verticalstructure}
P = P_{\rm{base}}e^{-\int_{R_{\rm{base}}}^{R_p}\frac{dr}{H(r)}}
\end{equation}
where $H(r) = kT_{\rm{eq}}/\mu g(r)$, the pressure scale height, governs the radius profile, and $g(r)=GM_p/r^2$ is the local gravity. $P_{\rm{base}}$ and $R_{\rm{base}}$ are the pressure and radius at the base of the atmosphere calculation, evaluated at the RCB here. The atmospheric scale height $H(r)$ is determined by three parameters: temperature, gravity and mean molecular weight.

However, the deep radiative atmosphere characterizes the location where the visible stellar radiation is mostly absorbed by the planet \citep{Guillot10}, thus having a higher temperature $T_{\rm{deep}}$ than $T_{\rm{eq}}$. In contrast, at a higher altitude, the optically thin region is determined by a colder ``skin temperature'' (see Figure \ref{rcb_comp}). This can lead to uncertainty in radii calculations if a simplified \emph{T-P} profile is used. Additionally, previous work often assumed local gravity to be constant, typically evaluated at the RCB, to further simplify computations. This neglect of variable gravity effects can result in an overestimation of local gravity at higher altitudes and, consequently, an underestimation of the planetary radius. Lastly, while the atmosphere is dominated by molecular hydrogen throughout the whole lower atmosphere as assumed in the previous models, a molecular-atomic transition occurring above the $\mu$bar level must be considered. This transition accounts for the decline in mean molecular weight; without considering it, the nanobar radius may be underestimated.

\subsubsection{Boil-Off-Constrained Variable Gravity Effect in an Isothermal Atmosphere}
The above discussion is all based on the hydrostatic atmosphere assumption. However, a young sub-Neptune with a hot interior can be subject to the boil-off mass loss, necessitating a steady-state atmosphere to self-consistently determine planetary radius. 

\begin{figure}
\centering
\includegraphics[width=0.5\textwidth]{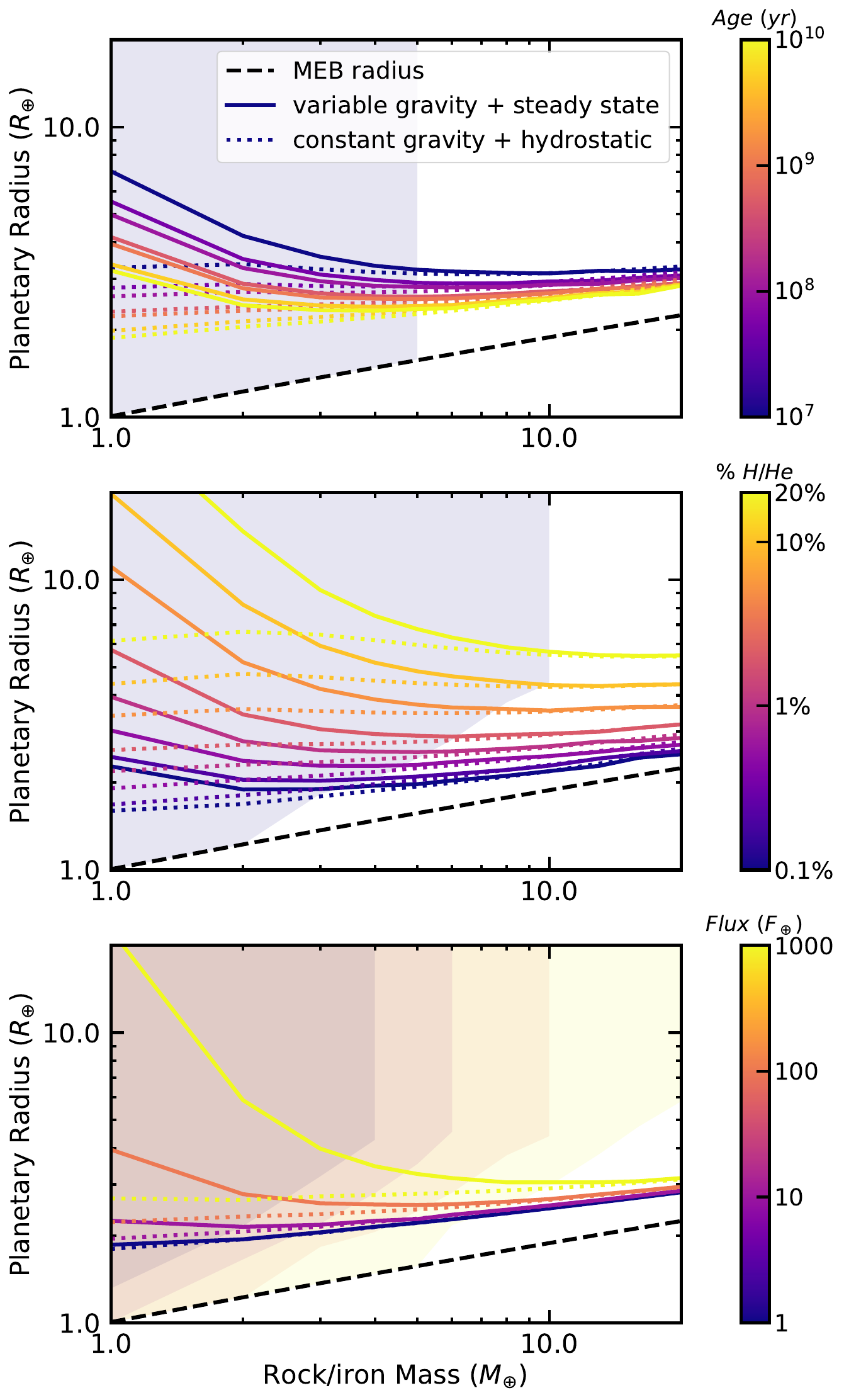} 
\caption{ We show the mass-radius diagrams for planets with a range of planetary age (top), H/He mass fraction (middle) and incident bolometric flux (bottom). Our default model has a H/He mass fraction of 1\%, incident bolometric flux of $100 F_\oplus$ and age of 1 Gyr. We compare models with variable gravity (solid) and constant gravity (dotted), both having an isothermal atmosphere. The rock/iron core radius is shown in dashed. For planets that are still boiling off, we mark the area with shaded color. At high rock/iron masses, the planetary radius is no longer constrained by boil-off (as discussed in Section \ref{post-boil-off}), resulting in the  truncation. In the bottom panel, note that the boil-off-prohibited region (shaded) extends to higher planetary masses with increasing flux. For the post-boil-off evolution, the difference between the two models is small.
}
\label{mass-radius}
\end{figure}

In Figure \ref{mass-radius}, we show mass-radius diagrams and compare the evolution of 20 mbar radius for planets with (solid) and without (dotted) the variable gravity effect. From top to bottom panels, we vary planetary age, H/He mass fraction and incident bolometric flux, respectively, while fixing other parameters. Note that for the variable gravity models, a steady state atmosphere is required to self-consistently assess planetary radius, because 20 mbar can be less than the confinement pressure needed to maintain a hydrostatic atmosphere for a very inflated planet. To standardize the comparison, all models are calculated with an isothermal \emph{T-P} profile. As a comparison, the MEB radius is marked in dashed black. The shaded areas indicate the planets that are still in the boil-off phase. Our results show that the differences in planetary radius between models with variable gravity (solid) and constant gravity (dashed) are small for the post-boil-off evolution if an isothermal atmosphere is assumed, meaning the steady-state Parker wind converges to the isothermal hydrostatic atmosphere after boil-off.

\begin{figure}
\centering
\includegraphics[width=0.5\textwidth]{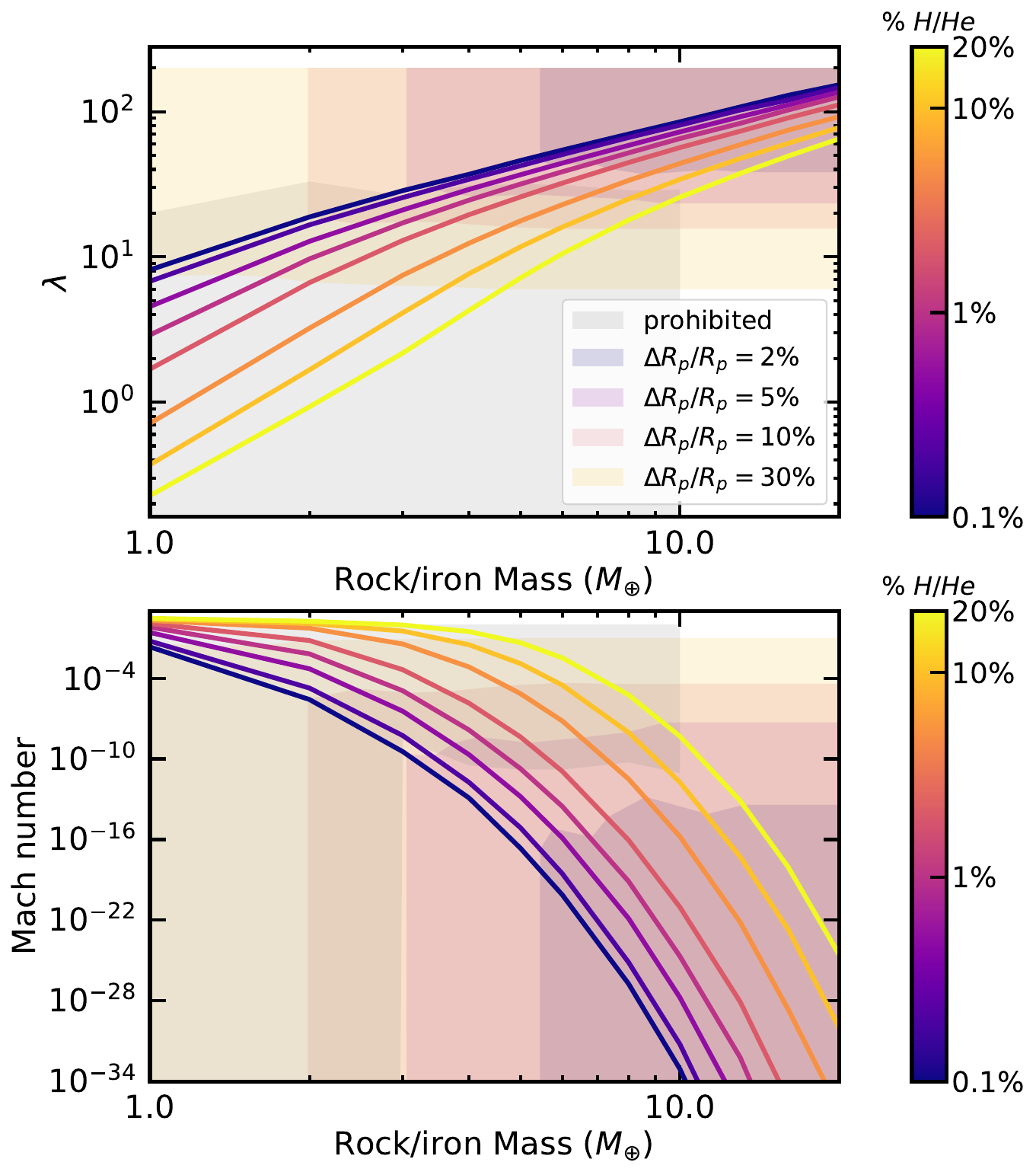} 
\caption{ The atmospheric escape parameter $\lambda$ (top panel) and Mach number (bottom panel) as a function of rock/iron mass and H/He mass fraction. Both are evaluated at a 20 mbar pressure level and an age of 10 Myr. We compute the relative difference in the optical radius between the variable gravity model and constant gravity model (same models in Figure \ref{mass-radius}), with regions in parameter space marked by color-coded contours, each associated with a specific tolerance value. Note that these models do not include a boil-off phase. The gray shaded area corresponds to boil-off conditions, which should lose a substantial amount of H/He mass if mass loss is switched on. Thus they are excluded from the discussion.
}
\label{lambda-velocity}
\end{figure}

To determine when the variable gravity effect becomes significant, we investigate two important parameters to quantify the difference between the variable gravity model and constant gravity model in Figure \ref {mass-radius}. Since the variable gravity is most important for warm interiors and is amplified by low planetary mass, this indicates a strong correlation to the vulnerability to atmospheric escape. In the top panel of Figure \ref{lambda-velocity}, we use the restricted Jeans parameter $\lambda$ \citep{Fossati17} to characterize this effect:
\begin{equation}
\label{lambda}
\lambda \equiv \frac{G M_{\rm{p}}\mu}{R_{\rm{p}} k T_{\rm{eq}}}= \frac{R_p}{H(R_p)}
\end{equation}
where $R_{p}$ is the planetary radius assessed at 20 mbar. $\lambda$ evaluates the ratio between the specific gravitational binding energy for the escaping matter to overcome and the thermal energy available to expand the atmosphere. A lower value of $\lambda$ correspond to a higher ratio between atmospheric scale height ($H(R_p)$) and radius, marking the planets that are susceptible to both boil-off and the variable gravity effect. For reference, at the sonic point of an isothermal Parker wind, $\lambda$ approaches unity.

The presented results are evaluated with 100 $F_\oplus$ at 10 Myr. We find that an atmosphere with $\lambda \leq 20$ leads to a boil-off mass loss, as indicated by the gray shaded area, which is consistent with hydrodynamic atmosphere modeling in \citet{Fossati17}. The color-coded contours represents regions with the relative difference smaller than a certain tolerance between the variable gravity model and constant gravity model. We find that the thresholds at any tolerance values are roughly constant with respect to planetary mass over a range of incident bolometric flux.

With this insensitive behavior, we argue that the threshold values can be used and generalized to a non-isothermal atmosphere case in the later sections but a hotter temperature $\sim T_{\rm{deep}}$ needs to be incorporated. Note that the threshold values weakly depend on the system age, with a greater $\lambda$ needed at a later age. For a tolerance smaller 2\%, we find that a constant gravity model can only be used to approximate a variable gravity model for $\lambda \geq 40$, which requires a core mass $\geq 6 M_\oplus$ for an envelope with $\geq$ 0.1\% of H/He mass fraction. However, a typical young, low-mass sub-Neptune, which tends to have a low $\lambda\sim20$, is beyond the range and thus needs to be evaluated with the variable gravity effect incorporated.

In the bottom panel, a similar treatment is applied to the Mach number, the ratio between the wind velocity (assuming a steady-state atmosphere) and the sound speed, at the 20 mbar pressure, and a similar behavior is found. 

\subsubsection{Enhanced Variable Gravity Effect in a Non-Isothermal Atmosphere}
However, we find the variable gravity can be much more pronounced (with a difference in radius up to 20\% compared to constant gravity model) if a more comprehensive two-stream radiative transfer model is assumed in the radiative atmosphere. This results from the higher temperature down to the RCB, which amplifies the local scale height in the radiative atmosphere, elevating the 20mbar radius to a higher altitude and reducing the gravity. 

\begin{figure}
\centering
\includegraphics[width=0.45\textwidth]{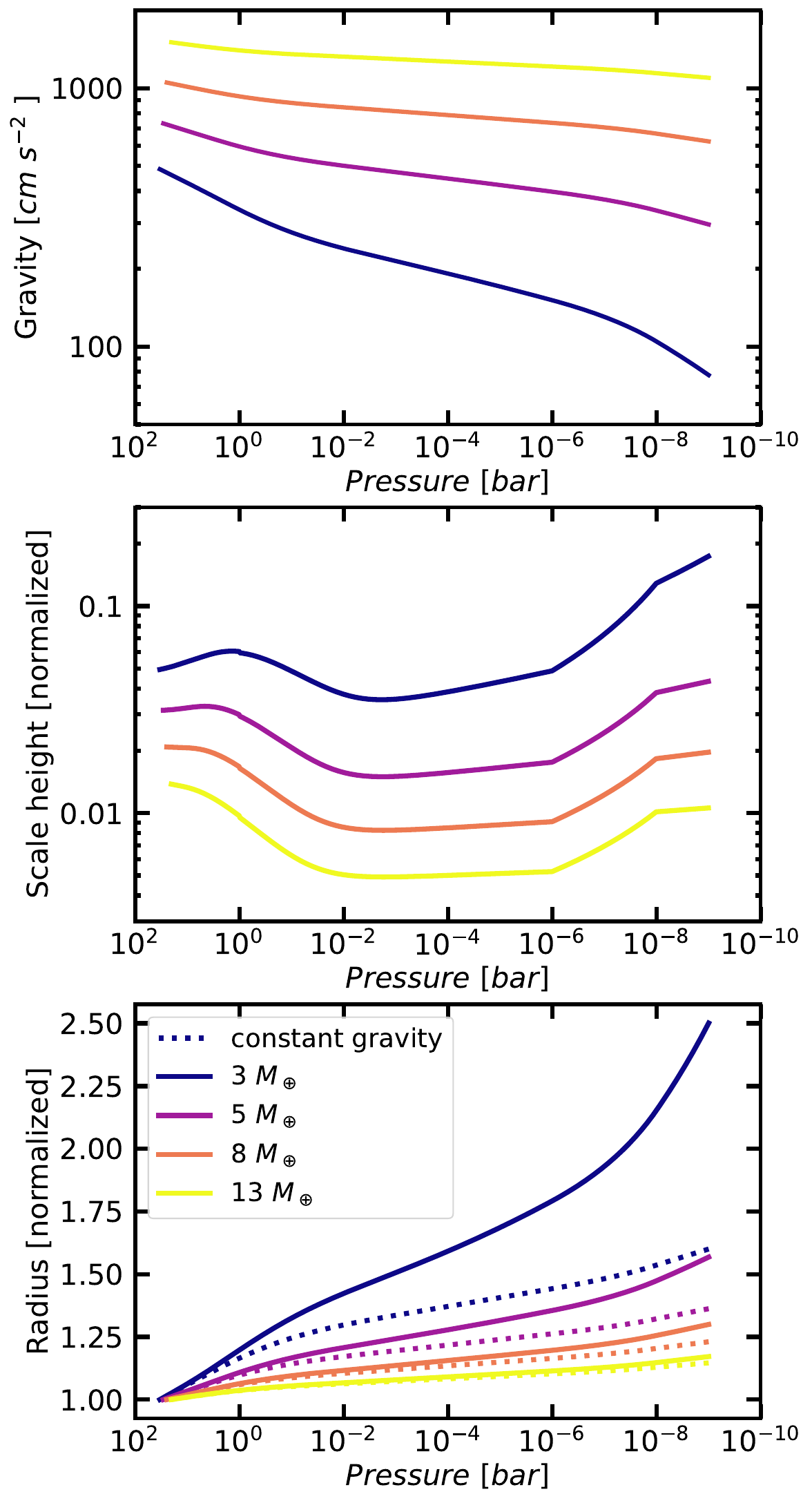} 
\caption{ The local gravity (top), scale height (middle) and radius (bottom) as a function of decreasing pressure (increasing altitude) of the radiative atmosphere. They are evaluated at 10 Myr for planets with a H/He mass fraction of 10\% and insolation of 10 $F_\oplus$. Our standard model with variable gravity and the two-stream temperature profile (solid) is compared to the constant gravity model (dotted), indicating a significant difference in radius between the two models for low-mass sub-Neptunes with $\leq 5 M_\oplus$. The strong variable gravity effect (top) for these planets is associated with the large scale heights (middle), exceeding 10\% of the total radius.
}
\label{gravity}
\end{figure}

In Figure \ref{gravity}, we show the local gravity (top), scale height (middle, normalized by $R_{\rm{RCB}}$) and radius (bottom) as a function of pressure (radius increases at a lower pressure in the radiative atmosphere), for a range of different rock/iron core masses. The constant gravity models (dashed, same models as in Figure \ref{mass-radius}) are included in the bottom panel for comparison. All models have the same H/He mass fraction of 1\%, which is allowed by a boil-off phase, an incident flux of 10 $F_\oplus$ and an age of 10 Myr. Planets with an inflated interior are found to be sensitive to the variable gravity effect, due to both the low gravity and large scale height above the RCB. Thus planets with low rock/iron mass, high stellar irradiation and/or high H/He envelope mass fraction are especially susceptible. Among planets constrained by boil-off, low-mass planets are more susceptible than high-mass planets, although the latter hold thicker envelopes. We find the secular thermal evolution of planets makes them less inflated, leading to a less important variable gravity effect. This effect, however, is modest compared to the influence from the interior composition and stellar heating. Moreover, the 1 nbar radius is usually a few scale heights above the optical radius, which consequently further reduces the gravity leading to a magnified variable gravity effect (bottom).

In the top and middle panels, we show that the gravity of the $3 M_\oplus$ planet (blue) decreases by a factor of 6-7, from the RCB to the 1 nbar radius, and the scale height can be up to 20\% of the planetary radius at a high altitude. As a comparison, for a heavier $13 M_\oplus$ planet (yellow) that has a gravity comparable to Earth's, the local gravity barely changes with altitude (decreasing pressure, top), showing a negligible difference in radius with and without the variable gravity (bottom). This is accompanied by the scale height that is only a small fraction of the radius. This is consistent with terrestrial planets and giant planets but largely differs from a typical sub-Neptune with $3-5 M_\oplus$. At a higher altitude above $\mu$bar, we find the decline in the mean molecular weight increase the scale height by more than a factor of 2, leading to a larger divergence in radius from the constant gravity model (dashed). 

\begin{figure}
\centering
\includegraphics[width=0.5\textwidth]{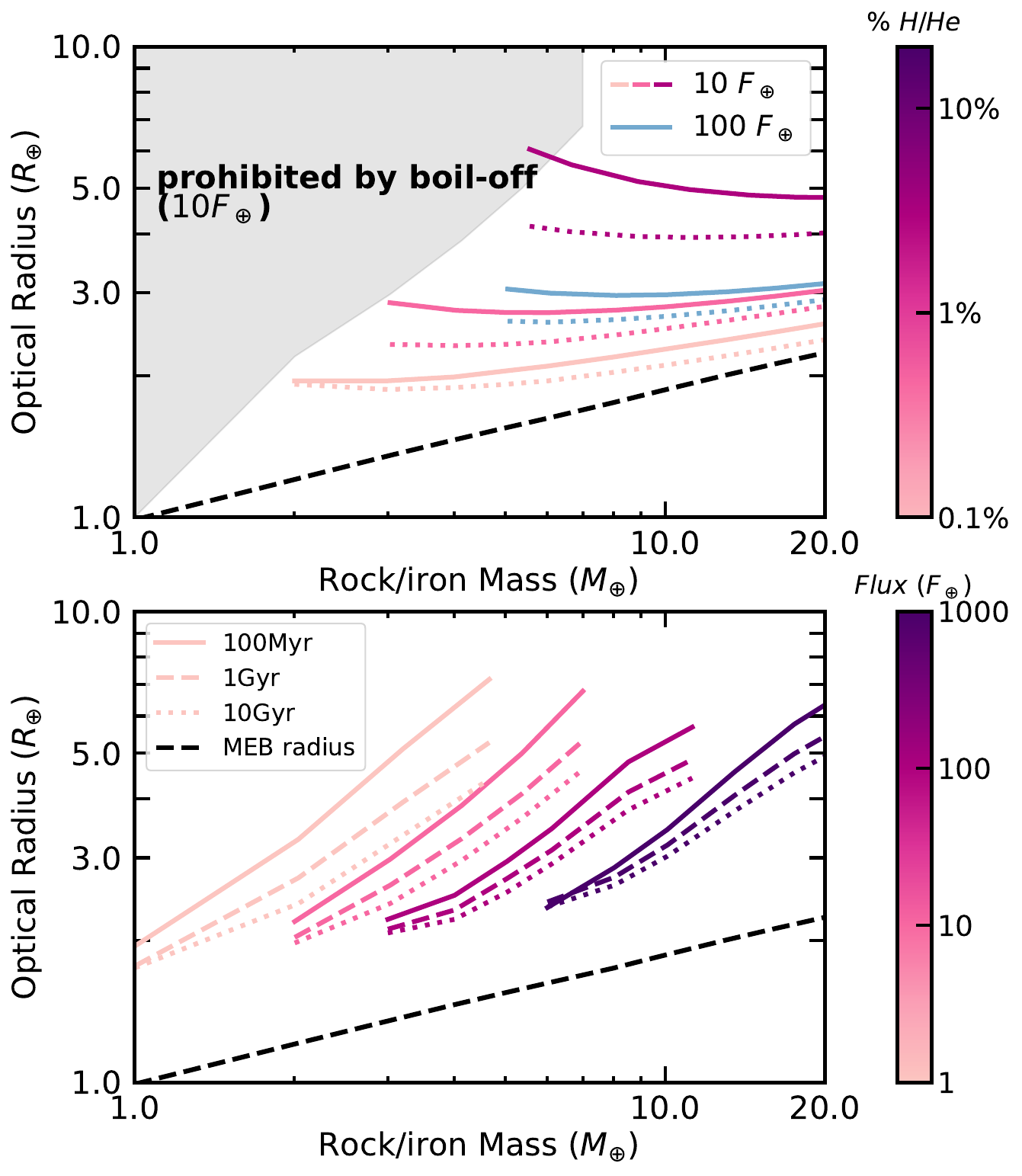} 
\caption{Mass-radius diagrams that account for the atmospheric \emph{T-P} profile and boil-off. In the top panel, we show isocompositional curves, comparing planets with different incident fluxes with those exposed to $10F_\oplus$ in red-purple and $100F_\oplus$ in blue (both having 1\%). For the $10F_\oplus$ models, the H/He mass fractions are varied, ranging from 0.1\% to 10\%, with lighter colors indicating lower fractions and darker colors indicating higher ones. Planetary ages are represented by solid for younger planets (100 Myr) and dotted for older planets (10 Gyr). The curves are truncated due to boil-off constraints, indicating that such planets would lose their H/He envelopes and not retain the parameters specified. The prohibited region is shown in gray (see discussion in Section \ref{post-boil-off}). In the bottom panel, the mass-radius relationship is shown for a fixed incident flux, with the H/He mass fractions determined by a boil-off phase (see the initially H/He-rich configuration in Section \ref{subsec:boil-off} and Table \ref{table:1x}). It is important to note that planets with small envelopes exhibit minimal shrinkage over 10 Gyr of thermal contraction due to the deep RCB effect, which results in a thicker radiative atmosphere. This behavior is discussed further in the text.
}
\label{mass-radius-full}
\end{figure}

A more detailed mass-radius diagram, which considers both boil-off and non-isothermal atmospheric \emph{T-P} profiles, is presented in Figure \ref{mass-radius-full}. It is important to note that the minimal decrease in radius over billions of years of thermal evolution for small envelopes ($\leq 0.2\%$) results from the persistent thickening of their radiative atmospheres (which starts to dominate the H/He mass). As the radiative atmosphere has a higher specific entropy than the envelope adiabat, those planets with increasing RCB pressure are expected to have a puffier atmosphere over time. 

\subsubsection{Consequence on the Radius Calculation}
\begin{figure}
\centering
\includegraphics[width=0.5\textwidth]{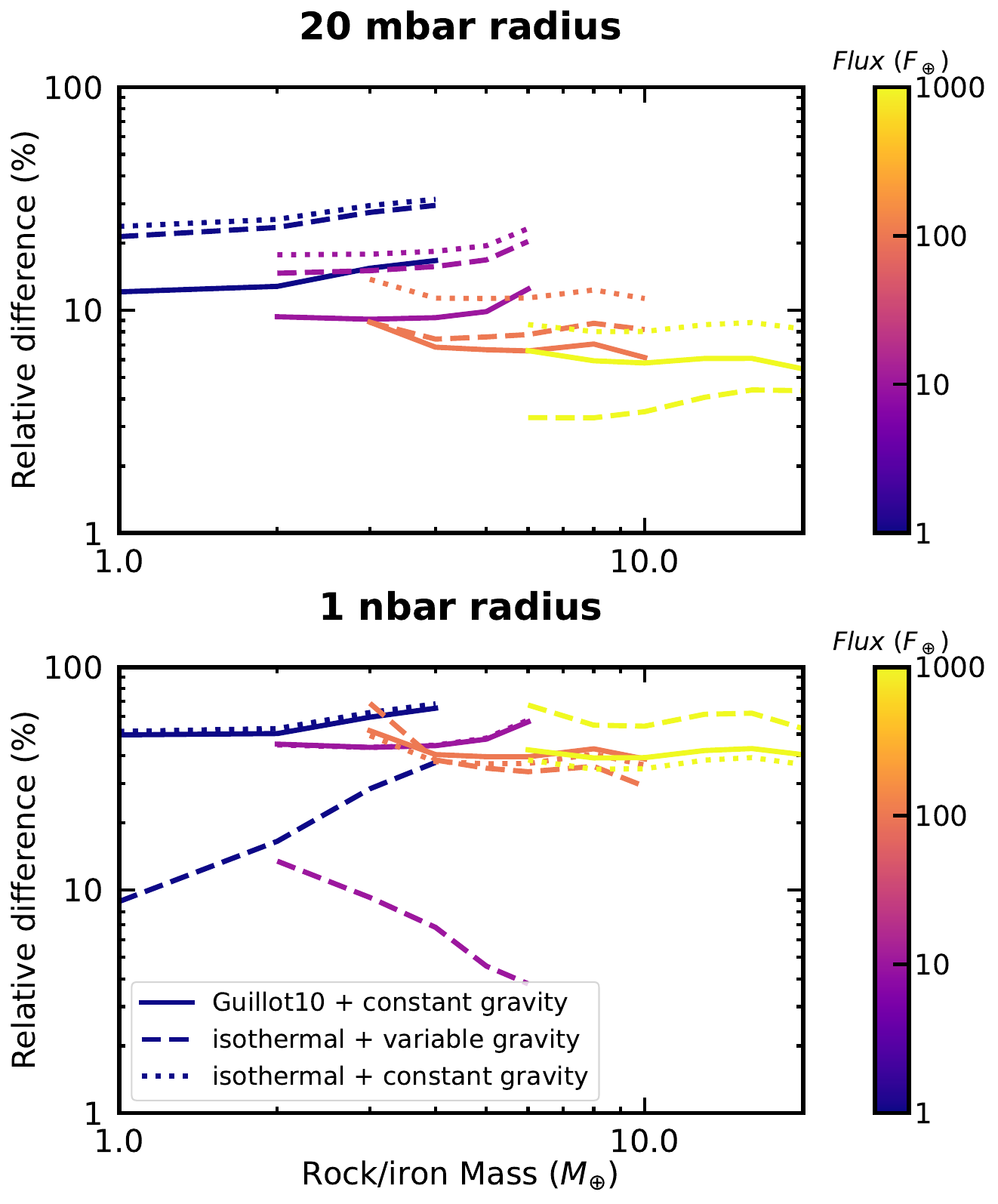} 
\caption{ We compare models with different combinations of assumptions (Guillot temperature profile/isothermal and whether the variable gravity effect is included) against our standard model (same as Figure \ref{gravity}). The plot illustrates the fractional difference in planetary radius relative to the standard model, evaluated at 10 Myr. The top panel shows the evaluation for the optical radius at 20 mbar, while the bottom panel is for the 1 nbar radius, a characteristic radius meaningful for photoevaporation. The H/He mass fraction is derived from a boil-off phase (Table \ref{table:1x}). Therefore, we do not include planets that are completely stripped or those invulnerable to boil-off due to their unbounded H/He mass fraction. However, a naive extrapolation at lower masses suggests that the relative difference is likely to be of the same order for envelope-free planets as for others, suggesting they potentially hold a thick radiative atmosphere.
}
\label{comparison}
\end{figure}

To provide a quantitative view of how different assumptions impact the planetary radius, we summarize the relative difference between models as a function rock/iron mass and incident bolometric flux in Figure \ref{comparison}. The H/He mass fraction is chosen to be the maximum amount allowed from the boil-off mass loss, as in Table \ref{table:1x}. Note that our results set the largest possible relative difference at a certain mass and bolometric flux, as a planet born with an initial H/He mass fraction smaller than the value in Table \ref{table:1x} is not vulnerable to boil-off and less sensitive to the variable gravity effect. Our standard model uses the \citet{Guillot10} T-P profile and includes the variable gravity effect. We compare the effects of variable gravity versus constant gravity, as well as an isothermal atmosphere versus the Guillot atmosphere. The relative differences are calculated at 10 Myr. However, we find that the relative differences vary only by a factor of order unity if we evolve the models to 1 Gyr. 

For the 20 mbar radius (top panel), we find that the variable gravity effect is more significant for low-mass planets than for heavier planets, leading to a relative difference on the order of 10\%. The influence of the choice of the T-P profile is equally significant. For the 1 nbar radius (bottom panel), the variable gravity effect results in a greater relative difference compared to the 20 mbar radius, reaching up to 60\%. The choice of T-P profile also plays an important role, except for planets with $10 F_\oplus$ (second dashed curve from the left in the bottom panel) because the average temperature of the Guillot profile below 1 nbar is closer to $T_{\rm{eq}}$ than in other models (as shown in Figure \ref{rcb_comp}, where the average temperature is hotter than $T_{\rm{eq}}$), which makes the isothermal atmosphere a reasonable approximation for this specific flux value. For this curve, the 1 nbar radius is slightly overestimated relative to our standard atmosphere model, resulting in an opposite slope compared to other curves. In all other cases, we find that these physical choices systematically underestimate atmospheric radii due to the underestimated temperature and overestimated gravity.

\begin{figure}
\centering
\includegraphics[width=0.5\textwidth]{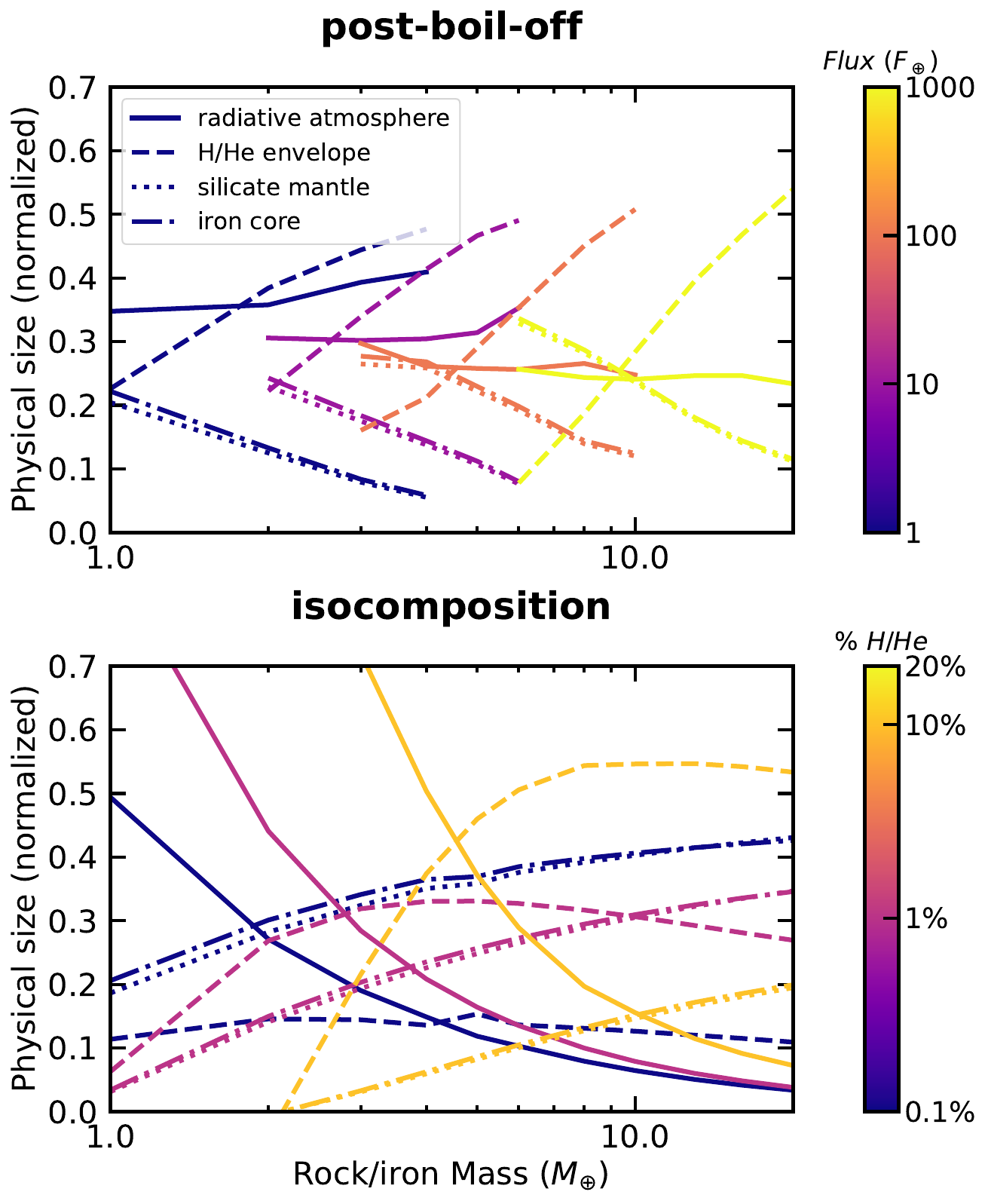} 
\caption{ We show the relative physical size of each structural layers at 10 Myr: radiative atmosphere below 20 mbar(solid), H/He envelope (dashed), silicate mantle (dotted) and iron core (dash-dot). They are normalized with the optical radius. In the top panel, the H/He mass fraction is calculated from a boil-off phase. We do not include planets that are not vulnerable to boil-off (with higher mass) and that are turned into a super-Earth (with lower mass). In the bottom panel, we vary the H/He mass fraction and fix the incident bolometric flux with $10 F_\oplus$ for each color-coded curve.
}
\label{composition}
\end{figure}

These effects highlight the important role of the radiative atmosphere in sub-Neptune structure modeling. We find that the radiative atmosphere constitutes a large fraction of a sub-Neptune's transit radius, up to 40\% for planets that experienced a boil-off with large atmospheric scale heights (Figure \ref{composition}). This effect is particularly significant for low-mass planets due to both the deep RCB and the large scale height of the radiative atmosphere. 

\citet{Howe14} also demonstrated the importance of the radiative atmosphere, assuming a constant scale height and constant RCB pressure at a fixed specific entropy, which finds a 20\% contribution to the transit radius of observed planets due to the atmosphere thickness. However, this scale height is evaluated at the observed transit radius with reduced gravity, overestimating the role of radiative atmosphere and underestimating the predicted envelope mass fraction.



Additionally, we present the optical radius at 1 Gyr calculated with the standard model as a function of rock/iron core mass, incident bolometric flux and H/He mass fraction in Table \ref{table:2}. Comprehensive radius grids, including the RCB radius, optical radius and 1 nbar radius for both pure H/He envelopes as well as 50$\times$ solar metallicity cases, across a broader range of planetary ages, are available online \footnote{https://doi.org/10.5281/zenodo.13985499} for further reference. In the 1 nbar and optical radius tables, planets susceptible to boil-off are excluded, as hydrostatic equilibrium no longer holds in the radiative atmosphere. A visualization tool is also available for interpolating the presented data \citep{Mardigras}.

In summary, we argue that variable gravity is a significant effect for sub-Neptunes and must be properly assessed with a comprehensive temperature profile. Other physical effects contribute substantially to the size of the radiative atmosphere, including the photodissociation in the upper atmosphere and the evolution of the RCB. Together, these factors make the radiative atmosphere a dominant component of a low- to -intermediate-mass sub-Neptune’s transit radius ($<10M_\oplus$).

\begin{deluxetable*}{c | c | cccccccc}
\tabletypesize{\scriptsize}
\tablewidth{0pt} 
\tablecaption{Sub-Neptune optical radius at 1 Gyr \label{table:2}}
\tablehead{
\colhead{Flux ($F_\oplus$)} & \colhead{ Mass ($M_\oplus$)}& \colhead{0.1\%} & \colhead{0.2\%} & \colhead{0.5\%} & \colhead{1.0\%} & \colhead{2.0\%} & \colhead{5.0\%} & \colhead{10.0\%} & \colhead{20.0\%}
} 
\startdata 
 1 & 1  & 1.63 & 1.82 & - & - & - & - & - & -\\      [0.5 ex]
 1 & 2 & 1.64 & 1.77 & 2.02	& 2.33 & 2.81 & - & - & -\\  [0.5 ex]
 1 & 3 & 1.73 & 1.83 & 2.04	& 2.28 & 2.66 & 3.56 & 4.88 & -\\  [0.5 ex]
 1 & 4 & 1.80 & 1.89 & 2.08 & 2.31 & 2.64 & 3.40 & 4.47 & 6.35\\   [0.5 ex]
 1 & 5 & 1.86 & 1.95 & 2.14	& 2.35 & 2.65 & 3.34 & 4.29	& 5.89\\   [0.5 ex]
 1 & 6 & 1.92 & 2.01 & 2.20 & 2.39 & 2.68 & 3.33 & 4.19	& 5.64\\   [0.5 ex]
 1 & 8 & 2.02 & 2.11 & 2.29	& 2.47 & 2.74 & 3.34 & 4.11	& 5.39\\   [0.5 ex]
 1 & 10 & 2.11 & 2.22 & 2.38 & 2.55	& 2.81 & 3.37 & 4.09 & 5.28\\   [0.5 ex]
 1 & 13 & 2.25 & 2.34 & 2.49 & 2.65	& 2.90 & 3.43 & 4.11 & 5.22\\   [0.5 ex]
 1 & 16 & 2.36 & 2.44 & 2.58 & 2.74	& 2.98 & 3.50 & 4.15 & 5.21\\   [0.5 ex]
 1 & 20	& 2.48 & 2.55 & 2.69 & 2.85	& 3.08 & 3.58 & 4.21 & 5.24\\   [1.5 ex]
\hline
 10	& 2 & 1.85 & 2.02 & 2.35 & - & - & - & -	& -\\     [0.5 ex]
 10	& 3 & 1.87 & 1.99 & 2.24 & 2.54	& 3.02 & - & - & -\\ [0.5 ex] 
 10	& 4 & 1.91 & 2.02 & 2.24 & 2.50	& 2.89 & 3.82 & -	& -\\ [0.5 ex]
 10	& 5 & 1.96 & 2.06 & 2.27 & 2.50	& 2.85 & 3.65 & 4.78 & -\\  [0.5 ex]
 10	& 6 & 2.00 & 2.10 & 2.31 & 2.52	& 2.84 & 3.57 & 4.57 & 6.25\\  [0.5 ex]
 10	& 8 & 2.09 & 2.19 & 2.38 & 2.57	& 2.87 & 3.52 & 4.37 & 5.79\\  [0.5 ex]
 10	& 10 & 2.17	& 2.28 & 2.45 & 2.63 & 2.91	& 3.51 & 4.30 & 5.57\\  [0.5 ex]
 10	& 13 & 2.30 & 2.40 & 2.55 & 2.72 & 2.98	& 3.55 & 4.26 & 5.44\\  [0.5 ex]
 10	& 16 & 2.41	& 2.49 & 2.64 & 2.80 & 3.05	& 3.59 & 4.28 & 5.39\\  [0.5 ex]
 10	& 20 & 2.52	& 2.59 & 2.74 & 2.90 & 3.14	& 3.66 & 4.32 & 5.39\\  [1.5 ex]
\hline
 100 & 3 & 2.15 & - & - & -	& - & - & - & -\\ [0.5 ex] 
 100 & 4 & 2.12	& 2.26 & 2.54 & - & - & - & - & -\\ [0.5 ex]
 100 & 5 & 2.13	& 2.25 & 2.51 & 2.80 & - & - & - & -\\  [0.5 ex]
 100 & 6 & 2.15	& 2.27 & 2.51 & 2.76 & 3.16 & - & - & -\\  [0.5 ex]
 100 & 8 & 2.21	& 2.33 & 2.54 & 2.76 & 3.10 & 3.87 & 4.90 & -\\  [0.5 ex]
 100 & 10 & 2.27 & 2.40 & 2.58 & 2.79 & 3.10 & 3.79 & 4.70 & 6.18\\  [0.5 ex]
 100 & 13 & 2.39 & 2.49	& 2.66 & 2.85 & 3.14 & 3.76	& 4.57 & 5.88\\  [0.5 ex]
 100 & 16 & 2.49 & 2.57	& 2.73 & 2.91 & 3.18 & 3.77	& 4.52 & 5.74\\  [0.5 ex]
 100 & 20 & 2.59 & 2.67	& 2.82 & 3.00 & 3.25 & 3.81	& 4.52 & 5.67\\  [1.5 ex]
\hline
 1000 & 6 & 2.54 & - & - & - & - & - & - & -\\  [0.5 ex]
 1000 & 8 & 2.52 & 2.69	& 2.95 & - & - & - & - & -\\  [0.5 ex]
 1000 & 10 & 2.55 & 2.69 & 2.92	& 3.20 & 3.63 & - & - & -\\  [0.5 ex]
 1000 & 13 & 2.61 & 2.72 & 2.93	& 3.17 & 3.54 & 4.36 & - & -\\  [0.5 ex]
 1000 & 16 & 2.67 & 2.77 & 2.96	& 3.19 & 3.52 & 4.26 & 5.23 & -\\  [0.5 ex]
 1000 & 20 & 2.74 & 2.83 & 3.02	& 3.22 & 3.53 & 4.21 & 5.09	& 6.52\\  [1.5 ex]
\enddata
\tablecomments{A dash symbol indicates that the planet cannot survive the boil-off phase with the corresponding mass fraction. The critical H/He mass fraction allowed by boil-off is based on Table \ref{table:1x}, rounded up.}
\end{deluxetable*}

\subsection{Implications for Inferred H/He Mass Fractions}
\label{res:comparison}
The prediction of larger planetary radii due to the radiative atmosphere in our model suggests that less H/He is needed to match a given planet's radius compared to previous models mentioned in the previous section (except for \citet{Howe14,Howe15}). To assess the implications of this, we applied our thermal evolution model to planets in the Kepler-11 system and directly compared our findings to those made by the \citet{Lopez12} model in Table \ref{table:kepler}.

We used the same planetary parameters as in \citet{Lopez12}, based on the Transit Timing Variations (TTV) measurements from \citet{Lissauer11}.  For consistency,we did not calculate error bars for our composition predictions. Our findings show that our model predicts envelope mass fractions usually 50\% lower for high- to intermediate-mass planets, with the largest differences observed for the low-mass Kepler-11f. This can be attributed to the stronger influence of variable gravity on planets with lower surface gravity. Notably, Kepler-11b likely lacks a convective envelope due to a cold interior and deep stellar heating penetration, where the radiative atmosphere, with a surface pressure $>$100 bar, constitutes approximately 25\% of its total radius. Our analysis show that these configuration may not be rare in the observed low-mass sub-Neptunes.

Another factor contributing to the lower H/He mass prediction in our model is the enhanced role of rock/iron luminosity as an energy source for inflating a planet. This energy primarily comes from the latent heat and the higher interior temperatures due to the adiabatic rock/iron profile. As illustrated in the bottom panels of Figure \ref{evolution}, our model's hotter interior contrasts significantly with previous models that assume an isothermal core, leading to a more inflated planetary radius. 

\begin{figure}
\centering
\includegraphics[width=0.45\textwidth]{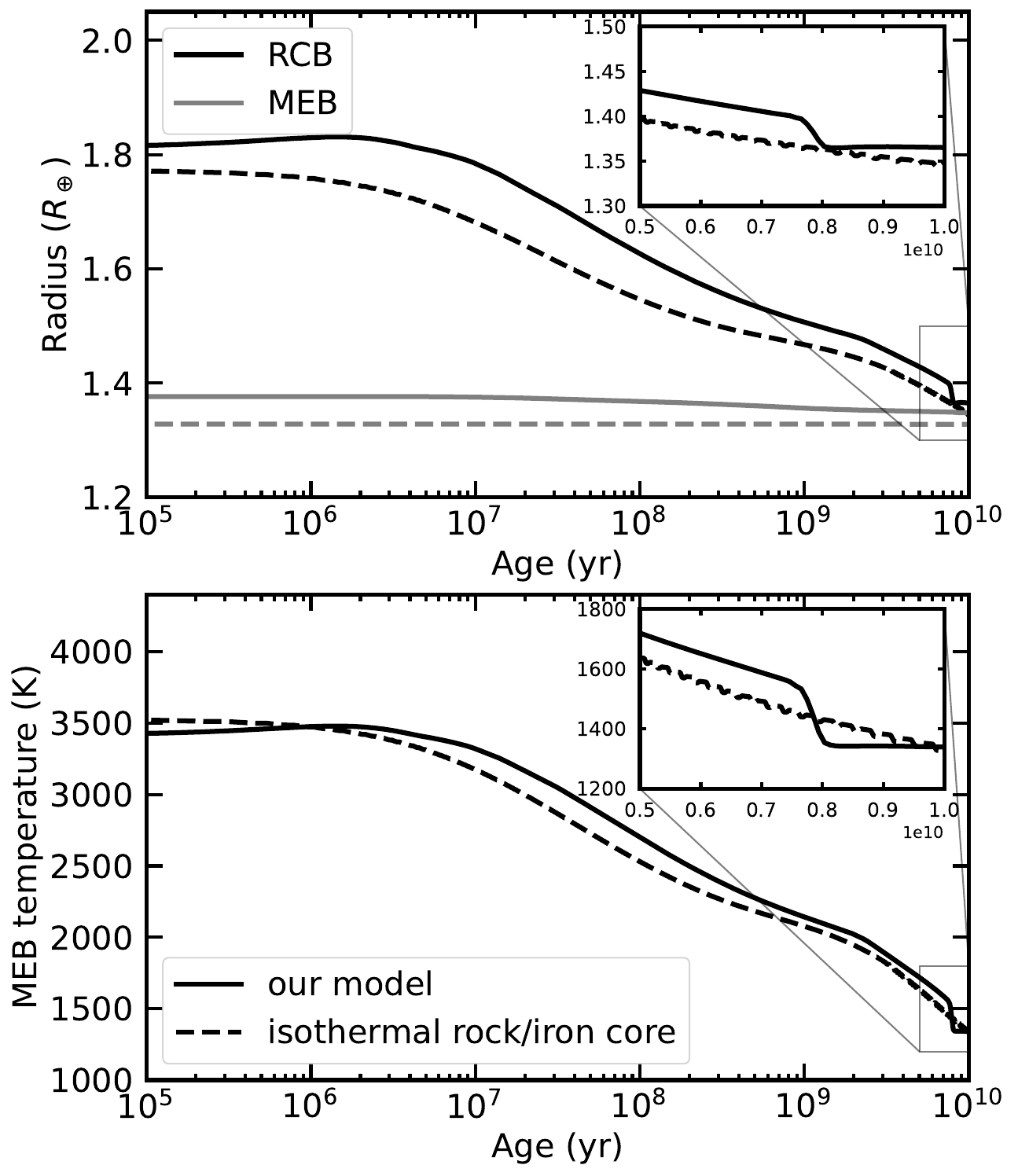} 
\caption{ A comparison of the evolution of the RCB radius (top panel) and core temperature (bottom panel) between our model (solid), which accounts for adiabatic rock/iron cores with parameterized convection, and previous models assuming an isothermal core (dashed). Both models have the same mass ($3M_\oplus$), incident flux ($100F_\oplus$) and H/He mass fraction (0.2\%), with identical initial envelope specific entropy. See text for detailed discussion of the differences and their implications.
}
\label{modelcomparison}
\end{figure}

In Figure \ref{modelcomparison}, we compare two models assessed with identical planetary parameters and initial conditions with our new model (solid) using parameterized convection and the isothermal assumption (dashed). The EOS for the rock/iron core for the isothermal model follows the work by \citet{Lopez12}. The top panel presents the radii evolution at both the RCB in black and the rock/iron core surface (MEB) in gray, while the bottom panel shows the evolution of the surface temperature of the rock/iron core. Note that our model predicts a much hotter central temperature, often several times higher than that of the isothermal model. Initially, our model predicts a larger RCB radius because the rock/iron core is molten, causing a 3\% size difference compared to the isothermal models. The cooler initial core temperature stems from the shallower location of the rock/iron surface. Early in the evolution, the planet in our model remains 10\% larger with a hotter interior due to the contribution of rock/iron luminosities. However, this effect diminishes over time as mantle convection becomes less efficient due to solidification. Notably, after 8 Gyr, the planet's envelope shrinks rapidly when the mantle fully solidifies, consistent with the processes described in Section \ref{res:core}. 

\begin{deluxetable*}{c | ccc | c | ccc}
\tabletypesize{\scriptsize}
\tablewidth{0pt} 
\tablecaption{Comparison of composition predictions for Kepler-11 planets \label{table:kepler}}
\tablehead{
\colhead{Planet name} & \colhead{$M_p$ ($M_\oplus$)} & \colhead{$R_p$ ($R_\oplus$)} & \colhead{$F_{bol}$ ($F_\oplus$)} & \colhead{$f_{\rm{env}}$ from Lopez12 (\%)} & \colhead{$f_{\rm{env}}$ (\%)} & \colhead{$f_{\rm{atm}}$ (\%)} & \colhead{$P_{\rm{surf}}$ (bar)}
} 
\startdata 
Kepler-11b & 4.3 & 1.97 & 137 & 0.3 & 0 & 0.01 & 220\\
Kepler-11c & 13.5 & 3.15 & 101 & 4.6 & 2.7 & 0.04 & 1453\\
Kepler-11d & 6.1 & 3.43 & 45 & 8.2 & 4.4 & 0.13 & 1210\\
Kepler-11e & 8.4 & 4.52 & 30 & 17.2 & 12.2 & 0.17 & 919\\
Kepler-11f & 2.3 & 2.61 & 18 & 4.1 & 0.97 & 0.13 & 1102\\
\enddata
\tablecomments{We present a comparison of planetary composition predictions between our model (rightmost columns) and the \citet{Lopez12} model for Kepler-11 planets. The key planetary parameters used for the evaluation, total mass $M_p$, optical radius $R_p$ and incident bolometric flux $F_{rm{bol}}$, are sourced from the TTV measurements by \citet{Lissauer11}. Our model predicts the envelope mass fraction $f_{\rm{env}}$ and the mass fraction of the radiative atmosphere $f_{\rm{atm}}$ independently. These combine to give the total H/He mass fraction $f_{\rm{HHe}} = f_{\rm{env}} + f_{\rm{atm}}$.  In most cases, the envelope mass dominates the total H/He mass, except for planets with a very thin envelope, where $f_{\rm{env}}\sim 0.1\%$. The surface pressure $P_{\rm{surf}}$ is evaluated at the RCB for H/He-rich sub-Neptunes and at the mantle surface for envelope-free super-Earths. In the case of super-Earths, we model the structure by assuming a radiative atmosphere on top of the rock/iron cores. This treatment allows us to extend our approach to these envelope-free planets, offering a more refined prediction for their surface conditions and interior structures.}
\end{deluxetable*}

\section{Discussion} \label{sec:disc}
\subsection{Enhanced Boil-Off Mass Loss}
\label{disc:boil-off}
The strength of boil-off mass loss is directly influenced by the planetary radius, which is primarily determined by the envelope specific entropy (TFM24). Planets with larger RCB radii exhibit a higher density at the sonic point, resulting in enhanced mass loss rates. This effect occurs when the rock/iron core delays the thermal contraction. However, TFM24 demonstrated that this core-enhanced mass loss effect is limited, as thermal evolution is largely decoupled from mass loss, resulting in negligible interior cooling rates during the boil-off process. In this study, we reexamine this behavior using an improved model for rock/iron evolution.

\begin{figure}
\centering
\includegraphics[width=0.45\textwidth]{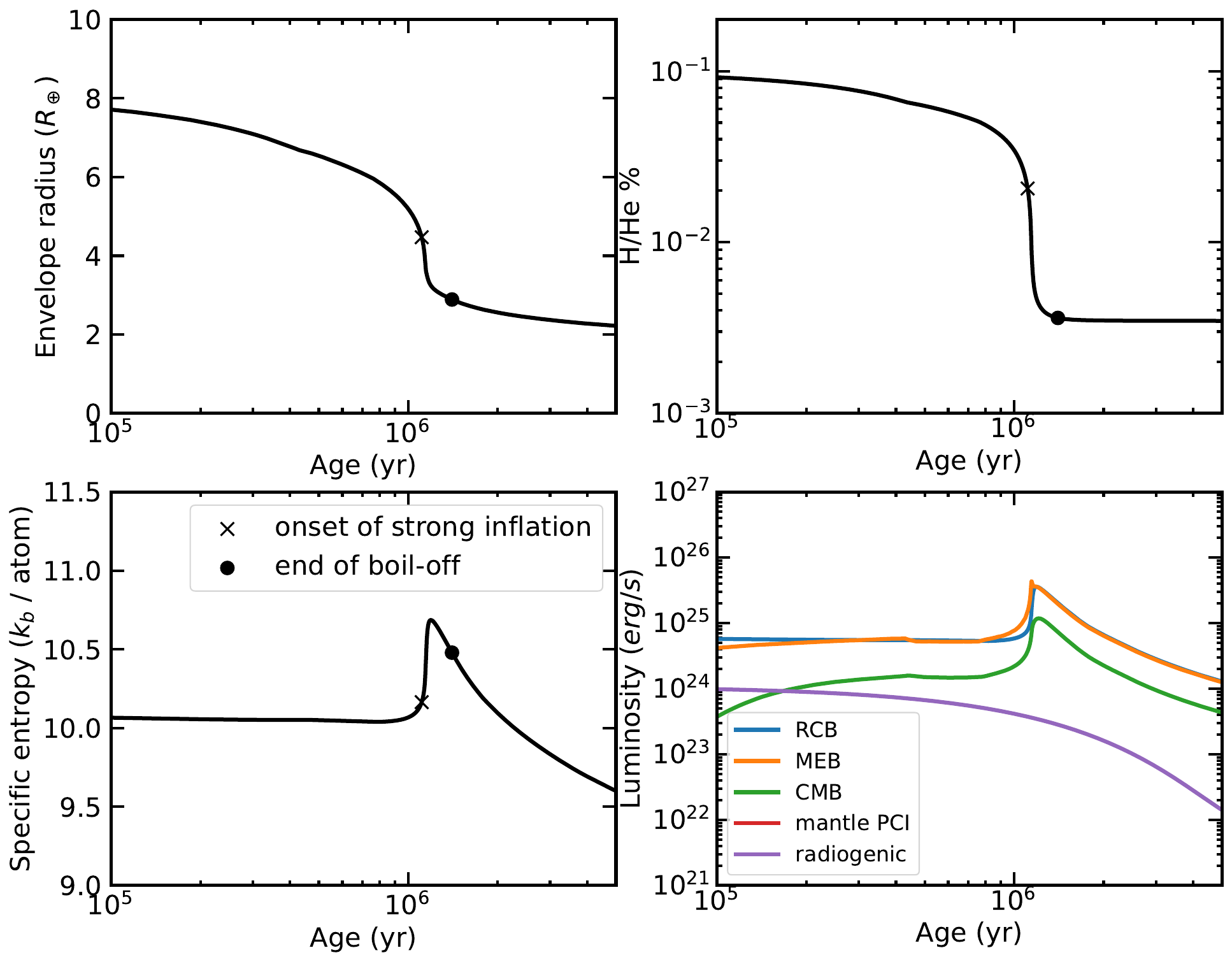} 
\caption{ We show the evolution tracks of the envelope radius (top left panel), H/He mass fraction (top right), envelope specific entropy (bottom left) and energy budget (bottom right) for a planet in boil-off. The planet has a core mass of 3$M_\oplus$, initial H/He mass fraction of 10\% and incident bolometric flux of 30 $F_\oplus$. The onset of strong inflation is marked with cross shapes, defined by Eq. \ref{coupling}. The end of boil-off is shown with circles. The wind is energetically constricted by the radiative cooling of the envelope, leading to a relatively long mass loss timescale. Note that, compared to Figure \ref{evolution},  the planet's mantle remains molten for the duration of evolution shown, so the mantle PCI (red) is not relevant here. For further details, refer to the discussion in the text.
}
\label{boiloff}
\end{figure}

We find that energy is released efficiently from the rock/iron interior. As an envelope loses mass and undergoes adiabatic expansion, the temperature at its bottom decreases rapidly, resulting in an increased temperature contrast at the MEB. This correspondingly increases the energy transport rate via convection at the MEB and in turn heats the envelope at a significant rate (orange in the bottom right panel of Figure \ref{boiloff}), leading to thermal inflation. This phenomenon occurs for planets that are highly susceptible to boil-off, characterized by their large atmospheric scale heights (meaning low-mass or highly irradiated). In contrast, this inflation is minimal in high-mass, less irradiated planets, because they possess larger energy reservoirs within their envelopes, along with insignificant changes in the MEB temperature due to their lower susceptibility to boil-off, consistent with the behavior reported in TFM24. We did not observe actual radius inflation in the boil-off phase (top left), as the rate of radius shrinkage due to mass loss (top right) exceeds the rate of thermal inflation (our term for H/He envelope entropy increase, bottom left).
 
Overall, we find thermal inflation contributes to a moderate mass loss enhancement due to the luminosity of the rock/iron core. The final mass fraction is reduced by more than a factor of 3 for low-mass planets ($\leq 2M_\oplus$), while the impact on massive planets ($\geq 5 M_\oplus$) is marginal. Compared to TFM24, the thermal inflation further decreases the final mass fraction by about 50\% for low- to intermediate-mass planets, as the new rock/iron modeling corresponds to a larger rock/iron energy reservoir.

However, the influence of thermal inflation and rock/iron luminosity is considered limited due to several reasons: (1) Since most of the initial H/He mass is lost for low-mass planets, a factor of 3 enhancement in the final mass fraction corresponds to a minimal change of less than 10\% in the time-integrated mass loss. (2) In many cases, boil-off alone is sufficient to completely strip a sub-Neptune's envelope, even without considering the contribution from core luminosity. (3) The reduction in final mass fraction due to rock/iron luminosity is balanced by an increase in envelope specific entropy, leading to a small difference (less than 10\%) in the post-boil-off radius. (4) The final mass fraction is determined by the specific entropy at the end of boil-off. After inflation, the increased radiative luminosity from the inflated interior accelerates cooling, effectively shutting off the boil-off. As a result, the final specific entropy is only slightly higher than the initial entropy, by less than $0.5 , k_b/\rm{atom}$. This leads to a limited enhancement of time-integrated mass loss. 
(5) Thermal inflation typically occurs late, usually after more than half of the initial H/He mass is lost. Its effects on mass loss become significant only after about 90\% of the initial H/He mass is lost (around 1 Myr, marked with cross shapes in Figure \ref{boiloff}). This timing aligns with the expectation that heating from the rock/iron interior becomes influential on the thermal evolution of the envelope once the thermal energy of the rock/iron exceeds the gravitational binding energy of the envelope, which correspond to an inconsiderable envelope mass. (6) While the energy-limited wind in Eq. \ref{mdot_elim} exhibits stronger coupling with thermal inflation compared to a non-energy-limited Parker wind, the coupling is still relatively weak, as the mass loss occurs about one order of magnitude faster than the thermal evolution timescale throughout most of the boil-off phase. Strong coupling only happens when the thermal inflation timescale is shorter than both the mass loss timescale $t_{\dot{M}}$ and the planetary age $t$, defined as:
\begin{equation}
\label{coupling}
t_{\rm{inf}} \equiv \frac{U}{L_{\rm{MEB}}} \lesssim \min(t,t_{\dot{M}})
\end{equation}
where $U$ is the total gravitational binding energy of the envelope. However, this strong coupling phase only lasts for approximately $10^5$ years as shown in the bottom left of Figure \ref{boiloff}.

In our model, we find that it is not necessary to incorporate disk dispersal modeling. For all of the planets modeled, their mass loss occurs in the intrinsic-limited regime (Section \ref{subsec:boil-off}). As a result, boil-off is inefficient until $\sim$ 1 Myr (Figure \ref{boiloff}). The boil-off phase typically continues for an additional one to two Myrs. Due to this behavior, we suggest that the boil-off mass loss is not sensitive to detailed disk evolution, given the typical $10^5$ year inner disk dispersal timescale. 
In a more comprehensive scenario with an extended disk dispersal time, a real planet would efficiently contract during the disk dispersal phase, adjusting its contraction timescale and specific entropy to match the dispersal time. In this case, the boil-off evolution can be modeled with a lower initial specific entropy set by the disk dispersal time, which results in a disk-dispersal limited scenario. Therefore, our assumptions of using a steady-state Parker wind model and neglecting disk dispersal are proven to be appropriate.

\subsection{Presence of Dynamo in the Liquid Iron}
\label{disc:dynamo}
In this section, we quantitatively discuss the conditions necessary for sustaining a dynamo in the liquid iron core. A common approach to characterize the strength of magnetic induction relative to magnetic diffusion is through the magnetic Reynolds number $R_m$:
\begin{equation}
\label{rm}
R_m = \frac{UD}{\lambda} = \mu_0\sigma UD
\end{equation}
where $\lambda$ is magnetic diffusivity, $U$ is the characteristic convective velocity and $D$ is the thickness of the liquid outer core. Studies have shown that for a dynamo to operate, $R_m$ must exceed a critical value of 50. Here, $\lambda$ is related to the vacuum permeability $\mu_0$ and electrical conductivity $\sigma$. Several scaling laws have been proposed for $U$, with \citet{Christensen10} demonstrating that the following relationship best fits simulation data:
\begin{equation}
U \propto \left(\frac{F_{\rm{conv}}}{\rho H_T}\right)^{2/5} \left(\frac{L}{\Omega}\right)^{1/5}
\end{equation}
We employ this form for $U$, noting that variations in scaling laws typically affect $R_m$ only by an order of unity. Here, $L$ is the length scale, taken to be the core radius $R_c$, and $\Omega$ is the rotational rate, assumed to correspond to a rotation period of 1 day. The precise value of $\Omega$ does not significantly affect our results. The temperature scale height $H_T=c_P/(\alpha g)$ is evaluated at each mass shell, with the average value taken for the analysis. The convective energy flux $F_{\rm{conv}}$ that triggers dynamo operation only occurs when the conductive flux along the adiabat is insufficient to transport the cooling energy of the core. This leads to:
\begin{equation}
\label{fluxconv}
F_{\rm{conv}} = F_{\rm{CMB}} - F_{\rm{cond}} = F_{\rm{CMB}} - k_c \frac{\partial T}{\partial r} \bigg |_{\rm_{CMB,c}}
\end{equation}
where $k_c$ is the thermal conductivity of the iron core. The last term is the adiabatic temperature gradient of the core evaluated at the CMB:
\begin{equation}
\frac{\partial T}{\partial r} \bigg |_{\rm_{CMB,c}}=\frac{\alpha_{\rm{CMB,c}}g_{\rm{CMB}}T_{\rm{CMB,c}}}{c_{\rm{Pc}}}
\end{equation}
where $\alpha_{\rm{CMB,c}}$, $g_{\rm{CMB}}$ and $T_{\rm{CMB,c}}$ are the thermal expansivity, gravity and temperature of the core at the CMB, respectively.

\begin{figure}
\centering
\includegraphics[width=0.5\textwidth]{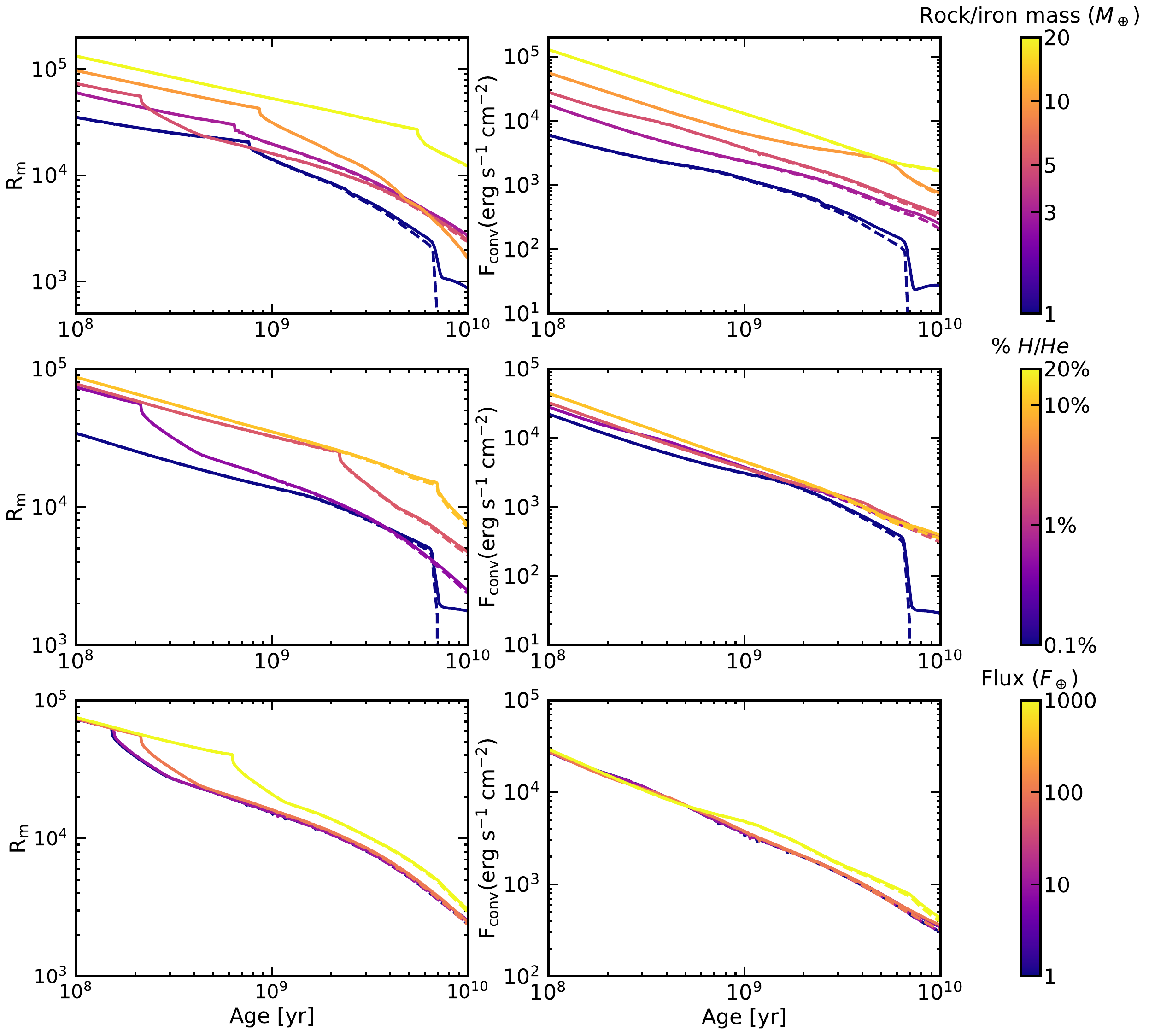} 
\caption{ The evolution of the magnetic Reynolds number $R_m$ (left panels) in the liquid outer iron core and the convective flux $F_{\rm{conv}}$ evaluated at the CMB (right). The models used are consistent with those in Figure \ref{solidification}. We assess two core thermal conductivity estimates: an intermediate $4\times10^6\ \rm{erg\ cm^{-1}\ K^{-1}}$ (solid lines) and a high $1\times10^7\ \rm{erg\ cm^{-1}\ K^{-1}}$ (dashed). We find that a dynamo exists in the iron core as long as the mantle surface remains in the liquid phase. Once the mantle surface solidifies (see Figure \ref{solidification}), the cooling rate of the iron core sharply declines, lowering the vigor of convection in the core. In this case, we find that dynamo operation starts to become sensitive to the conductivity choice.
}
\label{dynamo}
\end{figure}

In Figure \ref{dynamo}, we illustrate the evolution of $R_m$ on the left and $F_{\rm{conv}}$ on the right. Each panel varies rock/iron mass, H/He mass fraction, and incident bolometric flux, with different parameters indicated by distinct colors. A wide range of values for $k_c$ has been reported from simulation studies. Here, we evaluate an intermediate value of $4\times10^6\ \rm{erg\ s^{-1}\ cm^{-1}\ K^{-1}}$ \citep[solid]{Konopkov16} and a high value of $1\times10^7\ \rm{erg\ s^{-1}\ cm^{-1}\ K^{-1}}$ \citep[dashed]{Pozzo12} to assess their effects. There are two key limits that constrain $R_m$: if the iron core is completely solidified, dynamo action ceases. Additionally, the liquid outer core may stop being convective when conduction dominates energy transport, at which point dynamo action would also cease.

All modeled planets in the figure retain a liquid outer iron core by 10 Gyr. Moreover, order of magnitude calculations indicate that as long as a convective layer is present in the liquid iron, $R_m$ can readily exceed the critical value. Our numerical results show that $R_m$ typically ranges from $10^3-10^5$, well surpassing the critical threshold before $F_{\rm{conv}}$ drops below $10^{-3}\ \rm{erg\ cm^{-2}}$.

However, we find that the presence of convection in the liquid-phase iron core is highly sensitive to thermal conductivity. With the intermediate thermal conductivity $k_c$ (solid), the convective flux is always present to sustain a dynamo for all modeled planets. As $k_c$ increases (dashed), iron convection ceases at a transition time of a few Gyr  for the lowest-mass planets (1 $M_\oplus$, top) and those with the smallest H/He mass fraction (0.1\%, middle). This transition coincides with the complete solidification of the mantle surface (see right panels of Figure \ref{solidification}), as the slowed mantle cooling consequently decelerates the core's thermal evolution, reducing the CMB flux and ultimately quenching core convection. Notably, at both $k_c$ values, the conductive flux within the iron core remains negligible compared to the convective flux until the mantle surface solidifies.

In summary, we suggest that constraining the thermal conductivity of the iron core is crucial for improving our understanding of dynamo operation in low-mass, thin-envelope planets.

\subsection{Other Complications and Uncertainties} \label{disc:inter}
\subsubsection{Interaction between species}
As any modeling framework evolves towards a closer approximation of reality, some aspects can receive detailed attention, and other aspects are left to later work. Here, we did not consider the complex interaction between the magma ocean and the overlying envelope, which can dynamically modify both their compositions and temperature profiles. These processes, including volatile exchange at the MEB and the oxidation of hydrogen in the envelope that leads to water contamination, have been explored in numerous studies.  The evolution of water-rich planets demonstrates different observable features compared to that of water-free sub-Neptunes \citep{Zeng19,Luque22}. Such effects influence gas retention in the envelope, with feedback potentially impacting thermal evolution and solidification of the rock/iron core, particularly in planets with thin envelopes. However, it remains unclear whether the water dissolves into the iron core \citep{Luo24}, the mantle \citep{Schlichting22,Kite20,Vazan22}, or the envelope, or forms a distinct water layer \citep{Rogers10}. Such volatile impurities in the rock/iron core may alter its thermodynamic properties (e.g., melting temperature, heat capacity, and thermal conductivity), introducing uncertainties in the planet's thermal evolution. These complexities highlight future research directions.

\subsubsection{Differentiation}
At early ages during the formation phase, a sub-Neptune's rock/iron core is likely in a homogeneous mixture state, owing to frequent impacts during accretion. Over time, the iron content from impactors separates from the silicate material due to the density contrast, eventually forming a distinct iron core beneath a silicate mantle. This redistribution of mass releases energy comparable in magnitude to the total thermal energy of the rock/iron component $E_{\rm{ric}}$. However, core-mantle differentiation is not explicitly modeled in this work, nor in most previous sub-Neptune evolution models.

Although extensively studied in terrestrial planets, the differentiation process remains poorly understood in the context of sub-Neptunes. In this context, we consider two possible scenarios based on current theoretical understanding.

In the classical Earth-like models of differentiation, iron droplets rain out within the surface magma ocean, accumulating to form a liquid iron pond, and subsequently, sink through the overlaying solid mantle into the developing iron core \citep{Stevenson90,Rubie07}. However, our results suggest that sub-Neptunes typically retain a fully molten mantle for Gyr timescales (Figure \ref{solidification}), which implies that iron core formation may occur rapidly and directly, within a nearly instantaneous timescale. In this scenario, the energy released by differentiation contributes only modestly to the initial entropy of the envelope and interior, since the envelope energy $E_{\rm{env}}$ usually dominates over the energy of the rock/iron core $E_{\rm{ric}}$. We tested this effect by shortening the initial contraction timescale to 3 Myr, which results in a 10–20\% reduction in the post-boil-off envelope mass fraction. This alters the evolution of the envelope's specific entropy by only a negligible amount (see middle-left panel of Figure \ref{entropy-rcb}).

In contrast, \citet{Wahl15} propose a different differentiation mechanism based on the miscibility of iron and silicate/oxide mantle. They identify a critical temperature above which a homogeneous mixture of iron and mantle materials is stable. This critical temperature is relatively low and is reached within the solid mantle phase at high pressures, implying that differentiation could be delayed until later evolutionary stages. If this delayed scenario occurs, the differentiation energy would be released on Gyr timescales, when $E_{\rm{ric}}$ may exceed $E_{\rm{env}}$, potentially leading to elevated interior temperatures and delayed cooling. Given these uncertainties, the role of delayed differentiation in sub-Neptunes warrants further investigation in future studies.

\subsubsection{Silicate vapor}
Further uncertainty in radius evolution arises from the potential condensation of silicate vapor in the envelope, sourced from planetary accretion. Over Gyr timescales, silicate vapor may settle into the rock/iron core, releasing energy and potentially inflating the planetary radius \citep{Vazan23}. This process could decrease the estimated H/He mass fraction and enhance the mass loss rate \citep{Vazan24}.

\section{Conclusions} 
We develop a new python-based sub-Neptune evolution model that particularly focuses on both the interior physics of the rock/iron core and the detailed physical structure of radiative atmosphere. With the new model, we present the planetary structure from the molten center to nbar pressure levels and evolutionary trajectory and identify a region in parameter space that are prohibited by the boil-off mass loss. We emphasize on the following findings:
\begin{enumerate}
   \item{ The radiative atmosphere is a crucial layer in the physical structure of sub-Neptunes, contributing large fraction, up to 40\%, of their optical radius due to the large scale height in the layer. This phenomenon is particularly significant at lower pressure levels and for low-mass, highly irradiated planets. 
   }
   \item{The variable gravity effect (resulting in an outwardly increasing scale height) further amplifies the thickness of the radiative atmosphere. Radius assessment with this effect is sensitive to the temperature-pressure profile at each pressure level. The vertically variable mean molecular weight also plays an important role, which is critical for determining the nbar radius. These factors lead to a large relative difference (several tens of percent at 20 mbar, and much higher at 1 nbar) between models with and without them. See Section \ref{res:rad} for more details. Due to the large size of the radiative atmosphere, our model generally predicts a lower amount of H/He mass at a given radius compared to models that do not account for these effects. 
   }
   \item{ In the presence of a boil-off phase, we have constrained the maximum H/He mass fraction permissible for post-boil-off evolution, significantly reducing the dependence on and uncertainty associated with initial conditions. The inefficiency of radiative cooling energy to power the wind in the deep atmosphere results in a delayed boil-off (see Section \ref{disc:boil-off} for discussion). For planets that are highly susceptible to boil-off (losing more than 90\% of the initial mass), the envelope can experience thermal inflation driven by the luminosity of the rock/iron core, which is further enhanced by mass removal and the resulting increased temperature contrast at the MEB. However, this inflation is brief and typically reduces the final mass fraction by a factor of 2 to 3. For planets retaining more than 10\% of their initial envelope mass, a complete decoupling between thermal evolution and mass, as observed in TFM24, is recovered, leading to convergent final mass fractions.
   }
   
   \item{ Sub-Neptunes exhibit markedly different thermal evolutionary behaviors compared to terrestrial planets, primarily because their envelopes maintain the silicate mantle surface in a warm, liquid state for an extended duration. In sub-Neptunes, the cooling of the rock/iron core is largely constrained by radiative cooling at the top of the envelope. In contrast, terrestrial planets depend on their crusts for heat insulation. As a result, the onset of solidification processes in the mantle and core of sub-Neptunes occurs significantly later than in terrestrial planets.
   }
   
   \item{ Sub-Neptunes characterized by high insolation, substantial rock/iron core masses, and thick envelopes exhibit hotter interiors, resulting in solidification timescales that often exceed Gyrs. The solidification of the iron core slows considerably after the bottom of the mantle crystallizes, as solid convection is far less efficient than magma convection. Once the mantle surface solidifies, the thermal contraction of the envelope accelerates due to reduced heating from the rock/iron core, leading to a decoupling of thermal evolution between the rock/iron core and the envelope. While we are confident in the relative solidification timescales of the iron core among sub-Neptunes with various parameters, the absolute solidification timescale remains uncertain due to its high sensitivity to chemical impurities within the core. 
   }
   
   \item{ In low-mass planets, mantle solidification progresses strictly from the bottom up. In contrast, in heavier planets, it initiates in the middle of the lower mantle and quickly dominates the entire lower mantle. After this point, the lower mantle solidifies nearly simultaneously and the phase change interface ($\phi = 0.5$) advances upward. This behavior may be significantly influenced by the EOS employed, warranting further studies for confirmation.
   }

   \item{ Metal-enrichment in the envelope does not significantly impact its overall size of the H/He envelope, as the increased density from heavier elements is offset by the elevated temperature of the interior. However, this enrichment significantly reduces the optical radius by decreasing the radiative atmospheric scale height. Due to the higher mean molecular weight, a metal-rich envelope is less susceptible to boil-off, allowing it to retain two to three times more of its initial mass compared to pure H/He envelopes. These factors render the effects of contamination distinguishable in the context of planetary size distribution.
   }

   \item{ We conducted order-of-magnitude calculations, indicating that dynamo operations can persist for extended periods in sub-Neptunes, typically exceeding 10 Gyr, except for planets with masses $\leq 1M_\oplus$  and those with less substantial envelope masses. Dynamo activity continues as long as the mantle surface remains liquid, which ensures a sufficiently high heat transport rate at the CMB, thereby maintaining vigorous iron convection. However, after solidification at the mantle surface (MEB), the existence of the dynamo becomes highly sensitive to the thermal conductivity of the iron core.
   }
   
\end{enumerate}

\bibliography{reference}{}
\bibliographystyle{aasjournal}



\end{document}